\documentclass[fleqn,usenatbib]{rasti}

\usepackage{newtxtext,newtxmath}

\usepackage[T1]{fontenc}

\DeclareRobustCommand{\VAN}[3]{#2}
\let\VANthebibliography\thebibliography
\def\thebibliography{\DeclareRobustCommand{\VAN}[3]{##3}\VANthebibliography}

\usepackage{graphicx}	
\usepackage{amsmath}	
\usepackage{graphicx}
\usepackage{subcaption}
\usepackage{array}
\usepackage{booktabs}
\usepackage{siunitx}
\usepackage{multirow}
\usepackage{multicol}
\usepackage{mathrsfs}
\usepackage{rotating}
\usepackage{pdflscape}

\title[Transfer learning for transient search]{Transfer learning for transient search with small-field optical survey telescopes}

\author[Pranshu et al.]{
Kumar Pranshu,$^{1,2,3}$\thanks{E-mail: kumarpranshu86@gmail.com}
Kuntal Misra,$^{1}$
Rithesh A,$^{4,5}$
Jean Surdej,$^{1,3}$
and Sarvesh Kumar Yadav$^{1,2}$
\\
$^{1}$Aryabhatta Research Institute of Observational Sciences (ARIES), Manora Peak, Nainital, 263001, India\\
$^{2}$Department of Applied Optics and Photonics, University of Calcutta, Kolkata, 700106, India\\
$^{3}$Institute of Astrophysics and Geophysics, University of Li\`{e}ge, All\'{e}e du 6 Ao$\hat{\rm u}$t 19c, 4000 Li\`{e}ge, Belgium\\
$^{4}$Department of Physics, Loyola College, Nungambakkam, Chennai, 600034, Tamilnadu, India\\
$^{5}$Department of Physics, School of Advanced Sciences, Vellore Institute of Technology, Katpadi, Vellore, Tamilnadu, 632041, India\\
}

\date{Accepted 2026 June 05. Received 2026 April 10; in original form 2025 October 08}

\pubyear{\the\year{}}

\begin{document}
\label{firstpage}
\pagerange{\pageref{firstpage}--\pageref{lastpage}}
\maketitle

\begin{abstract}
The advent of optical sky surveys has enabled several automated programs for searching transients. Many of these programs extensively use supervised machine learning (ML) algorithms to automate these searches. Effective implementation of such a strategy has an advantage over non-automated methods of transient search in terms of reduced manual labour and reporting latency. Training the relevant ML algorithms often requires extensive labelled training datasets that might not be readily available for new or small field-of-view survey telescopes. 
Transfer Learning (TL) is an ML technique that is often employed to address this issue by transferring knowledge from a pre-trained model, trained on an extensive dataset for a related task, to enhance performance on a new task with a limited dataset available. This paper demonstrates TL for a Convolutional Neural Network (CNN)-based real/bogus classifier model for transient detection between extensive publicly available image data from the Zwicky Transient Facility (ZTF) and a small and labelled dataset from the 4-m International Liquid Mirror Telescope (ILMT). The same technique was employed to train two different types of transient alert classifiers to characterise the detected candidates based on detection image stamps into 3 and 4 classes, respectively. The resulting model for the real/bogus classifier achieved an accuracy of 97.3\% on the test dataset. Additionally, an accuracy of 92.9\% was achieved for the 3-class classifier and 85.6\% for the 4-class classifier. Furthermore, the statistical significance of the effectiveness of this technique was established with an unpaired t-test between TL models and baseline models trained without TL.         
\end{abstract}




\section{Introduction}

The advent of automated sky surveys has revolutionised time-domain astronomy by enabling discoveries of numerous and often rare transient events across diverse classifications. These transient events--ranging from supernovae (SNe) to optical counterparts of gamma-ray bursts (GRB)--can benefit from immediate follow-up observations to capture the critical phase of their evolution. Several large-scale survey programs like the Zwicky Transient Facility\footnote{https://www.ztf.caltech.edu/} \citep[ZTF;][]{2019PASP..131a8002B}, Panoramic Survey Telescope and Rapid Response System\citep[Pan-STARRS;][]{2016arXiv161205560C}, Asteroid Terrestrial-impact Last Alert System \citep[ATLAS;][]{2018PASP..130f4505T}, Gravitational-wave Optical Transient Observer \citep[GOTO;][]{10.1093/mnras/stac013} and recently Vera Rubin Observatory (VRO) to conduct the Legacy Survey of Space and Time \citep[LSST;][]{2019ApJ...873..111I} have been established to that end. Alongside the larger surveys, smaller optical surveys can utilise high survey cadence to contribute towards more specific science cases like searching and probing optical counterparts of fast-fading transients such as fast blue optical transients \citep[FBOTs;][]{2014ApJ...794...23D, 2018ApJ...865L...3P, 2019MNRAS.484.1031P}, fast radio bursts \citep[FRBs;][]{2007Sci...318..777L, 2013Sci...341...53T}, GRBs \citep{1973ApJ...182L..85K,1992Natur.355..143M,1997Natur.386..686V}, X-ray transients \citep{1997Natur.387..783C}, and gravitational wave transients \citep{Soares_Santos_2017}.

Specifically, an optical survey telescope acquires large volumes of image data per observation night. Many automated transient search programs implement machine learning (ML)--based algorithms to facilitate the search for transients in data from such surveys. Techniques like support vector machines (SVM), artificial neural networks (ANN), and random forest classifiers (RFC) have been implemented in past and present surveys \citep{2007ApJ...665.1246B} like Nearby Supernova Factory \citep{2004NewAR..48..637W}, Palomar Transient Factory \citep[PTF;][]{2009PASP..121.1395L}, Intermediate PTF \citep[iPTF;][]{10.1093/mnras/stt1306} and ZTF \citep{2019PASP..131c8002M}. These techniques separate real transient candidates from artefact/noise using candidate-level descriptive features like shape parameters, magnitudes, astrometric accuracy, etc, and hence, have to be trained with datasets of such features. More recent works have explored the implementation of convolutional neural networks \citep[CNN;][]{726791} on image data to find transients \citep{2017PASA...34...37A,Cabrera_Vives_2017,10.1093/mnras/stx2161,Duev_2019,Mahabal_2019, 2020MNRAS.497.2641T, refId0, 2023AJ....166..115A, S-PLUS}. These CNN-based implementations use datasets consisting of image cutouts containing real transients and non-transient artefacts.

A common approach for isolating transient sources in optical images involves image subtraction \citep{1998ApJ...503..325A,Bramich_2008,Cao}. This technique removes obscuring objects like host galaxies and field stars by subtracting a template/reference image from the science image, resulting in a difference image. The transient source appears as a point source in the difference image, which also contains many artefacts. In a CNN-based strategy for transient detection, a trained CNN is leveraged to separate the `real' transient events from the `bogus' detections or artefacts. The CNNs are trained with an extensive ground truth dataset of detected transients and observed artefacts within the image repository of the respective transient survey. This usually assumes the availability of manually labelled, large-scale pre-existing images corresponding to such events, acquired with the survey telescope. The dataset should appropriately reflect the extensive diversity in the nature of such detections. Such vast and diverse datasets might be difficult to compile for small-scale transient search programs or newly commissioned telescopes for conducting optical surveys. Training a deep learning-based classifier like CNN with an insufficient dataset risks it towards overfitting, where it performs well on the training data or other very similar data but fails to generalise on the `unseen' test data. 

One such survey is the ongoing transient search using the 4-m International Liquid Mirror Telescope\footnote{https://www.aries.res.in/facilities/astronomical-telescopes/ilmt}\textsuperscript{,}%
\footnote{http://www.ilmt.ulg.ac.be/home/} \citep[ILMT;][]{2025A&A...694A..80S}. It is a zenith-pointing, optical survey telescope in India with a field-of-view (FoV) of 22$'$. One of the objectives of the survey is to detect and discover transient sources in its FoV. The current framework \citep{2025MNRAS.538..133P} employs CNN models trained on synthetic datasets comprising \textit{stellar} point sources extracted from science frames, rather than genuine transient events. While the method has demonstrated considerable efficacy, the use of artificially constructed training data can introduce systematic biases into the model \citep{electronics13173509}. Specifically, such datasets may fail to fully encapsulate the complexity and variability in image data corresponding to real transient events, potentially limiting the model’s generalisation capability in operational settings. But the small size of such a training dataset, owing to the small FoV of the ILMT, poses challenges concerning CNN model training with the \textit{real-world} data, as discussed earlier.   

Transfer Learning \citep[TL;][]{5288526, Ribani_TL_CNN} is an ML strategy that is often useful in the absence of an extensive training dataset for a deep learning-based task (e.g. CNNs for image classification). Previous works have explored the use of TL for astronomical cases \citep{Vilalta_2018}, like classification of galaxy morphology \citep{10.1093/mnras/sty3497}, variable sources \citep{2021A&A...653A..22K}, transients \citep{2025MNRAS.542L.132G}, and star clusters \citep{10.1093/mnras/stad2238}. In this strategy, the existence of a model trained on a large dataset from a related task is assumed. The knowledge acquired from the original (source) task can be reused to enhance the model's performance on the new (target) task. For this work, the source real/bogus classifier model was constructed with a large and labelled training dataset derived from the ZTF transient alerts. This model was then retrained using TL on a smaller dataset of transients detected with the ILMT survey. This ensured the transfer of knowledge learnt from the vast and diverse set of ZTF-detected transients to the model trained for the ILMT. 

Furthermore, performing a sub-classification of the `real' candidates is desirable to facilitate early follow-up of a preferred class of detected candidates. Previous works have explored the use of photometric data to classify the candidates into categories like various types of variable stars (VS), SNe, active galactic nuclei (AGNs), etc. Such methods involve either computing features from the lightcurve (e.g. \citealt{Richards_2011,Pichara_2016,Martínez-Palomera_2018,Boone_2019,S_nchez_S_ez_2021}) or using lightcurves directly (e.g. \citealt{8280984,2018NatAs...2..151N,2019PASP..131k8002M}) as inputs to the classifier. \citealt{Carrasco_Davis_2021} proposed the \textit{single-shot} `stamp' classifier method for ZTF alerts that uses information available in a single alert, including a triplet of science, reference and difference images (also referred to as `stamps' in their work),  associated with any candidate to classify them into 5 categories. The present scheme \citep{2025MNRAS.538..133P} implemented with the ILMT data stream performs a more rudimentary classification of the `real' candidates into 3 classes based only on the reference stamps using a CNN. The three classes are: (i) \textit{extended-host} candidates (e.g., SNe, AGN with a visible galaxy, variable multiply imaged quasars superimposed on the lens galaxy, etc), (ii) \textit{point-host} candidates (e.g. VS, quasars, etc.) and (iii) \textit{hostless/orphan} candidates (mainly asteroids). For this work, two different types of CNN-based transient candidate classifiers were trained using the TL technique. The first type is a 3-class classifier which classifies the detected candidate into three classes, same as the present scheme adopted for the ILMT, while the second type (called the 4-class classifier) extends this scheme with an additional \textit{bogus} class. The purpose of the fourth class in the candidate classifier is to filter out bogus/artefacts that might have passed through the initial candidate detection stage. Either type of alert classifier can be cascaded with the real/bogus classifier: the 3-class model performs straightforward classification of candidates that pass the real/bogus stage, while the 4-class model enables more refined filtering of artefacts. All the trained models were thoroughly validated and demonstrated a significant improvement in performance with this technique. This work uses an image-stamp TL-based approach, which is different from \citealt{2025MNRAS.542L.132G} that uses lightcurves. 
     
This study aims to serve as a \textit{proof of concept} for future enhancements in transient detection and classification with the ILMT, rather than as an immediate replacement for the current scheme.

The paper is structured as follows. Section~\ref{sec:transfer_learning} gives a general discussion on TL for CNN. Section~\ref{sec:data} discusses strategies and considerations adopted for dataset construction. A discussion on model training for the real/bogus and transient alert classifiers is given in Section~\ref{sec:model_training}. The test results are summarised in Section~\ref{sec:results}. Deployment of the trained models on full frame ILMT images is discussed in Section \ref{sec:deployment}. Finally, the conclusion and discussion about this work is presented in Section \ref{sec:conclusion_discussion}.

\section{Transfer Learning}
\label{sec:transfer_learning}

Training deep learning models like the CNNs for image classification requires an extensive training dataset. Construction of such datasets has two important prerequisites: (i) the availability of large volumes of data samples to create the dataset, and (ii) a dedicated human resource to label the data samples. In situations where the data is very user-specific and small-scale, as in the case of the ILMT, training such models entirely with a custom dataset becomes difficult. CNNs often contain a large number of tunable parameters, which makes them more susceptible to overfitting, especially when trained on limited datasets — a manifestation of the \textit{bias-variance tradeoff} \citep{Geman_1992}. Such models perform well on training data but fail to generalise on test data. Furthermore, small datasets often fail to represent the full variability or complexity of the problem domain, resulting in trained models that lack versatility. This limits the possibilities of training robust models for image-based transient detection and candidate classification for ILMT data.

TL is an ML technique that is well-suited to overcome the challenges discussed above. A common approach in TL involves training a model on an existing large dataset, also called the \textit{source} dataset, and then adapting it to the new task on the \textit{target} dataset. In the problem being discussed, the source dataset can comprise labelled detection images from other transient survey telescopes like the ZTF, which already has a sufficiently extensive set of detected transients. The resulting model (source model) possesses knowledge regarding the general features relevant to images of astronomical transients. It should be noted that `knowledge' here effectively represents the trainable model parameters of the CNN-based classifiers. However, the source model lacks the feature knowledge specific to the ILMT image dataset (which will be referred to as the \textit{target} dataset for the remaining discussion), which can significantly affect the accuracy of classification. 

\begin{figure}
\centering
    \includegraphics[width=\columnwidth]{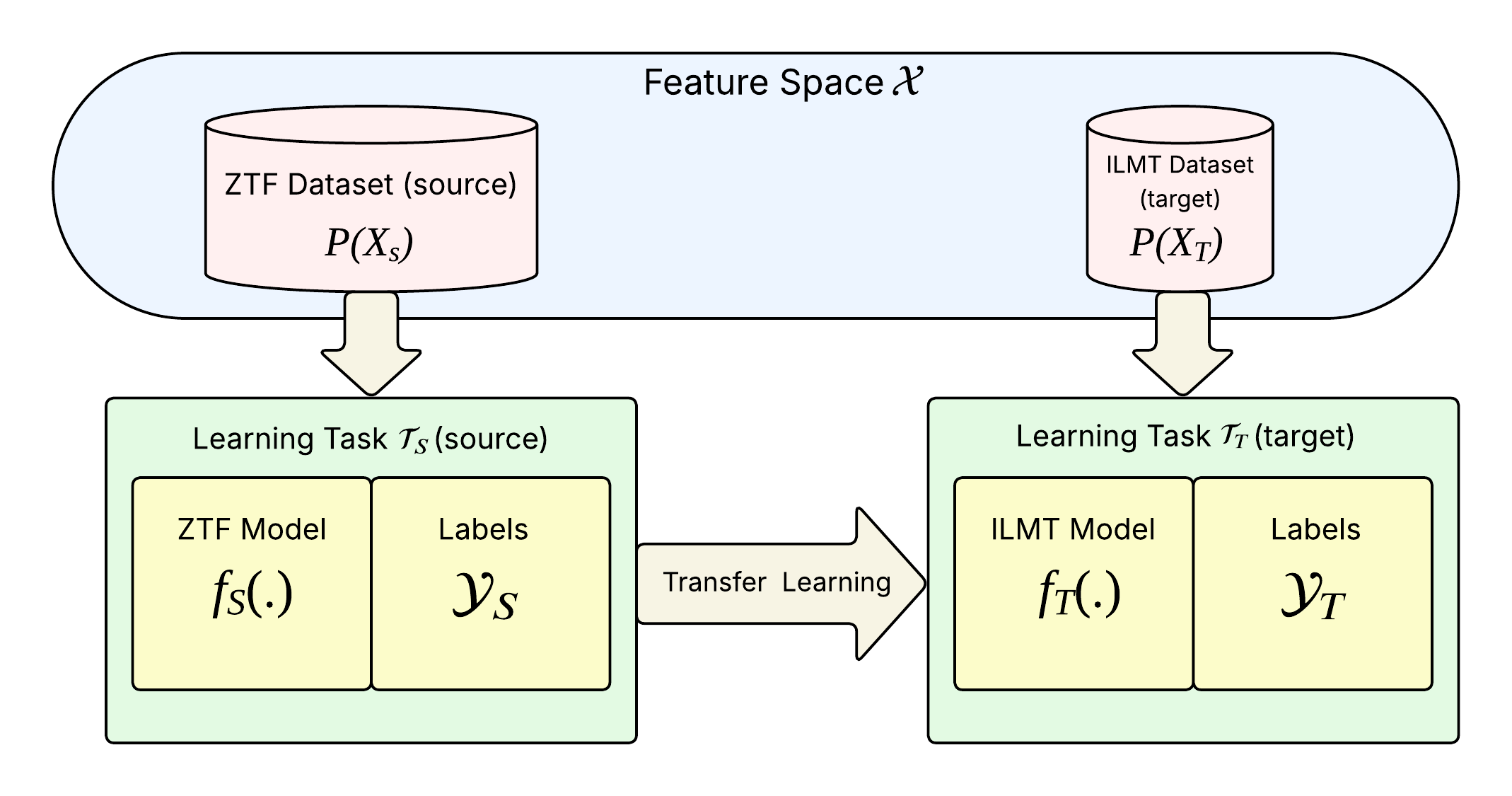}
    \caption{A representation of Transfer Learning (TL) between source and target datasets with a shared feature space. The source dataset typically has a larger size. The knowledge acquired from the source model, trained on the source dataset, is transferred to the target dataset. Figure concept and design adapted from \citet{Ribani_TL_CNN}.}
    \label{fig:TL_CNN}
\end{figure}

To better understand TL, the concepts of domain and task are reviewed here (refer to \citealt{5288526} and \citealt{Ribani_TL_CNN} for a more detailed discussion). A domain $\mathcal{D}$ is composed of two components: a feature space $\mathcal{X}$ and marginal probability distribution $P(X)$ where $X = \{x_1, x_2, x_3,....,x_n\} \in \mathcal{X}$. Here, $x_i$ is the $i^{th}$ instance of training data in the learning sample $X$. Therefore, a domain can be formally represented as $\mathcal{D} = \{\mathcal{X}, P(X)\}$. Considering a domain of images, the feature space $\mathcal{X}$ spans a range of possible pixel values, edges, lines, textures, etc. Given a domain, a task consists of a label space $\mathcal{Y}$ and a predictive function $f(.)$ that predicts a label for a data sample in $X$. Hence, a task is formally defined as $\mathcal{T} = \{\mathcal{Y}, f(.)\}$. The function $f(.)$ is not directly observed and is learned from training data, which consists of pairs $\{x_i, y_i\}$, where $x_i \in X$ and $y_i \in \mathcal{Y}$. Corresponding to the source dataset, there exists a source domain $\mathcal{D}_S$ and for the target dataset, there exists a target domain $\mathcal{D}_T$, with their respective predictive functions $f_S$ and $f_T$. TL is formally defined as a process that will help to improve the learning of the function $f_T$ in domain $\mathcal{D_T}$, given the existence of function $f_S$ in domain $\mathcal{D}_S$, where $\mathcal{D}_S \neq \mathcal{D}_T$, or $\mathcal{T}_S \neq \mathcal{T}_T$. The process effectively transfers knowledge from $f_S$ to $f_T$. For example, the predictive functions $f_S$ and $f_T$ can be source and target CNN models for transient detection and alert classification, with the respective classification tasks $\mathcal{T}_S$ and $\mathcal{T}_T$. For an effective transfer of knowledge, it is desirable to have maximum similarity between source domain $\mathcal{D}_S$ and target domain $\mathcal{D}_T$. A schematic representation of TL under the above-defined paradigm is shown in Figure~\ref{fig:TL_CNN}.

\begin{figure*}
    \includegraphics[width=\textwidth]{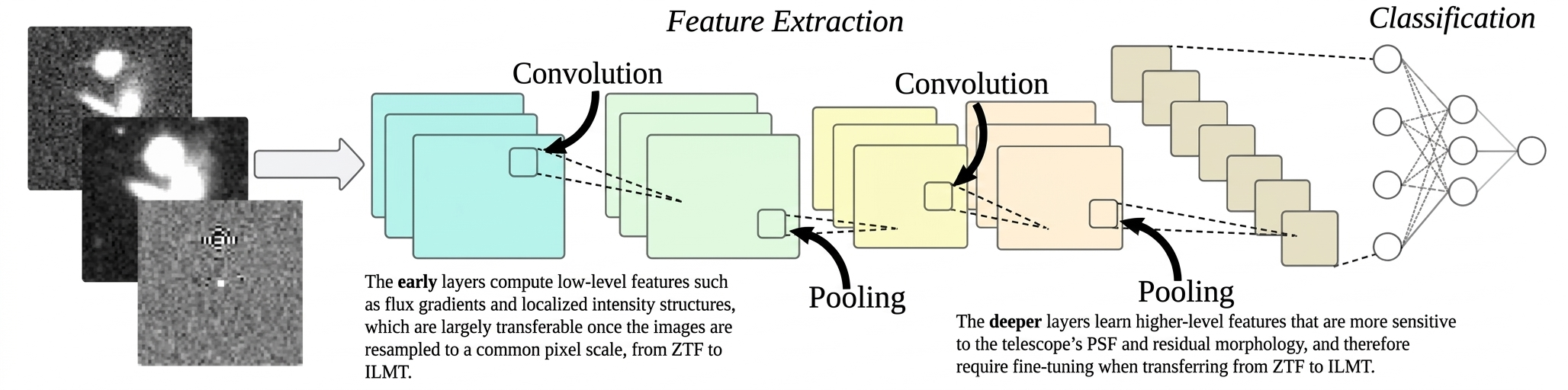}
    \caption{Schematic illustration of various layers in a CNN and their relevance in transfer learning. The input tensor is a science, reference and difference image cutout triplet. The feature extraction region extracts general and specific features from the input, and the fully-connected classifier classifies the input (e.g. real/bogus) based on extracted features.}
    \label{fig:CNN_layers_TL}
\end{figure*}

Fine-tuning is a TL-based approach often used for CNNs where the layers in the source model are retrained on the target dataset, but with a lowered learning rate. This ensures that the model parameters do not vary significantly from their source/pre-trained values, which encode the source knowledge to be transferred. Another approach to preserve this source/pre-trained knowledge is by making some layers untrainable (also called layer \textit{freezing}) while retraining. Usually, under this approach, the early layers of the convolutional base, which represent low-level features such as edges, textures, and gradients, are frozen. The deeper layers, which learn more domain-specific high-level features, are kept trainable. For example, the high-level features in the telescope images can be influenced by the point spread function (PSF), which is specific to the telescope. Figure~\ref{fig:CNN_layers_TL} gives a schematic representation of various regions in a CNN and their relevance in TL. These techniques were used to train the real/bogus classifier, 3-class and 4-class transient alert classifiers for the ILMT data. The subsequent sections give a detailed discussion of dataset organisation, model training and performance metrics of the trained models.  

\section{Training Data}
\label{sec:data}

\begin{figure*}
    \centering
    \begin{subfigure}[t]{0.49\textwidth}
        \centering
        \includegraphics[width=\linewidth]{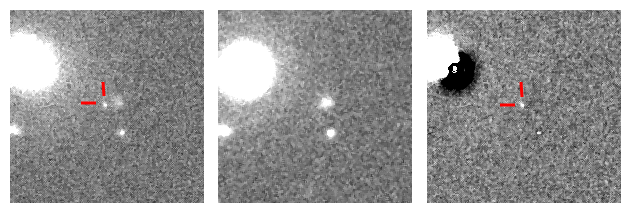}\\[-1pt]
        \includegraphics[width=\linewidth]{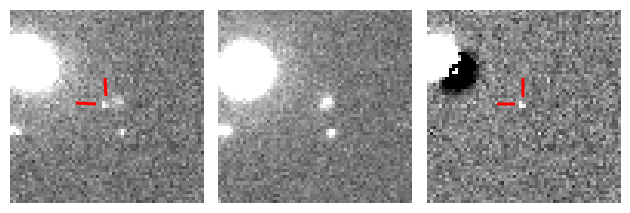}
        \caption{Genuine transient detection}
        \label{fig:real_example}
    \end{subfigure}
    \hfill
    \begin{subfigure}[t]{0.49\textwidth}
        \centering
        \includegraphics[width=\linewidth]{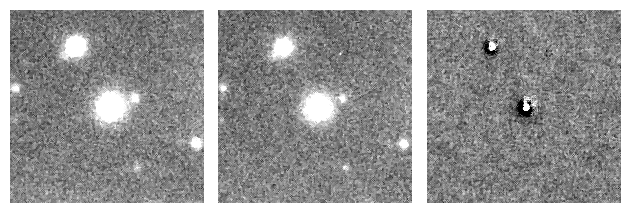}\\[-1pt]
        \includegraphics[width=\linewidth]{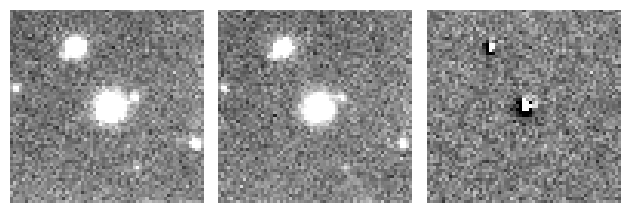}
        \caption{Spurious source}
        \label{fig:bogus_example}
    \end{subfigure}

    \caption{Examples of ILMT detections: (a) a genuine transient adjacent to a host galaxy (highlighted with a red marker) and (b) a spurious source. 
    For each case, the top row shows 195 × 195 pixel cutouts of the science, reference, and difference frames, and the bottom row shows their downsampled 63 × 63 pixel versions. 
    The native pixel scale is 0$''$.323/pixel, while the downsampled stamps are at 1$''$/pixel, consistent with ZTF.}
    \label{fig:downsampling_examples}
\end{figure*}

\citet{Duev_2019} demonstrated a technique to separate a real astrophysical transient from bogus alerts in the ZTF alert stream by implementing a CNN-based real/bogus classifier on 63$\times$63 pixel cutout image triplets of science, reference and difference stamps. The real class (Figure~\ref{fig:real_example}) represents a variety of events like SNe, AGNs, VS, asteroid detections, and other exotic transients. The bogus class (Figure~\ref{fig:bogus_example}) represent spurious detections in difference stamps like cosmic hits, random background fluctuations, and `bad' image subtraction due to misalignment of science and reference images, failure to match PSFs of science and reference images, etc. \citet{Carrasco_Davis_2021} further proposed an alert classification framework for Automatic Learning for the Rapid Classification of Events\footnote{https://alerce.science/} \citep[ALeRCE;][]{F_rster_2021}, which is a ZTF alert broker. This scheme utilised the same triplet image cutouts combined with additional descriptive features available with ZTF alerts to sub-classify them into 5 classes, viz. SNe, AGNs, VSs, asteroids and bogus. 
The first four classes can be conveniently grouped as \textit{real} alerts. 
The ALeRCE broker also provides a more detailed classification of the real ZTF alerts based on respective lightcurves \citep{S_nchez_S_ez_2021}. Both these techniques use ML-based methodology to classify the transient alerts by assigning respective class probability scores.

\begin{figure}
    \includegraphics[width=0.49\textwidth]{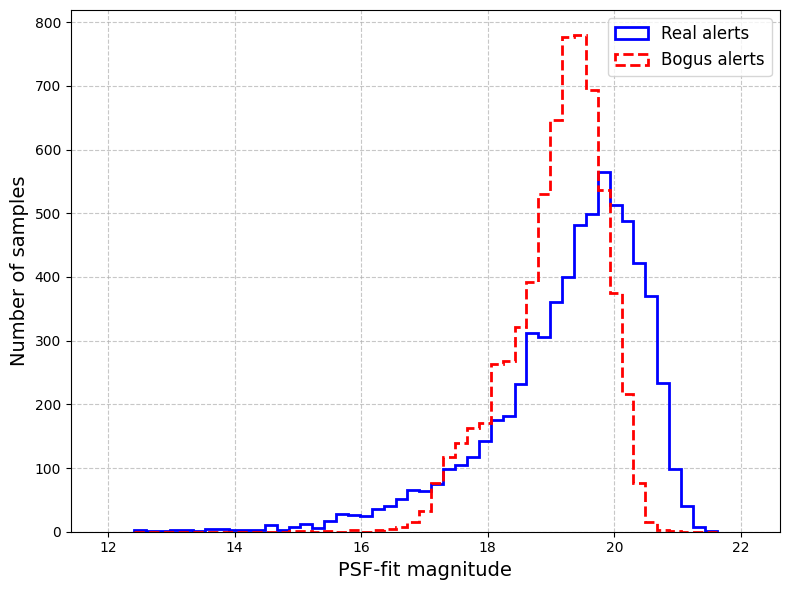}
    \caption{Histogram depicting the magnitude distribution of ZTF transient alerts. The ALeRCE broker executed a real/bogus classification of these alerts, facilitating the incorporation of selected samples into the source dataset based on the classification results.}
    \label{fig:alert_mag}
\end{figure}

In the present work, the science, reference, and difference stamps corresponding to the ALeRCE-classified ZTF alerts were collected to create the source training datasets for the real/bogus and transient alert classifiers. The \texttt{ALeRCE client\footnote{https://alerce.readthedocs.io/en/latest/} Python} library was used to download the alert stamp data. The samples in the datasets were additionally vetted manually and were used to train the source CNN models. Every sample in the datasets is a 3-channel (representing the three stamps) array and has a shape of 63$\times$63$\times$3 pixels. These samples span a broad spectrum of features and magnitudes (Figure~\ref{fig:alert_mag}). For the target dataset, stamps corresponding to transients detected with the ILMT during its third observation cycle (October 2023--May 2024) were pooled. 

\begin{figure*}
    \centering
    \begin{subfigure}{0.245\textwidth}
        \centering
        Extended-host
    \end{subfigure}
    \hfill
    \begin{subfigure}{0.245\textwidth}
        \centering
        Point-source
    \end{subfigure}
    \hfill
    \begin{subfigure}{0.245\textwidth}
        \centering
        Orphan
    \end{subfigure}
    \hfill
    \begin{subfigure}{0.245\textwidth}
        \centering
        Bogus
    \end{subfigure}
    \begin{subfigure}{0.245\textwidth}
        \centering
        \includegraphics[width=\textwidth]{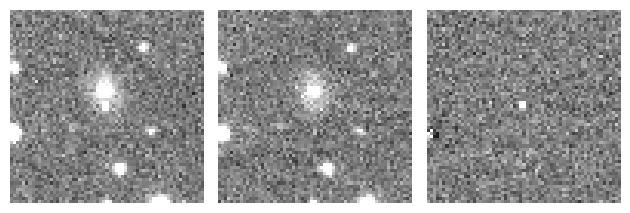}
    \end{subfigure}
    \hfill
    \begin{subfigure}{0.245\textwidth}
        \centering
        \includegraphics[width=\textwidth]{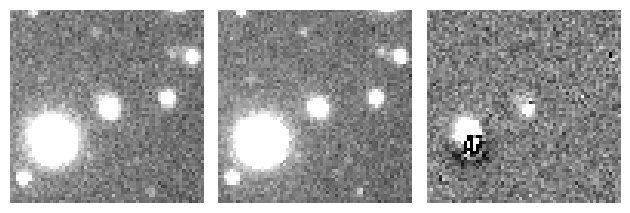}
    \end{subfigure}
    \hfill
    \begin{subfigure}{0.245\textwidth}
        \centering
        \includegraphics[width=\textwidth]{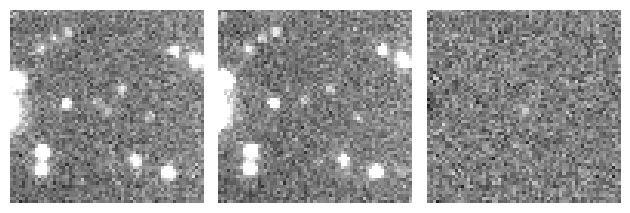}
    \end{subfigure}
    \hfill
    \begin{subfigure}{0.245\textwidth}
        \centering
        \includegraphics[width=\textwidth]{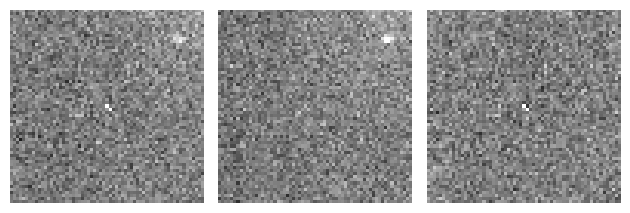}
    \end{subfigure}
    \begin{subfigure}{0.245\textwidth}
        \centering
        \includegraphics[width=\textwidth]{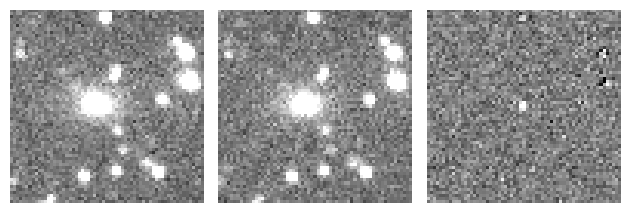}
    \end{subfigure}
    \hfill
    \begin{subfigure}{0.245\textwidth}
        \centering
        \includegraphics[width=\textwidth]{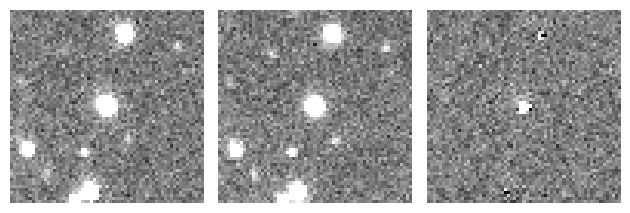}
    \end{subfigure}
    \hfill
    \begin{subfigure}{0.245\textwidth}
        \centering
        \includegraphics[width=\textwidth]{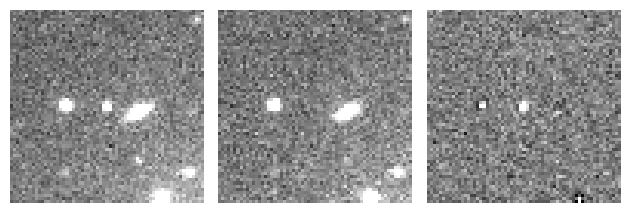}
    \end{subfigure}
    \hfill
    \begin{subfigure}{0.245\textwidth}
        \centering
        \includegraphics[width=\textwidth]{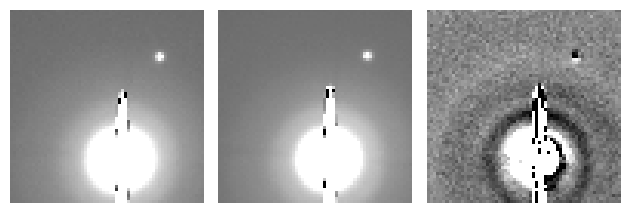}
    \end{subfigure}
    \begin{subfigure}{0.245\textwidth}
        \centering
        \includegraphics[width=\textwidth]{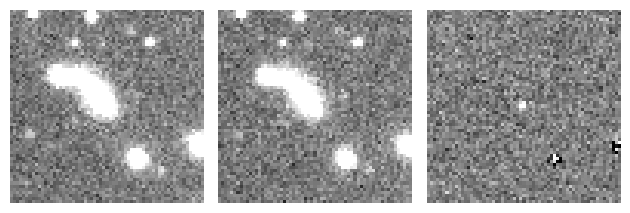}
    \end{subfigure}
    \hfill
    \begin{subfigure}{0.245\textwidth}
        \centering
        \includegraphics[width=\textwidth]{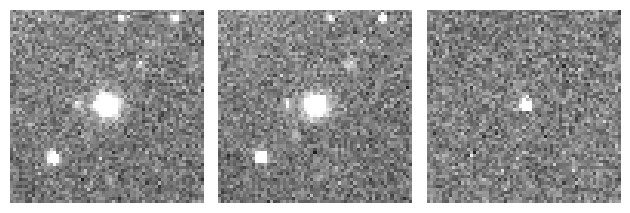}
    \end{subfigure}
    \hfill
    \begin{subfigure}{0.245\textwidth}
        \centering
        \includegraphics[width=\textwidth]{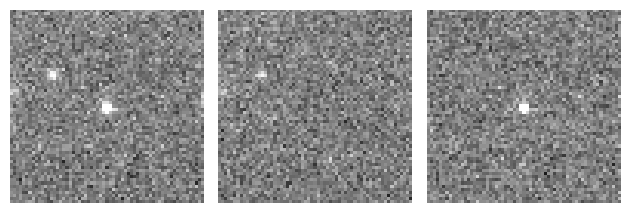}
    \end{subfigure}
    \hfill
    \begin{subfigure}{0.245\textwidth}
        \centering
        \includegraphics[width=\textwidth]{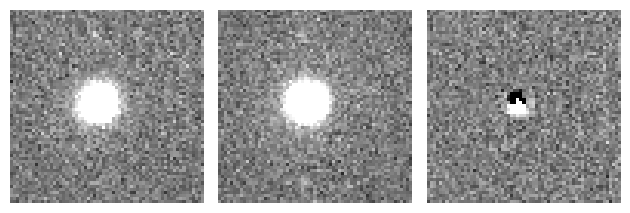}
    \end{subfigure}
    \begin{subfigure}{0.245\textwidth}
        \centering
        \includegraphics[width=\textwidth]{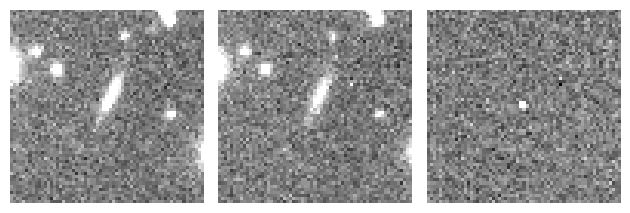}
    \end{subfigure}
    \hfill
    \begin{subfigure}{0.245\textwidth}
        \centering
        \includegraphics[width=\textwidth]{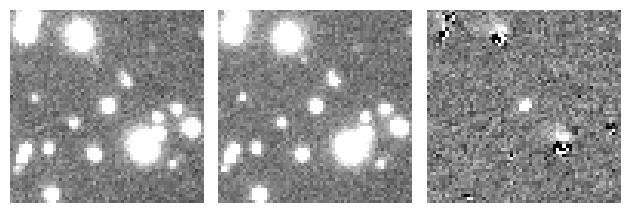}
    \end{subfigure}
    \hfill
    \begin{subfigure}{0.245\textwidth}
        \centering
        \includegraphics[width=\textwidth]{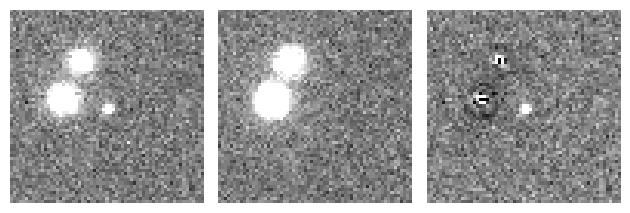}
    \end{subfigure}
    \hfill
    \begin{subfigure}{0.245\textwidth}
        \centering
        \includegraphics[width=\textwidth]{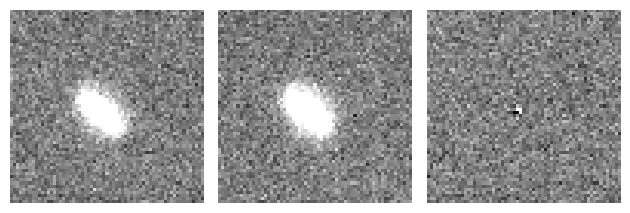}
    \end{subfigure}
    \caption{Examples of sources belonging to the four possible candidate classes. A triplet of science, reference and difference stamps for every source is shown. The sources were correctly classified into their relevant classes by the real/bogus classifier, 3-class and 4-class candidate alert classifiers.}
    \label{fig:triplet_scores}
\end{figure*} 

The pixel scale of the ILMT is around 0.323 arcseconds/pixel, while that for the ZTF is 1 arcsecond/pixel. For this reason, instead of using the ILMT cutout stamps directly, the stamps were downsampled to match their pixel scale to that of the ZTF (illustrated in  Figure~\ref{fig:downsampling_examples}). This was achieved using a \texttt{Python} module called \texttt{zoom}, which is a spline interpolation-based image resampling method available with \texttt{scipy} \citep{Virtanen_2020}. This was done to enhance the morphological similarity between the source and target domains, hence facilitating an optimal knowledge transfer from the source model to the target model. 
The science, reference and difference cutouts were then combined to create 3-channel arrays with which the target models were trained. Both the source and target datasets for the real/bogus classifier and transient alert classifiers were further split in an 85:15 ratio between the training and validation sets. This split was dynamic, and was performed afresh for every version of a model. This ensured that the training was unaffected by specific training samples.  

\subsection{The real/bogus classifier dataset}
\label{sec:real/bogus dataset}

The source real dataset was compiled by pooling stamps primarily for the candidates, which were classified into different categories by the ALeRCE lightcurve classifier and with multiple detections. This reduced the likelihood of the detection being an artefact. Also, it was ensured that the ALeRCE-assigned classification score for a selected class was sufficiently high (preferably between 0.6--0.8). The classifications considered were of the types ranging from SNe-like transients (nearly 2850 samples), variable stars (nearly 1200 samples), and AGNs (nearly 450 samples). Since the asteroids do not have a well-defined positional lightcurve, they had to be collected by only selecting samples with high classification score (greater than 0.8) as asteroid from the ALeRCE stamp classifier. Nearly 2000 samples of asteroids were selected. The different types of candidates reflect the various observational scenarios expected in a transient survey. Finally, a representative subset of the source real dataset was inspected visually.

The source bogus dataset was prepared by combining datasets from different data sources. The first batch of around 300 bogus stamps came from the candidates classified as bogus by the ALeRCE stamp classifier. Another batch of stamps with around 3300 samples was curated by manually differencing ZTF images using the custom-built \texttt{ILMTDiff} image subtraction software \citep{2025MNRAS.538..133P} and extracting artefacts from them. The third batch of bogus stamps (having nearly 1500 samples) consisted of subtraction artefacts cropped out from the difference images made available by the ZTF itself. The bogus dataset also includes samples with an associated galaxy ($\sim$ 900 samples) in science and reference stamps. This was done to ensure that the model does not correlate the presence of a galaxy with a real transient. The final dataset comprised 6640 real samples and 6041 bogus samples. For the source bogus dataset, all the samples were inspected visually.

The target real/bogus dataset was created by subtracting ILMT frames using the \texttt{ILMTDiff} image subtraction algorithm and subsequently extracting different types of transient/variable candidates and bogus candidates. The extended-host transient candidates were significantly under-represented, so data augmentation by 90$^{\circ}$, 180$^{\circ}$, and 270$^{\circ}$ degree rotations was applied to enhance their representation. This final target dataset comprised 367 real sources and 380 bogus sources. All the samples in the target dataset were manually vetted. The class-wise breakdown of the datasets is shown in Table~\ref{table:sample_counts_rb}.

\subsection{The transient alert classifier dataset}
\label{sec:transient alert classifier dataset}
To train the source candidate alert classifier models, a dataset comprising 4 classes, viz. \textit{bogus}, \textit{extended-host objects}, \textit{point-source objects}, and \textit{orphans} were compiled. The extended-host objects comprise transient events with an associated host galaxy. Examples of such events can include SNe, AGNs with a visible extended galaxy, tidal disruption events (TDEs), GRB afterglows with clear host galaxies, and variable multiply imaged quasars superimposed on the lens galaxy, etc. Point-source objects include point-like transient/variable sources, viz. variable stars, quasars, etc., while orphans have no visible host. Examples of orphan transients are asteroids, some cataclysmic variables (CVs), transients like GRBs afterglow, SNe with no detectable host galaxy, orphan SNe, etc. Examples of the four defined classes of transients are shown in Figure~\ref{fig:triplet_scores}. For the extended-host class, nearly 2850 ZTF candidates identified as SNe by both the ALeRCE stamp classifier and the lightcurve classifier were included. The point-source class was constructed with nearly 1200 ALeRCE-classified variable star samples, and the orphan class consisted of 2000 asteroid samples. To avoid ambiguity, the sample of AGNs included in the real dataset was not included in the alert classifier dataset, as they can present both like an extended-host object (e.g. in case of some Seyfert galaxies) or point sources (e.g quasars). The same bogus class was used as that for the real/bogus dataset. A representative subset of the dataset samples were also visually inspected.   

 The target transient alert classifier dataset consisted of 167 extended-host objects, 140 orphan objects, and 60 point-source objects, all derived from the target real dataset. The bogus class again consisted of the same 380 samples as that in the target real/bogus dataset. The class-wise breakdown of the datasets is shown in Table~\ref{table:sample_counts_tr}.  

\begin{table}
\centering
\captionof{table}{Sample counts for real and bogus classes from the source (ZTF) and the target (ILMT) real/bogus datasets.}
\begin{tabular}{lcc}
\hline
         & \textbf{ZTF} & \textbf{ILMT} \\
\hline
\textbf{Real}  & 6640       & 367        \\
\textbf{Bogus} & 6041       & 380        \\
\hline
\end{tabular}
\label{table:sample_counts_rb}
\end{table}

\begin{table}
\centering
\captionof{table}{Sample counts for different classes from the source (ZTF) and the target (ILMT) transient alert classifier datasets. The source dataset does not include samples of ZTF alerts classified as AGNs by ALeRCE.}
\begin{tabular}{lcc}
\hline
         & \textbf{ZTF} & \textbf{ILMT} \\
\hline
\textbf{Extended-host}  & 2857       & 167        \\
\textbf{Point-source}  & 1216       & 60        \\
\textbf{Orphans}  & 1955       & 140        \\
\textbf{Bogus} & 6041       & 380        \\
\hline
\end{tabular}
\label{table:sample_counts_tr}
\end{table}

\section{Model Training}
\label{sec:model_training}

The model training was undertaken in two steps. The first step involved training the source models for the real/bogus and candidate alert classifiers. The source models were then readapted to the target datasets for the respective classifiers using TL in the second step. Before model training, dataset samples were pre-processed. The NaN values in multiple ZTF cutouts were handled by replacing them with the median value of the sample. The samples were then channel-wise mean-centred and standardised. All relevant CNN model architectures were created and trained using Google \texttt{TensorFlow} \citep{abadi2016tensorflow}. All the models were trained using the NVIDIA T4$\times$2 GPU available with the \texttt{kaggle}$^{\textregistered}$\footnote{https://www.kaggle.com/} platform. The training methodology for the real/bogus and transient alert classifier is explained in detail in the following subsections.  

\subsection{The real/bogus classifier}

\begin{figure}
    \includegraphics[width=\columnwidth]{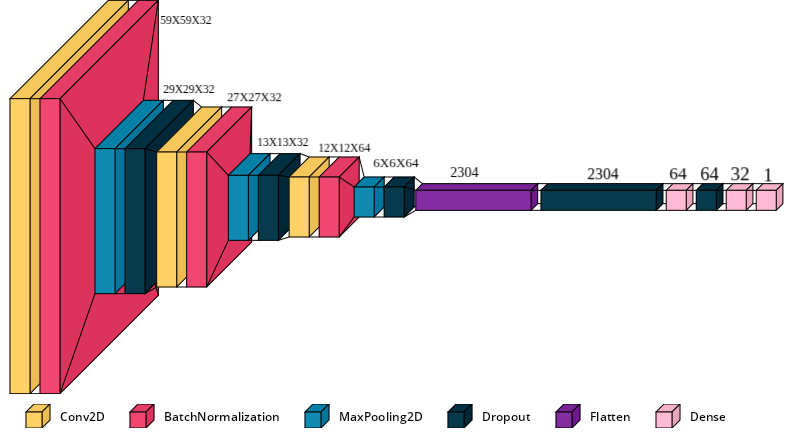}
    \caption{CNN architecture of the real/bogus classifier. The same architecture was used for the 3-class and 4-class transient alert classifiers, except for the softmax output layer (instead of sigmoid) with 3 and 4 dimensions, respectively.}
    \label{fig:arch_RB}
\end{figure}

The CNN architecture (Figure~\ref{fig:arch_RB}) designed for the real/bogus classification task comprises three convolutional blocks, each paired with a 2$\times$2 max-pooling layer to downsample feature maps. From the early to deeper layers (shown from left to right), the filter sizes are 5$\times$5, 3$\times$3, and 2$\times$2, respectively. The \texttt{ReLU} activation function \citep{inproceedings_nair} is applied in the intermediate layers, while a \texttt{sigmoid} activation function is utilised in the output layer to produce a probability score, consistent with the binary classification objective. Dropout regularisation \citep{JMLR:v15:srivastava14a}, L2 regularisation, and batch normalisation \citep{2015arXiv150203167I} were explored during training to reduce overfitting and enhance generalisability. The parameter for L2 regularisation was set to 0.02, and the dropout rate was set to 0.5 for all layers. The loss function to be minimised while training was binary-cross entropy, and the optimiser was \texttt{adam} \citep{adam}. The model architecture is based on that presented in \citet{2025MNRAS.538..133P}, with necessary modifications implemented to accommodate differences in the input tensor shape, while preserving the core design that demonstrated robust performance under operational conditions. The source model was trained with the ZTF dataset for about 80 epochs with a learning rate of 10$^{-3}$, with which it attained a validation accuracy of around 98\% on the ZTF validation data. Early stopping was used while training with the maximum validation accuracy kept as the stopping criteria. The accuracy and loss curves are shown in Figure~\ref{fig:CLF4_ACC_LOSS}. 

\begin{figure}
    \centering \includegraphics[width=0.5\textwidth]{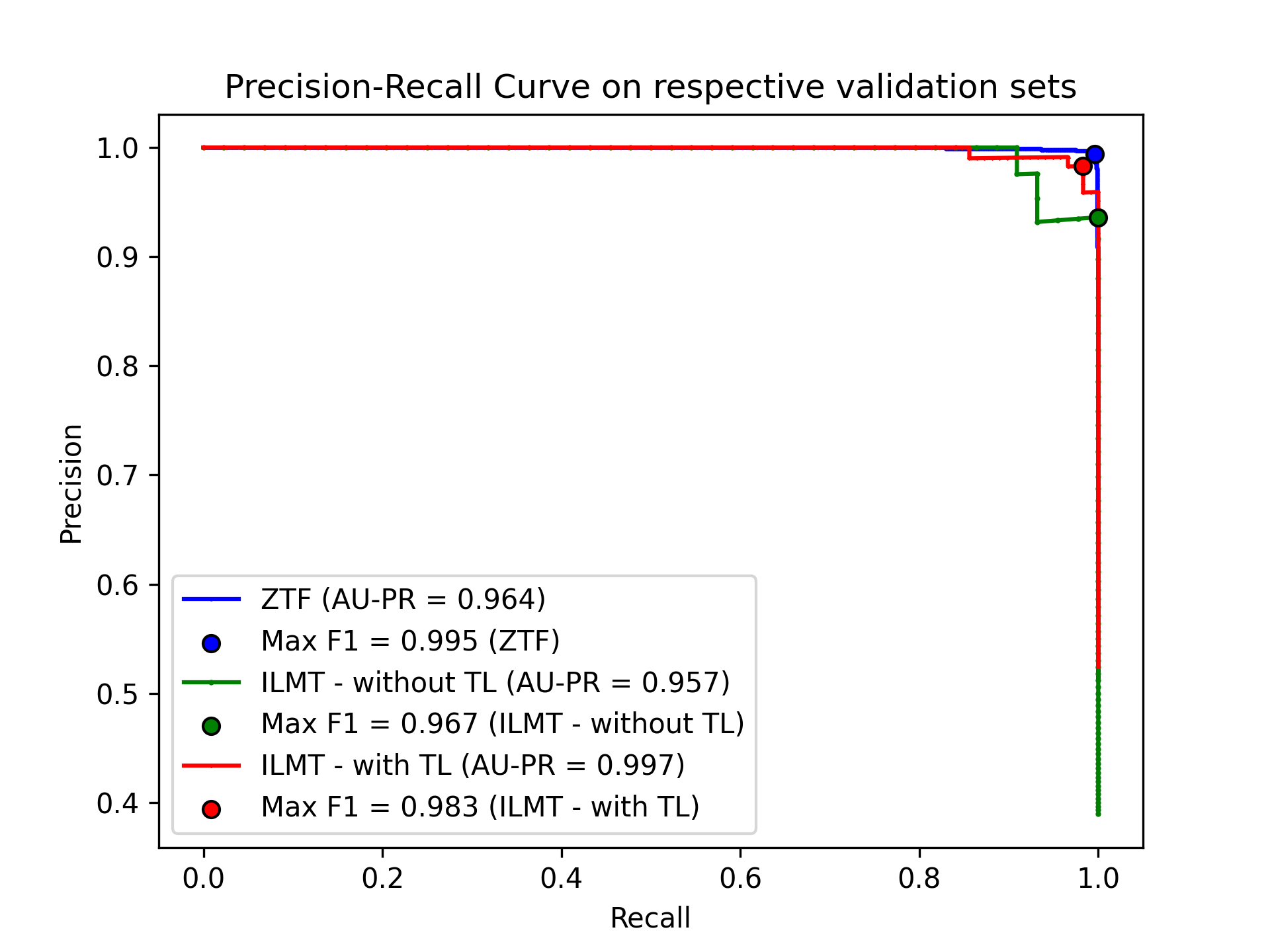}
    \caption{Evaluation of optimal threshold for CNN-based real/bogus classifier using F1 score maximisation on the validation dataset.}
    \label{fig:F1_val}
\end{figure}

The source and target real/bogus classification tasks share similarities in that both are binary classification problems and utilise stamps with the same effective pixel scale to perform similar tasks, ensuring consistency in the input data format. The fine-tuning approach of TL was used to train the target models. This ensured that the training did not modify the existing weights significantly, preserving the broader characteristic knowledge acquired from the previous training performed for the source real/bogus classification task. Only the incremental modifications necessary to adapt to the new target dataset were performed. Therefore, the source ZTF model was trained on the smaller ILMT dataset after reducing the learning rate to 10$^{-5}$. The resulting validation accuracy achieved after nearly 900 epochs of training was around 97\% (Figure~\ref{fig:CLF4_ACC_LOSS}). Early stopping was used again while training with the maximum validation accuracy kept as stopping criteria. For the real/bogus classifiers, the optimal classification threshold for was selected after maximising the F1 score on validation dataset (Figure~\ref{fig:F1_val}). The F1 score is defined as follows:

\begin{equation}
F_1 = 2 \cdot \frac{\text{Precision} \cdot \text{Recall}}{\text{Precision} + \text{Recall}}
\end{equation}
where:
\[
\text{Precision} = \frac{\text{True Positives}}{\text{True Positives + False Positives}}
\]
\[
\text{Recall} = \frac{\text{True Positives}}{\text{True Positives + False Negatives}}
\]  

\subsection{The transient alert classifier}

The architecture for the 3-class and 4-class classifiers was kept the same as the real/bogus classifier, except for the output layer (Figure~\ref{fig:arch_RB}). The output layers for the 3-class and 4-class transient alert classifiers have 3 and 4-dimensional output tensors, respectively, and the output activation function used for both was \texttt{softmax}. The 4-class source alert classifier was trained for about 90 epochs on the ZTF 4-class classifier dataset with a learning rate of 10$^{-3}$ and attained a validation accuracy of around 95\%. This trained model was then retrained on the target ILMT 4-class classifier dataset for about 180 epochs with a lower learning rate of 5$\times$10$^{-5}$, attaining a validation accuracy of around 92\% (Figure~\ref{fig:CLF4_ACC_LOSS}).

For the 3-class classifier, it was observed that the accuracy of the source model on the target ILMT data was already quite high. So, it was decided to use the conventional TL technique by freezing the convolutional base and training only the fully connected classifier layers, again with a reduced learning rate of $5\times10^{-5}$. This model was then trained for nearly 300 epochs on the target dataset, and the validation accuracy achieved was 96\% (Figure~\ref{fig:CLF4_ACC_LOSS}). Early stopping was used while training the source and target alert classifier models, with the maximum validation accuracy kept as the stopping criteria. For both the alert classifier models, class weighting was used to address class imbalance.

\section{Model evaluation on The test data}
\label{sec:results}

The training datasets used for training the transient detection and alert classifier models were curated by extracting cutouts of sources from the images of the third observation cycle (October 2023--May 2024) of the ILMT. On the other hand, the test dataset was curated using images from the first observation cycle (October 2022--November 2022). This was done to ensure an unbiased benchmarking of the trained models. Figure~\ref{fig:triplet_scores} shows examples of candidate triplets from the test dataset classified correctly into respective classes by the candidate detection and alert classification models. Additionally, the test dataset for the transient alert classifier was balanced using the random undersampling technique for more faithful assessment. The test real/bogus dataset was left unchanged as the imbalance in that case was not very severe. Detailed inferences are given in the following subsections.

\subsection{The real/bogus classifier}
The real test set was prepared by extracting triplets of science, reference, and difference stamps corresponding to known minor planets/asteroids and variable stars in the images. The bogus test dataset was prepared by extracting triplets corresponding to subtraction artefacts in the difference images. Special care was taken to ensure diversity in the types of samples chosen to prepare the test dataset. The final dataset comprised 269 real samples and 334 bogus samples. The pre-processing steps for the test samples were the same as the training samples. The test accuracy for the TL real/bogus model was determined to be 97.3\%. The same for the model trained with a smaller ILMT (target) dataset without TL was 90.5\%. When the source real/bogus model was implemented on the test data, the accuracy was 91.5\%. This test accuracy for the source model is reasonably high and indicates good transferability to the target dataset.

\begin{figure*}
    \centering
    \begin{subfigure}[b]{0.49\textwidth}
        \centering
        \includegraphics[width=\textwidth]{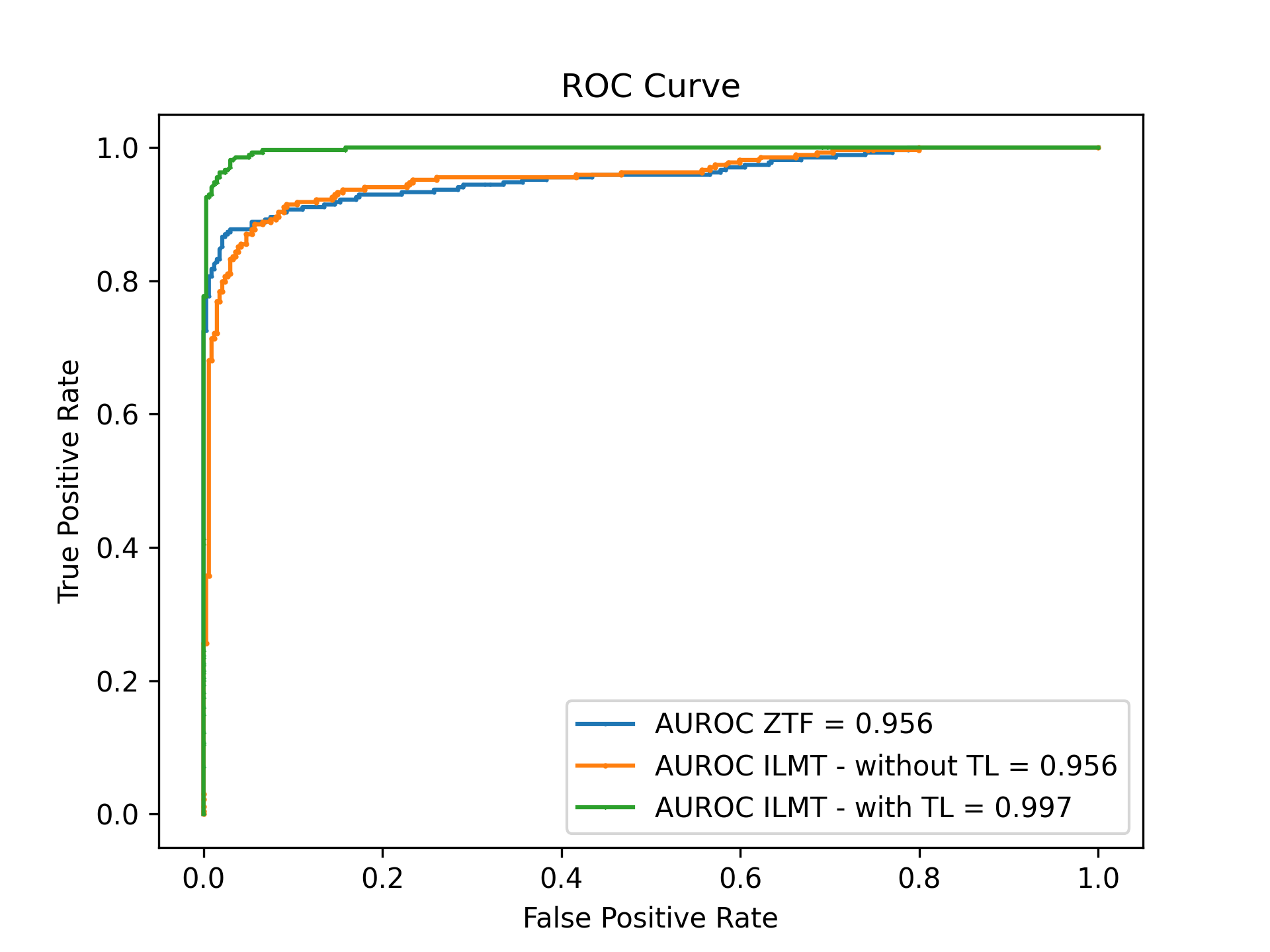}
        \label{fig:ROC_RB}
    \end{subfigure}
    \hfill
    \begin{subfigure}[b]{0.49\textwidth}
        \centering
        \includegraphics[width=\textwidth]{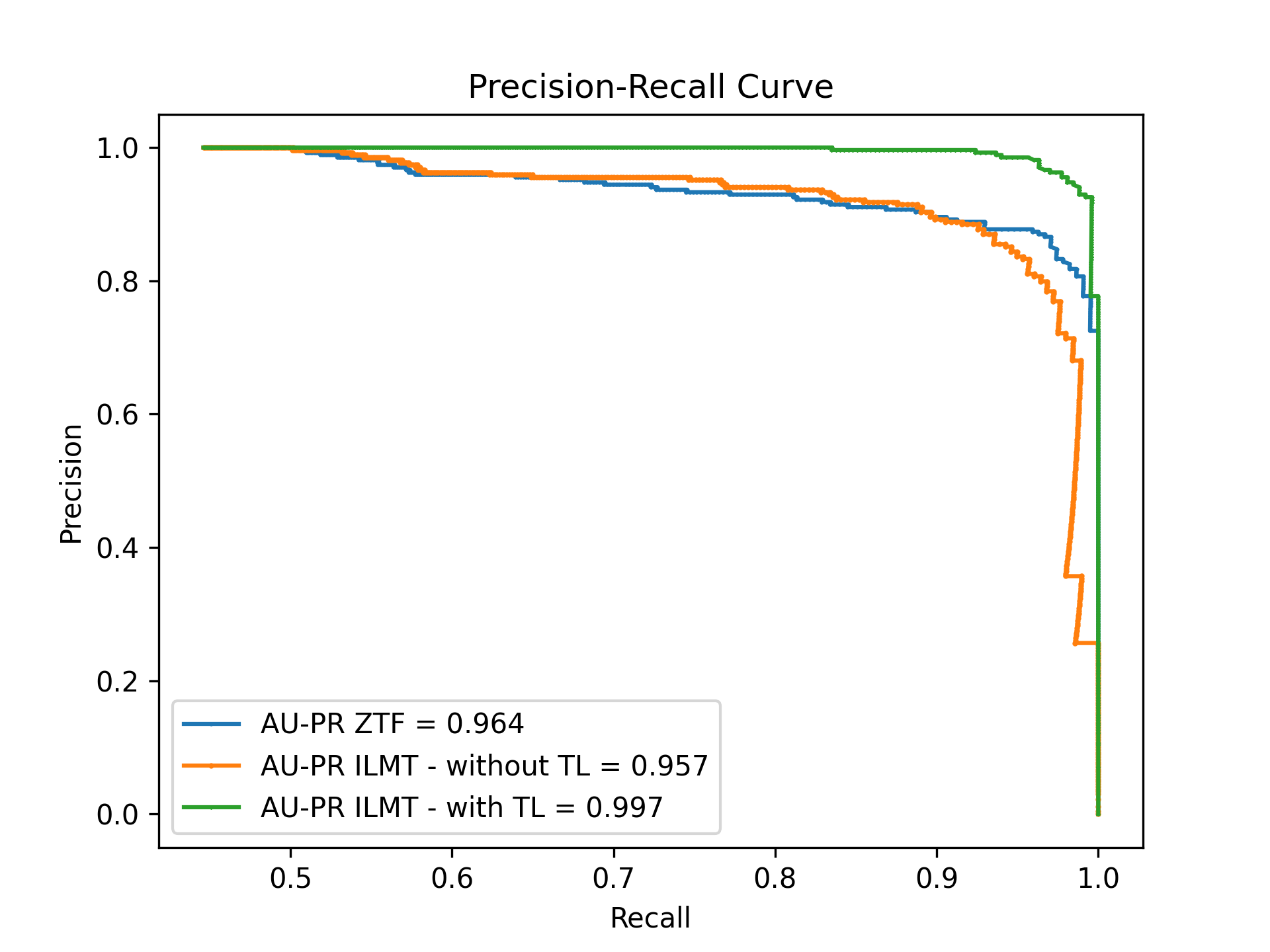}
        \label{fig:PR_RB}
    \end{subfigure}%
    \caption{Receiver operating characteristic (ROC) curves and precision-recall curves for the source (ZTF) model, model trained without TL (baseline) and target (ILMT) model trained with TL for the real/bogus classifier.}
    \label{fig:ROC_PR}
\end{figure*}

The receiver operating characteristic (ROC) curves and the precision-recall (PR) curves for the real/bogus classifier models evaluated on test dataset are shown in Figure~\ref{fig:ROC_PR}. These curves provide a threshold-independent evaluation of classifier performance and illustrate trade-offs between key metrics (such as recall vs false-positive rate for ROC curves, and precision vs recall for PR curves). The areas under the ROC and precision–recall curves (AUROC and AU-PR) offer concise summaries of the model’s discriminative ability, with higher values indicating better performance. From the figures, it is evident that the transfer-learned real/bogus model achieve higher AUROC and AU-PR compared to both the source and baseline models, demonstrating improved classification efficiency. The confusion matrices for the source, baseline and target real/bogus classifiers on the test dataset are shown in Figure~\ref{tab:CM_RB}.

\begin{figure*}
\setlength{\extrarowheight}{5pt}
\centering
\Large
\hspace{-50pt}
\begin{tabular}{cc|c|c|}
  & \multicolumn{1}{c}{} & \multicolumn{2}{c}{Predicted} \\
  & \multicolumn{1}{c}{} & \multicolumn{1}{c}{Bogus}  & \multicolumn{1}{c}{Real} \\\cline{3-4}
  \hspace{-5pt}
  \multirow{2}{*}{\rotatebox[origin=c]{90}{\hspace{-20pt} Actual}} & Bogus & 328 & 6 \\[10pt] \cline{3-4}
                          & Real & 45 & 224 \\[10pt] \cline{3-4}
\end{tabular}
\hspace{0.5cm}
\begin{tabular}{cc|c|c|}
  & \multicolumn{1}{c}{} & \multicolumn{2}{c}{Predicted} \\
  & \multicolumn{1}{c}{} & \multicolumn{1}{c}{Bogus}  & \multicolumn{1}{c}{Real} \\\cline{3-4}
  \hspace{-5pt}
  \multirow{2}{*}{\rotatebox[origin=c]{90}{\hspace{-20pt} Actual}} & Bogus & 323 & 11 \\[10pt] \cline{3-4}
                          & Real & 5 & 264 \\[10pt] \cline{3-4}
\end{tabular}
\hspace{0.5cm}
\begin{tabular}{cc|c|c|}
  & \multicolumn{1}{c}{} & \multicolumn{2}{c}{Predicted} \\
  & \multicolumn{1}{c}{} & \multicolumn{1}{c}{Bogus}  & \multicolumn{1}{c}{Real} \\\cline{3-4}
  \hspace{-5pt}
  \multirow{2}{*}{\rotatebox[origin=c]{90}{\hspace{-20pt} Actual}} & Bogus & 299 & 35 \\[10pt] \cline{3-4}
                          & Real & 22 & 247 \\[10pt] \cline{3-4}
\end{tabular}
\caption{Confusion matrices obtained for real/bogus classifier source model (left) and target model (centre). The same is also shown for a model trained with the small ILMT dataset (baseline) without using TL (right).}
\label{tab:CM_RB}
\end{figure*}

\begin{figure}
    \includegraphics[width=\columnwidth]{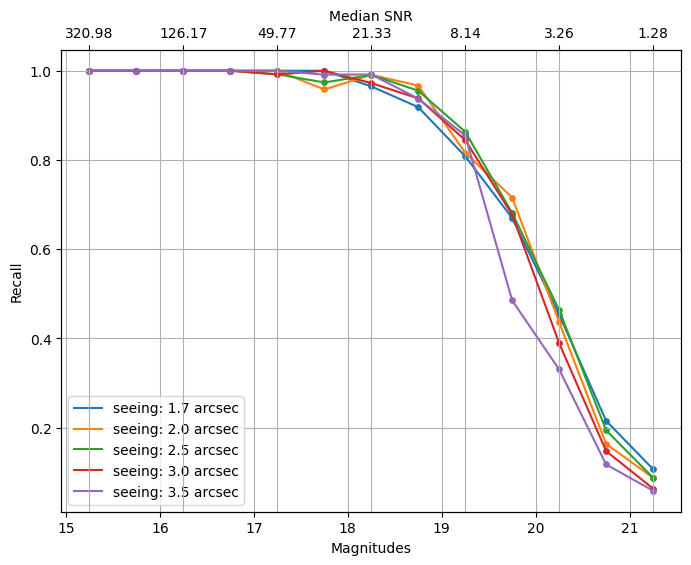}
    \caption{Recall values evaluated for simulated `orphan' sources at various seeing/FWHM and magnitude ranges. The plot also shows the median S/N ratio (here defined as the ratio between the amplitude of the simulated Gaussian source and the local noise) for different magnitude ranges.}
    \label{fig:simulated_recall}
\end{figure}

To evaluate the detection completeness of the real/bogus classifier trained using the TL technique, it was implemented on simulated \textit{orphan} sources injected in a set of photometrically calibrated ILMT images. The sources were simulated with 2D Gaussian profiles and covered a range of magnitudes and FWHMs. Figure~\ref{fig:simulated_recall} shows median recall (fraction of sources identified as real) for various magnitude ranges. It also shows the median signal-to-noise (S/N) ratio across different FWHMs for each magnitude range. The plot shows the expected decrease in recall for fainter detection magnitudes, while no significant dependence on FWHM is observed for the trained real/bogus models. It should be noted, however, that the simulated Gaussian profiles differ slightly from the actual PSF of sources in the ILMT images. The difference is more pronounced near the `wings', which are broader for the observed PSF than the simulated sources.  

\subsection{The transient alert classifier}

\begin{figure*}
\setlength{\extrarowheight}{5pt}

\includegraphics[width=0.8\textwidth]{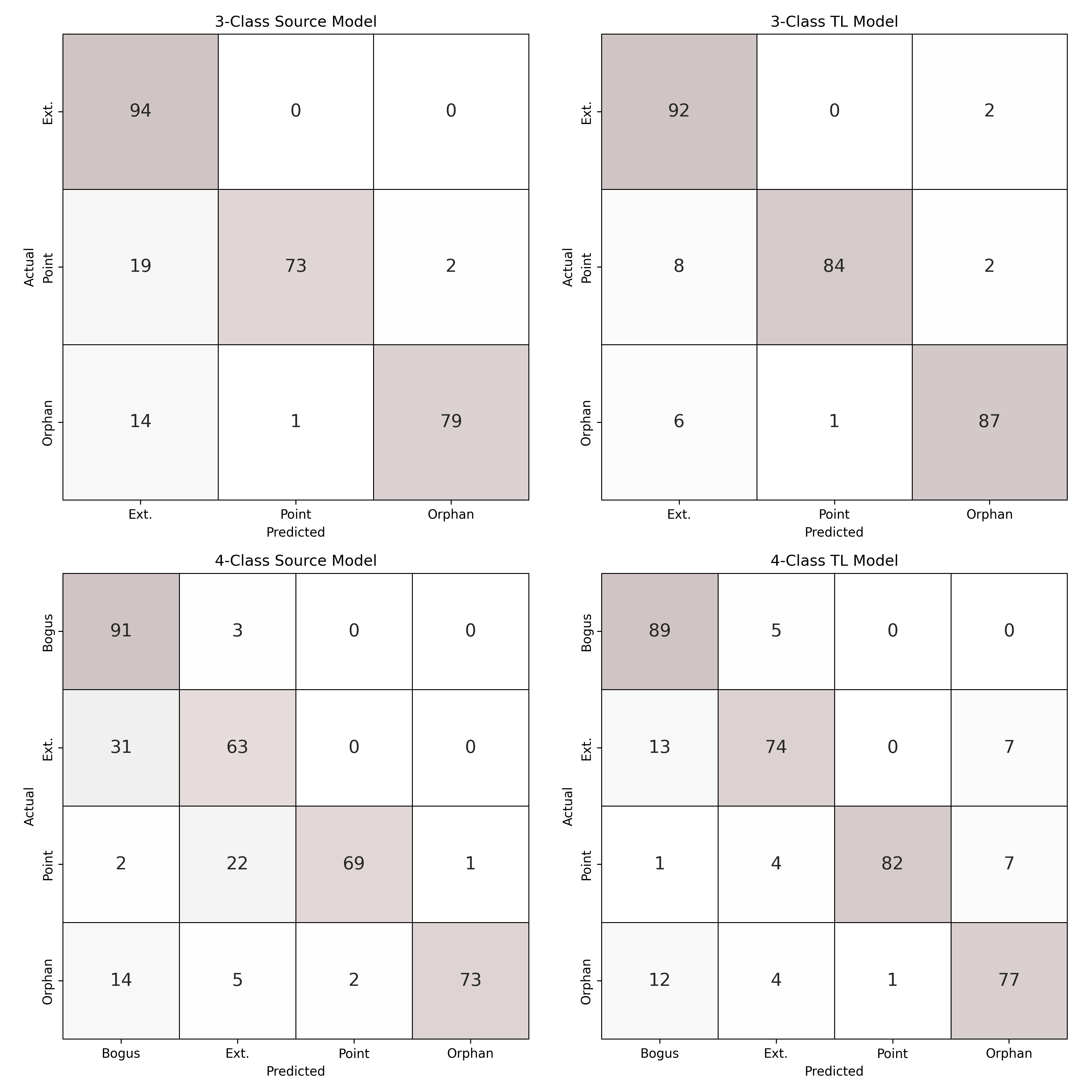}

\caption{\label{demo-table}Comparison between confusion matrices of the source (left) 3-class (top) and 4-class (bottom) transient alert classifier models and the models modified with TL (right) on the same test dataset.}

\label{tab:CM_3_4_class}
\end{figure*}

The first ILMT observation cycle (October 2022--November 2022), which was used to create the test dataset, did not detect `SN-like' sources with a clear extended host galaxy. Therefore, the original test dataset for the real/bogus classifier was supplemented with artificial SN-like sources prepared by injecting ILMT PSFs and simulated Gaussian profiles within and around a few galaxies present in the images. The ILMT PSF was created by extracting stellar profiles from the ILMT frames. The FWHM of the simulated profiles was around 2$''$, which is close to the median seeing of the ILMT images. The test dataset was partitioned into 4 classes as discussed earlier for the source candidate classifier model. 
The 3 and 4 class candidate classifiers were separately evaluated on respective test datasets, which resulted in classification accuracies of 92.9\% and 85.6\%, respectively. The confusion matrices for the source and target transient alert classifier performances on the test dataset are shown in Figure~\ref{tab:CM_3_4_class}. Table~\ref{tab:test_results} summarises all the test results. 

\begin{table}
\centering
\captionof{table}{Test accuracies achieved for the source model, target model and the baseline model trained without TL on a small training dataset. The accuracies were evaluated for the real/bogus classifier, 3-class and 4-class transient alert classifiers.}
\begin{tabular}{|m{1.7cm}|m{1.7cm}|m{1.7cm}|m{1.7cm}|}
    \hline
    \textbf{Classifier type} & \textbf{Source model} & \textbf{Target model} & \textbf{Baseline model} \\
    \hline
     Real/bogus & 91.5\% & 97.3\% & 90.5\% \\
     3-class & 86.9\% & 92.9\% & 77.3\% \\
     4-class & 78.7\% & 85.6\% & 75.5\% \\
     \hline
\end{tabular}
\hspace{3cm}
\label{tab:test_results}
\end{table}

\subsection{Unpaired sample t-test to compare mean accuracies of TL and baseline models}

The statistical significance of the effectiveness of transfer learning for training the three types of models was determined by performing an unpaired sample t-test between TL and baseline models. The mutual independence between the populations of TL and baseline models justified the use of the unpaired t-test. The two samples comprised 20 TL and baseline models each. Each of the 20 TL models was trained over a separately trained source model. The accuracies of all the trained models were evaluated on the same test dataset for real/bogus and transient alert classifiers. The threshold for the real bogus binary classifier was again selected using F1 score maximisation on validation sets. Classification in the transient alert classifier was determined by selecting the class with the maximum probability (a.k.a argmax) from the CNN output. For the sample of baseline models, the mean accuracies were determined to be 91.8\% (standard deviation (SD) =1.5\%), 77.2\% (SD=3.9\%) and 76.4\% (SD=2.5\%) for real/bogus, 3-class classifiers and 4-class classifiers, respectively. For the TL models, the mean accuracies were determined to be 95.6\% (SD=1.2\%), 92.9\% (SD=1.4\%) and 81.7\% (SD=4.3\%) for real/bogus, 3-class classifiers and 4-class classifiers, respectively. Under the null hypothesis, which posits no difference in the mean accuracies between the two samples (baseline models and TL models), the calculated t-statistics were 8.90, 16.75, and 4.82 for the three model types. The corresponding p-values for each test were significantly below the predefined significance level ($\alpha = 0.05$), providing sufficient evidence to reject the null hypothesis and conclude that the mean accuracy of TL models is better than baseline models, thereby demonstrating the efficacy of the TL method. The result of the t-test is summarised in Table~\ref{tab:t_test_results}.

\begin{table*}
\centering
\caption{Comparison of baseline and TL model accuracies using a two-sample unpaired t-test. The standard deviation of model accuracies in the selected sample is mentioned with the mean accuracies. The resulting p-values help conclude that the effectiveness of TL is statistically significant.}
\label{tab:t_test_results}
\sisetup{
  detect-all,
  input-symbols = (),
  table-align-text-post = false
}
\begin{tabular}{@{}l
                >{\centering\arraybackslash}m{3.5cm}
                >{\centering\arraybackslash}m{3.5cm}
                S[table-format=2.2]
                S[table-format=1.1e-1]
                @{}}
\toprule
\textbf{Classifier} &
\textbf{Baseline Mean Accuracy (SD)} &
\textbf{TL Mean Accuracy (SD)} &
\textbf{t-statistic} &
\textbf{p-value} \\
\midrule
Real/Bogus &
\multicolumn{1}{c}{91.8\% (1.5\%)} &
\multicolumn{1}{c}{95.6\% (1.2\%)} &
8.90 &
1.2e-10 \\
3-Class &
\multicolumn{1}{c}{77.2\% (3.9\%)} &
\multicolumn{1}{c}{92.9\% (1.4\%)} &
16.75 &
1.4e-14 \\
4-Class &
\multicolumn{1}{c}{76.4\% (2.5\%)} &
\multicolumn{1}{c}{81.7\% (4.3\%)} &
4.82 &
3.8e-5 \\
\bottomrule
\end{tabular}
\end{table*}

\section{Model deployment with the \texttt{PyLMT} pipeline}
\label{sec:deployment}

\begin{figure}
    \centering
    \includegraphics[width=0.47\textwidth]{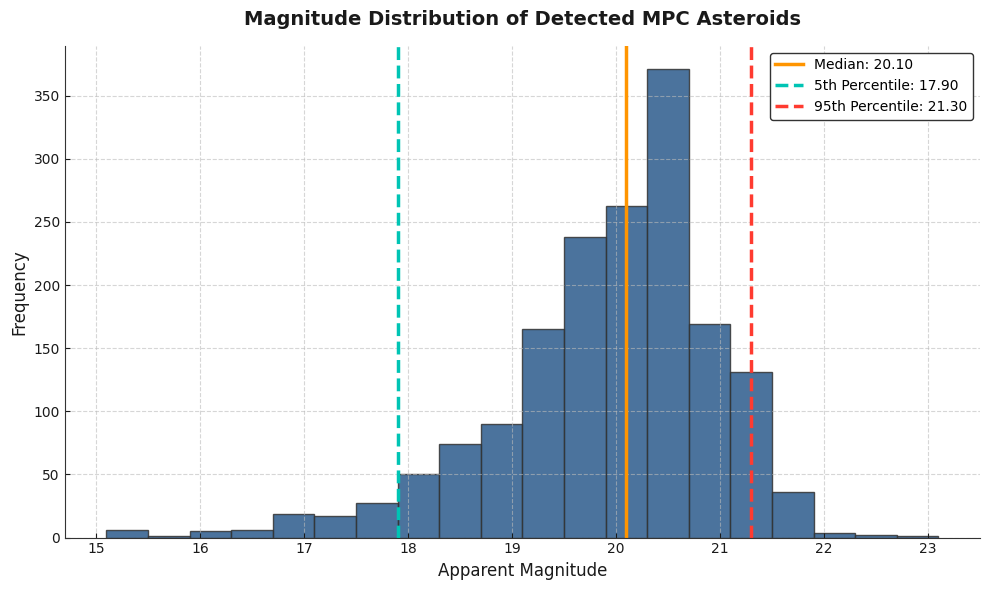}
    \caption{Distribution of MPC obtained V-band magnitudes of asteroids detected with the pipeline by only using the TL models for real/bogus classifiers.}
    \label{fig:mag-asteroid}
\end{figure}

The existing \texttt{PyLMT} transient detection pipeline \citep{2025MNRAS.538..133P} was modified by integrating it with the newly trained TL classifiers. Nearly 300 full-frame science-ready ILMT images of size 36864$\times$4096 pixels were passed through the resulting pipeline to obtain the transient candidates. The final real/bogus classification scheme required positive classification from two different models to further reduce the false positive detections in full-frame images. The final result was generated as sets of \texttt{PDF} files, which were subsequently vetted by visual inspection. Positive detections corresponded to multiple types of astrophysical phenomena/objects, including asteroids, variable stars, AGNs, and SNe candidates. Figure~\ref{fig:mag-asteroid} shows the distribution of the V-band magnitudes of the detected asteroids. These detections were confirmed using the \texttt{SkyBoT} API service provided by the Institute of Celestial Mechanics and Computation of Ephemerides (IMCCE). From the magnitude distribution, the median V-band magnitude of the asteroid detections was determined to be around 20, while the 5$^{th}$ and 95$^{th}$ percentile magnitudes were determined to be around 17.9 and 21.3, respectively. This distribution gives a practical view of the magnitude range over which the pipeline is able to detect asteroids, set by both the underlying asteroid population and the sensitivity of the ILMT images. However, cross-matched detections with SkyBoT serve only as a practical diagnostic of the pipeline performance, leveraging the well-characterized and abundant asteroid population. Since the cross-match is performed only on detected candidates, it does not provide a measure of completeness. Consequently, successful matches should not be interpreted as evidence that all asteroids present in the images have been recovered by the model. 

\begin{figure*}
\centering
\begin{subfigure}[b]{0.48\textwidth}
  \centering
  \begin{tabular}{|c@{\hspace{0.1cm}}c@{\hspace{0.1cm}}c@{\hspace{0.1cm}}|}
    \hline
    \textbf{Science} & \textbf{Reference} & \textbf{Difference} \\
    \includegraphics[width=0.33\linewidth]{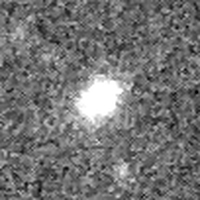} &
    \includegraphics[width=0.33\linewidth]{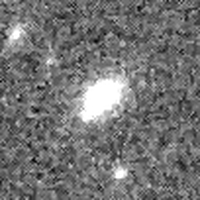} &
    \includegraphics[width=0.33\linewidth]{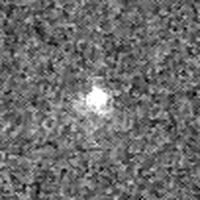} \\
    \includegraphics[width=0.33\linewidth]{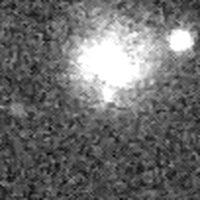} &
    \includegraphics[width=0.33\linewidth]{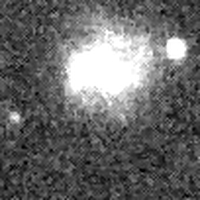} &
    \includegraphics[width=0.33\linewidth]{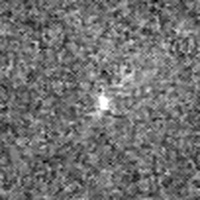} \\
    \includegraphics[width=0.33\linewidth]{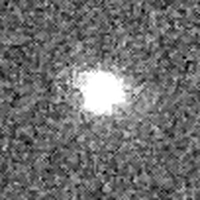} &
    \includegraphics[width=0.33\linewidth]{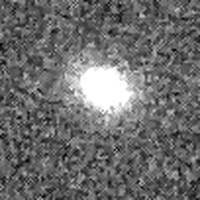} &
    \includegraphics[width=0.33\linewidth]{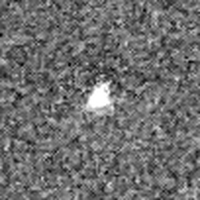} \\
    \includegraphics[width=0.33\linewidth]{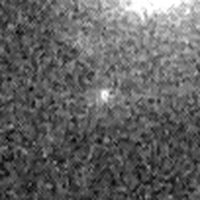} &
    \includegraphics[width=0.33\linewidth]{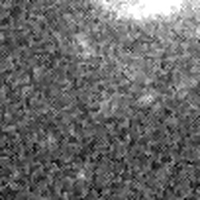} &
    \includegraphics[width=0.33\linewidth]{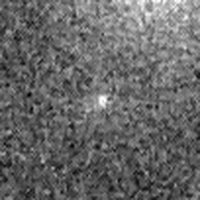} \\
    \hline
  \end{tabular}
\end{subfigure}
\hfill
\begin{subfigure}[b]{0.48\textwidth}
  \centering
  \begin{tabular}{|c@{\hspace{0.1cm}}c@{\hspace{0.1cm}}c@{\hspace{0.1cm}}|}
    \hline
    \textbf{Science} & \textbf{Reference} & \textbf{Difference} \\
    \includegraphics[width=0.33\linewidth]{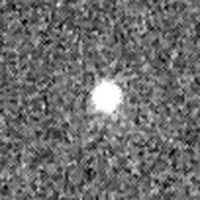} &
    \includegraphics[width=0.33\linewidth]{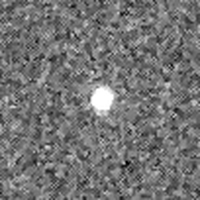} &
    \includegraphics[width=0.33\linewidth]{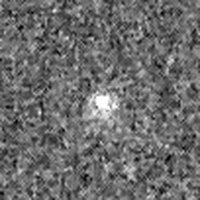} \\
    \includegraphics[width=0.33\linewidth]{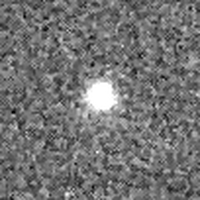} &
    \includegraphics[width=0.33\linewidth]{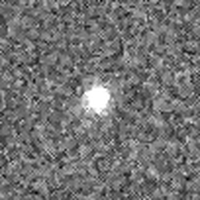} &
    \includegraphics[width=0.33\linewidth]{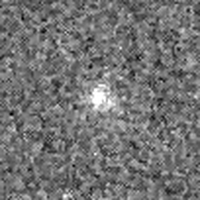} \\
    \includegraphics[width=0.33\linewidth]{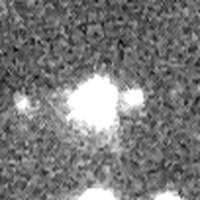} &
    \includegraphics[width=0.33\linewidth]{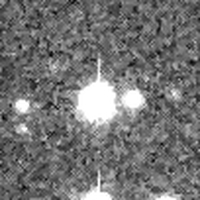} &
    \includegraphics[width=0.33\linewidth]{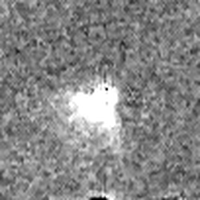} \\
    \includegraphics[width=0.33\linewidth]{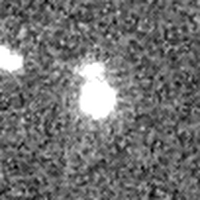} &
    \includegraphics[width=0.33\linewidth]{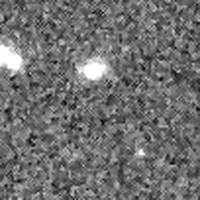} &
    \includegraphics[width=0.33\linewidth]{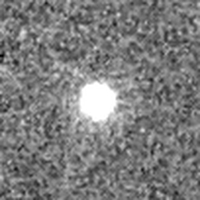} \\
    \hline
  \end{tabular}
  
\end{subfigure}

  \caption{Left side: From top to bottom: science, reference and difference image cutouts of supernova candidates AT2024abso, AT2025chp, AT2025dgl, and AT2025re, detected in the ILMT science images using TL models for real/bogus classifiers. All the 4 candidates have a clear and extended-host galaxy associated with them. The first three candidates were classified correctly as \textit{extended-host} objects by the 3 and 4-class classifiers. AT2025re was misclassified as \textit{hostless} by the 4-class classifier but correctly classified as \textit{extended-host} by the 3-class classifier. The size of the cutouts shown is 102$\times$102 pixel ($\sim31'' \times 31''$). Right side: From top to bottom: science, reference and difference image cutouts of variability detected in the Seyfert 1 galaxy 2MASS J11061222+2927307, the QSO J0937+2937, and the variable star ATO J084.9485+29.3722. The three candidates are clearly point sources. The bottom panel shows detection of the asteroid 1999 ND50, appearing hostless. The first 3 candidates were classified as \textit{point-source} objects by the 3-class and 4-class classifiers. The asteroid candidate was classified as \textit{hostless}. The size of the cutouts shown is 102$\times$102 pixel ($\sim31'' \times 31''$).}

\label{tab:PyLMT_detections}

\end{figure*}

Detections of SN candidates AT~2024abso, 2025chp, 2025dgl and 2025re are shown in Figure~\ref{tab:PyLMT_detections} (left panel). Detection of flux variations in the Seyfert galaxy 2MASS~J11061222+2927307, the QSO~J0937+2937, the variable star ATO~J084.9485+29.3722, and the asteroid 1999~ND50 is shown in Figure~\ref{tab:PyLMT_detections} (right panel).

\section{Conclusion and Discussion}
\label{sec:conclusion_discussion}

In this paper, we demonstrated a proof of concept to use TL and related fine-tuning techniques to train CNNs for real/bogus classifiers and transient alert classifiers for detecting and classifying transient candidates, respectively, with small-field survey telescopes like the ILMT. This technique can potentially help overcome the limitations of small datasets for training such models, which can very likely be present with such surveys. 

For this work, a CNN was trained to classify the sources in the difference images as real or bogus/artefacts using this method. Another type of model, known as the transient alert classifier, were trained to subclassify the detected `real' candidates based on morphological features in the images associated with candidates. The transient alert classifier is of two types: (i) a 3-class classifier and (ii) a 4-class classifier. The respective source models for all three classifiers were trained using a large dataset of ZTF-detected transients, subclassified with the ALeRCE broker. The target models were trained on the smaller ILMT datasets. The accuracies achieved with this technique were 97.3\% for real/bogus classifier, 92.9\% for the 3-class classifier, and 85.6\% for the 4-class classifier. The trained models were evaluated on an independent test dataset resulting in an improved performance over non-TL/baseline models. The trained models were incorporated in the working pipeline and were deployed on full-frame ILMT images, demonstrating their performance capability in operational setting.

Beyond the quantitative improvements in accuracy, the results indicate that morphological features learned from large-scale surveys such as ZTF can generalise effectively to smaller surveys when domain differences are appropriately mitigated. This demonstrates that the fundamental image-level signatures of genuine transients and subtraction artefacts are largely transferable across optical surveys, making transfer learning a practical and scalable strategy for newly commissioned or small-field facilities. In particular, small FoV surveys typically lack the volume of labelled detections required to train deep CNNs from scratch; the approach presented here enables such facilities to leverage knowledge distilled from larger surveys, thereby accelerating the deployment of reliable transient pipelines without waiting for years of data accumulation. Operationally, even modest gains in real/bogus discrimination can substantially reduce false positives in nightly processing, which is especially valuable for small surveys with limited human resources for candidate vetting and follow-up coordination. 

However, the relatively small size of the ILMT target dataset and the partial reliance on injected sources for certain subclasses highlights the need for continued expansion of real labelled data. Furthermore, deep-learning models trained over limited temporal baselines are susceptible to data drift caused by evolving observing conditions and instrumental variations. In future, to ensure long-term robustness, a continual learning framework \citep{PARISI201954} with periodic retraining should be adopted, where techniques such as elastic weight consolidation \citep[EWC;][]{doi:10.1073/pnas.1611835114} can mitigate catastrophic forgetting while preserving previously acquired knowledge.

\section*{Acknowledgements}
The authors thank the referees for critically reviewing the manuscript which has improved the presentation of the paper. The authors acknowledge the use of the \href{https://www.kaggle.com}{Kaggle} platform for model training and experimentation. We also thank the \href{https://alerce.science/}{ALeRCE} broker service for providing access to ZTF transient alert data, which was used in the creation of training datasets. The training dataset is based on observations obtained with the ZTF, a project supported by the National Science Foundation under Grant No. AST-2034437 and a collaboration including Caltech, IPAC, the Weizmann Institute for Science, the Oskar Klein Center at Stockholm University, the University of Maryland, Deutsches Elektronen-Synchrotron and Humboldt University, the TANGO Consortium of Taiwan, the University of Wisconsin at Milwaukee, Trinity College Dublin, Lawrence Livermore National Laboratories, and IN2P3, France. Operations are conducted by the COO, IPAC, and UW. The 4-m International Liquid Mirror Telescope (ILMT) project results from a collaboration between the Institute of Astrophysics and Geophysics (the University of Li\`{e}ge, Belgium), the Universities of British Columbia, Laval, Montreal, Toronto, Victoria and York University, and Aryabhatta Research Institute of observational sciencES (ARIES, India). 
KP acknowledges the support from the Erasmus+ Programme of the European Union for a research visit to the Institute of Astrophysics and Geophysics, University of Liège, Belgium (Allée du 6 Août 19c, 4000 Liège, Belgium).
KM acknowledges the support from the BRICS grant DST/ICD/BRICS/Call-5/CoNMuTraMO/2023 (G) funded by the DST, India. RA acknowledges the assistance from the National Initiative on Undergraduate Science (NIUS) Program of the Homi Bhabha Science Education and Research (HBCSE)-TIFR, Mumbai (India).

\section*{Data Availability}

To perform this work, the authors extensively used the images acquired with the ILMT during its first three observation cycles. The raw and processed images with astrometric calibrations are routinely made available to the public domain and can be accessed using the URL \url{https://cloud.aries.res.in/index.php/s/xPER9Y3XuaCsTL9}. The publicly available survey images from the ZTF were used to perform initial testing of the pipeline scheme, including key algorithms like image subtraction. The images can be accessed through the IRSA platform using the URL \url{https://irsa.ipac.caltech.edu/applications/ztf/?__action=layout.showDropDown&}. Other relevant data can be made available upon request to the authors.

\bibliographystyle{mnras}
\bibliography{Transfer_Learning} 

@ARTICLE{Cao,
       author = {{Cao}, Yi and {Nugent}, Peter E. and {Kasliwal}, Mansi M.},
        title = "{Intermediate Palomar Transient Factory: Realtime Image Subtraction Pipeline}",
      journal = {\pasp},
         year = 2016,
        month = nov,
       volume = {128},
       number = {969},
        pages = {114502},
          doi = {10.1088/1538-3873/128/969/114502},
archivePrefix = {arXiv},
       eprint = {1608.01006},
 primaryClass = {astro-ph.IM},
       adsurl = {https://ui.adsabs.harvard.edu/abs/2016PASP..128k4502C}
}

@article{Mahabal_2019,
	doi = {10.1088/1538-3873/aaf3fa},
  
	url = {https://doi.org/10.1088%2F1538-3873%2Faaf3fa},
  
	year = 2019,
	month = {jan},
  
	publisher = {{IOP} Publishing},
  
	volume = {131},
  
	number = {997},
  
	pages = {038002},
  
	author = {Ashish Mahabal and Umaa Rebbapragada and Richard Walters and Frank J. Masci and Nadejda Blagorodnova and Jan van Roestel and Quan-Zhi Ye and Rahul Biswas and Kevin Burdge and Chan-Kao Chang and Dmitry A. Duev and V. Zach Golkhou and Adam A. Miller and Jakob Nordin and Charlotte Ward and Scott Adams and Eric C. Bellm and Doug Branton and Brian Bue and Chris Cannella and Andrew Connolly and Richard Dekany and Ulrich Feindt and Tiara Hung and Lucy Fortson and Sara Frederick and C. Fremling and Suvi Gezari and Matthew Graham and Steven Groom and Mansi M. Kasliwal and Shrinivas Kulkarni and Thomas Kupfer and Hsing Wen Lin and Chris Lintott and Ragnhild Lunnan and John Parejko and Thomas A. Prince and Reed Riddle and Ben Rusholme and Nicholas Saunders and Nima Sedaghat and David L. Shupe and Leo P. Singer and Maayane T. Soumagnac and Paula Szkody and Yutaro Tachibana and Kushal Tirumala and Sjoert van Velzen and Darryl Wright},
  
	title = {Machine Learning for the Zwicky Transient Facility},
  
	journal = {Publications of the Astronomical Society of the Pacific}
}

@ARTICLE{726791,
  author={Lecun, Y. and Bottou, L. and Bengio, Y. and Haffner, P.},
  journal={Proceedings of the IEEE}, 
  title={Gradient-based learning applied to document recognition}, 
  year={1998},
  volume={86},
  number={11},
  pages={2278-2324},
  doi={10.1109/5.726791}}

@ARTICLE{1998ApJ...503..325A,
       author = {{Alard}, C. and {Lupton}, Robert H.},
        title = "{A Method for Optimal Image Subtraction}",
      journal = {\apj},
         year = 1998,
        month = aug,
       volume = {503},
       number = {1},
        pages = {325-331},
          doi = {10.1086/305984},
archivePrefix = {arXiv},
       eprint = {astro-ph/9712287},
 primaryClass = {astro-ph},
       adsurl = {https://ui.adsabs.harvard.edu/abs/1998ApJ...503..325A}
}

@article{Bramich_2008,
	doi = {10.1111/j.1745-3933.2008.00464.x},
  
	url = {https://doi.org/10.1111%2Fj.1745-3933.2008.00464.x},
  
	year = 2008,
	month = {may},
  
	publisher = {Oxford University Press ({OUP})},
  
	volume = {386},
  
	number = {1},
  
	pages = {L77--L81},
  
	author = {D. M. Bramich},
  
	title = {A new algorithm for difference image analysis},
  
	journal = {Monthly Notices of the Royal Astronomical Society: Letters}
}

@article{Duev_2019,
	doi = {10.1093/mnras/stz2357},
  
	url = {https://doi.org/10.1093%2Fmnras%2Fstz2357},
  
	year = 2019,
	month = {aug},
  
	publisher = {Oxford University Press ({OUP})},
  
	volume = {489},
  
	number = {3},
  
	pages = {3582--3590},
  
	author = {Dmitry A Duev and Ashish Mahabal and Frank J Masci and Matthew J Graham and Ben Rusholme and Richard Walters and Ishani Karmarkar and Sara Frederick and Mansi M Kasliwal and Umaa Rebbapragada and Charlotte Ward},
  
	title = {Real-bogus classification for the Zwicky Transient Facility using deep learning},
  
	journal = {Monthly Notices of the Royal Astronomical Society}
}

@ARTICLE{2019PASP..131a8002B,
       author = {{Bellm}, Eric C. and {Kulkarni}, Shrinivas R. and {Graham}, Matthew J. and {Dekany}, Richard and {Smith}, Roger M. and {Riddle}, Reed and {Masci}, Frank J. and {Helou}, George and {Prince}, Thomas A. and {Adams}, Scott M. and {Barbarino}, C. and {Barlow}, Tom and {Bauer}, James and {Beck}, Ron and {Belicki}, Justin and {Biswas}, Rahul and {Blagorodnova}, Nadejda and {Bodewits}, Dennis and {Bolin}, Bryce and {Brinnel}, Valery and {Brooke}, Tim and {Bue}, Brian and {Bulla}, Mattia and {Burruss}, Rick and {Cenko}, S. Bradley and {Chang}, Chan-Kao and {Connolly}, Andrew and {Coughlin}, Michael and {Cromer}, John and {Cunningham}, Virginia and {De}, Kishalay and {Delacroix}, Alex and {Desai}, Vandana and {Duev}, Dmitry A. and {Eadie}, Gwendolyn and {Farnham}, Tony L. and {Feeney}, Michael and {Feindt}, Ulrich and {Flynn}, David and {Franckowiak}, Anna and {Frederick}, S. and {Fremling}, C. and {Gal-Yam}, Avishay and {Gezari}, Suvi and {Giomi}, Matteo and {Goldstein}, Daniel A. and {Golkhou}, V. Zach and {Goobar}, Ariel and {Groom}, Steven and {Hacopians}, Eugean and {Hale}, David and {Henning}, John and {Ho}, Anna Y.~Q. and {Hover}, David and {Howell}, Justin and {Hung}, Tiara and {Huppenkothen}, Daniela and {Imel}, David and {Ip}, Wing-Huen and {Ivezi{\'c}}, {\v{Z}}eljko and {Jackson}, Edward and {Jones}, Lynne and {Juric}, Mario and {Kasliwal}, Mansi M. and {Kaspi}, S. and {Kaye}, Stephen and {Kelley}, Michael S.~P. and {Kowalski}, Marek and {Kramer}, Emily and {Kupfer}, Thomas and {Landry}, Walter and {Laher}, Russ R. and {Lee}, Chien-De and {Lin}, Hsing Wen and {Lin}, Zhong-Yi and {Lunnan}, Ragnhild and {Giomi}, Matteo and {Mahabal}, Ashish and {Mao}, Peter and {Miller}, Adam A. and {Monkewitz}, Serge and {Murphy}, Patrick and {Ngeow}, Chow-Choong and {Nordin}, Jakob and {Nugent}, Peter and {Ofek}, Eran and {Patterson}, Maria T. and {Penprase}, Bryan and {Porter}, Michael and {Rauch}, Ludwig and {Rebbapragada}, Umaa and {Reiley}, Dan and {Rigault}, Mickael and {Rodriguez}, Hector and {van Roestel}, Jan and {Rusholme}, Ben and {van Santen}, Jakob and {Schulze}, S. and {Shupe}, David L. and {Singer}, Leo P. and {Soumagnac}, Maayane T. and {Stein}, Robert and {Surace}, Jason and {Sollerman}, Jesper and {Szkody}, Paula and {Taddia}, F. and {Terek}, Scott and {Van Sistine}, Angela and {van Velzen}, Sjoert and {Vestrand}, W. Thomas and {Walters}, Richard and {Ward}, Charlotte and {Ye}, Quan-Zhi and {Yu}, Po-Chieh and {Yan}, Lin and {Zolkower}, Jeffry},
        title = "{The Zwicky Transient Facility: System Overview, Performance, and First Results}",
      journal = {\pasp},
         year = 2019,
        month = jan,
       volume = {131},
       number = {995},
        pages = {018002},
          doi = {10.1088/1538-3873/aaecbe},
archivePrefix = {arXiv},
       eprint = {1902.01932},
 primaryClass = {astro-ph.IM},
       adsurl = {https://ui.adsabs.harvard.edu/abs/2019PASP..131a8002B}
}

@article{S_nchez_S_ez_2021,
	doi = {10.3847/1538-3881/abd5c1},
  
	url = {https://doi.org/10.3847%2F1538-3881%2Fabd5c1},
  
	year = 2021,
	month = {feb},
  
	publisher = {American Astronomical Society},
  
	volume = {161},
  
	number = {3},
  
	pages = {141},
  
	author = {P. S{\'{a}}nchez-S{\'{a}}ez and I. Reyes and C. Valenzuela and F. F{\"o}rster and S. Eyheramendy and F. Elorrieta and F. E. Bauer and G. Cabrera-Vives and P. A. Est{\'{e}}vez and M. Catelan and G. Pignata and P. Huijse and D. De Cicco and P. Ar{\'{e}}valo and R. Carrasco-Davis and J. Abril and R. Kurtev and J. Borissova and J. Arredondo and E. Castillo-Navarrete and D. Rodriguez and D. Ruz-Mieres and A. Moya and L. Sabatini-Gacit{\'{u}}a and C. Sep{\'{u}}lveda-Cobo and E. Camacho-I{\~{n}}iguez},
  
	title = {Alert Classification for the {ALeRCE} Broker System: The Light Curve Classifier},
  
	journal = {The Astronomical Journal}
}

@article{F_rster_2021,
   title={The Automatic Learning for the Rapid Classification of Events (ALeRCE) Alert Broker},
   volume={161},
   ISSN={1538-3881},
   url={http://dx.doi.org/10.3847/1538-3881/abe9bc},
   DOI={10.3847/1538-3881/abe9bc},
   number={5},
   journal={The Astronomical Journal},
   publisher={American Astronomical Society},
   author={Förster, F. and Cabrera-Vives, G. and Castillo-Navarrete, E. and Estévez, P. A. and Sánchez-Sáez, P. and Arredondo, J. and Bauer, F. E. and Carrasco-Davis, R. and Catelan, M. and Elorrieta, F. and Eyheramendy, S. and Huijse, P. and Pignata, G. and Reyes, E. and Reyes, I. and Rodríguez-Mancini, D. and Ruz-Mieres, D. and Valenzuela, C. and Álvarez-Maldonado, I. and Astorga, N. and Borissova, J. and Clocchiatti, A. and De Cicco, D. and Donoso-Oliva, C. and Hernández-García, L. and Graham, M. J. and Jordán, A. and Kurtev, R. and Mahabal, A. and Maureira, J. C. and Muñoz-Arancibia, A. and Molina-Ferreiro, R. and Moya, A. and Palma, W. and Pérez-Carrasco, M. and Protopapas, P. and Romero, M. and Sabatini-Gacitua, L. and Sánchez, A. and Martín, J. San and Sepúlveda-Cobo, C. and Vera, E. and Vergara, J. R.},
   year={2021},
   month=apr, pages={242} }

@article{Virtanen_2020,
	doi = {10.1038/s41592-019-0686-2},
  
	url = {https://doi.org/10.1038%2Fs41592-019-0686-2},
  
	year = 2020,
	month = {feb},
  
	publisher = {Springer Science and Business Media {LLC}
},
  
	volume = {17},
  
	number = {3},
  
	pages = {261--272},
  
	author = {Pauli Virtanen and Ralf Gommers and Travis E. Oliphant and Matt Haberland and Tyler Reddy and David Cournapeau and Evgeni Burovski and Pearu Peterson and Warren Weckesser and Jonathan Bright and St{\'{e}}fan J. van der Walt and Matthew Brett and Joshua Wilson and K. Jarrod Millman and Nikolay Mayorov and Andrew R. J. Nelson and Eric Jones and Robert Kern and Eric Larson and C J Carey and {\.{I}}lhan Polat and Yu Feng and Eric W. Moore and Jake VanderPlas and Denis Laxalde and Josef Perktold and Robert Cimrman and Ian Henriksen and E. A. Quintero and Charles R. Harris and Anne M. Archibald and Ant{\^{o}}nio H. Ribeiro and Fabian Pedregosa and Paul van Mulbregt and Aditya Vijaykumar and Alessandro Pietro Bardelli and Alex Rothberg and Andreas Hilboll and Andreas Kloeckner and Anthony Scopatz and Antony Lee and Ariel Rokem and C. Nathan Woods and Chad Fulton and Charles Masson and Christian Häggström and Clark Fitzgerald and David A. Nicholson and David R. Hagen and Dmitrii V. Pasechnik and Emanuele Olivetti and Eric Martin and Eric Wieser and Fabrice Silva and Felix Lenders and Florian Wilhelm and G. Young and Gavin A. Price and Gert-Ludwig Ingold and Gregory E. Allen and Gregory R. Lee and Herv{\'{e}} Audren and Irvin Probst and Jörg P. Dietrich and Jacob Silterra and James T Webber and Janko Slavi{\v{c}} and Joel Nothman and Johannes Buchner and Johannes Kulick and Johannes L. Schönberger and Jos{\'{e}} Vin{\'{\i}}cius de Miranda Cardoso and Joscha Reimer and Joseph Harrington and Juan Luis Cano Rodr{\'{\i}}guez and Juan Nunez-Iglesias and Justin Kuczynski and Kevin Tritz and Martin Thoma and Matthew Newville and Matthias Kümmerer and Maximilian Bolingbroke and Michael Tartre and Mikhail Pak and Nathaniel J. Smith and Nikolai Nowaczyk and Nikolay Shebanov and Oleksandr Pavlyk and Per A. Brodtkorb and Perry Lee and Robert T. McGibbon and Roman Feldbauer and Sam Lewis and Sam Tygier and Scott Sievert and Sebastiano Vigna and Stefan Peterson and Surhud More and Tadeusz Pudlik and Takuya Oshima and Thomas J. Pingel and Thomas P. Robitaille and Thomas Spura and Thouis R. Jones and Tim Cera and Tim Leslie and Tiziano Zito and Tom Krauss and Utkarsh Upadhyay and Yaroslav O. Halchenko and Yoshiki V{\'{a}}zquez-Baeza and},
  
	title = {{SciPy} 1.0: fundamental algorithms for scientific computing in Python},
  
	journal = {Nature Methods}
}

@misc{abadi2016tensorflow,
      title={TensorFlow: Large-Scale Machine Learning on Heterogeneous Distributed Systems}, 
      author={Martín Abadi and Ashish Agarwal and Paul Barham and Eugene Brevdo and Zhifeng Chen and Craig Citro and Greg S. Corrado and Andy Davis and Jeffrey Dean and Matthieu Devin and Sanjay Ghemawat and Ian Goodfellow and Andrew Harp and Geoffrey Irving and Michael Isard and Yangqing Jia and Rafal Jozefowicz and Lukasz Kaiser and Manjunath Kudlur and Josh Levenberg and Dan Mane and Rajat Monga and Sherry Moore and Derek Murray and Chris Olah and Mike Schuster and Jonathon Shlens and Benoit Steiner and Ilya Sutskever and Kunal Talwar and Paul Tucker and Vincent Vanhoucke and Vijay Vasudevan and Fernanda Viegas and Oriol Vinyals and Pete Warden and Martin Wattenberg and Martin Wicke and Yuan Yu and Xiaoqiang Zheng},
      year={2016},
      eprint={1603.04467},
      archivePrefix={arXiv},
      primaryClass={cs.DC}
}

@article{Carrasco_Davis_2021,
	doi = {10.3847/1538-3881/ac0ef1},
  
	url = {https://doi.org/10.3847%2F1538-3881%2Fac0ef1},
  
	year = 2021,
	month = {nov},
  
	publisher = {American Astronomical Society},
  
	volume = {162},
  
	number = {6},
  
	pages = {231},
  
	author = {R. Carrasco-Davis and E. Reyes and C. Valenzuela and F. F{\"o}rster and P. A. Est{\'{e}
}vez and G. Pignata and F. E. Bauer and I. Reyes and P. S{\'{a}}nchez-S{\'{a}}ez and G. Cabrera-Vives and S. Eyheramendy and M. Catelan and J. Arredondo and E. Castillo-Navarrete and D. Rodr{\'{\i}}guez-Mancini and D. Ruz-Mieres and A. Moya and L. Sabatini-Gacit{\'{u}}a and C. Sep{\'{u}}lveda-Cobo and A. A. Mahabal and J. Silva-Farf{\'{a}}n and E. Camacho-I{\~{n}}iguez and L. Galbany},
  
	title = {Alert Classification for the {ALeRCE} Broker System: The Real-time Stamp Classifier},
  
	journal = {The Astronomical Journal}
}

@ARTICLE{2007ApJ...665.1246B,
       author = {{Bailey}, S. and {Aragon}, C. and {Romano}, R. and {Thomas}, R.~C. and {Weaver}, B.~A. and {Wong}, D.},
        title = "{How to Find More Supernovae with Less Work: Object Classification Techniques for Difference Imaging}",
      journal = {\apj},
         year = 2007,
        month = aug,
       volume = {665},
       number = {2},
        pages = {1246-1253},
          doi = {10.1086/519832},
archivePrefix = {arXiv},
       eprint = {0705.0493},
 primaryClass = {astro-ph},
       adsurl = {https://ui.adsabs.harvard.edu/abs/2007ApJ...665.1246B}
}

@ARTICLE{2004NewAR..48..637W,
       author = {{Wood-Vasey}, W.~M. and {Aldering}, G. and {Lee}, B.~C. and {Loken}, S. and {Nugent}, P. and {Perlmutter}, S. and {Siegrist}, J. and {Wang}, L. and {Antilogus}, P. and {Astier}, P. and {Hardin}, D. and {Pain}, R. and {Copin}, Y. and {Smadja}, G. and {Gangler}, E. and {Castera}, A. and {Adam}, G. and {Bacon}, R. and {Lemonnier}, J. -P. and {P{\'e}contal}, A. and {P{\'e}contal}, E. and {Kessler}, R.},
        title = "{The Nearby Supernova Factory}",
      journal = {\nar},
         year = 2004,
        month = may,
       volume = {48},
       number = {7-8},
        pages = {637-640},
          doi = {10.1016/j.newar.2003.12.056},
archivePrefix = {arXiv},
       eprint = {astro-ph/0401513},
 primaryClass = {astro-ph},
       adsurl = {https://ui.adsabs.harvard.edu/abs/2004NewAR..48..637W}
}

@ARTICLE{2009PASP..121.1395L,
       author = {{Law}, Nicholas M. and {Kulkarni}, Shrinivas R. and {Dekany}, Richard G. and {Ofek}, Eran O. and {Quimby}, Robert M. and {Nugent}, Peter E. and {Surace}, Jason and {Grillmair}, Carl C. and {Bloom}, Joshua S. and {Kasliwal}, Mansi M. and {Bildsten}, Lars and {Brown}, Tim and {Cenko}, S. Bradley and {Ciardi}, David and {Croner}, Ernest and {Djorgovski}, S. George and {van Eyken}, Julian and {Filippenko}, Alexei V. and {Fox}, Derek B. and {Gal-Yam}, Avishay and {Hale}, David and {Hamam}, Nouhad and {Helou}, George and {Henning}, John and {Howell}, D. Andrew and {Jacobsen}, Janet and {Laher}, Russ and {Mattingly}, Sean and {McKenna}, Dan and {Pickles}, Andrew and {Poznanski}, Dovi and {Rahmer}, Gustavo and {Rau}, Arne and {Rosing}, Wayne and {Shara}, Michael and {Smith}, Roger and {Starr}, Dan and {Sullivan}, Mark and {Velur}, Viswa and {Walters}, Richard and {Zolkower}, Jeff},
        title = "{The Palomar Transient Factory: System Overview, Performance, and First Results}",
      journal = {\pasp},
         year = 2009,
        month = dec,
       volume = {121},
       number = {886},
        pages = {1395},
          doi = {10.1086/648598},
archivePrefix = {arXiv},
       eprint = {0906.5350},
 primaryClass = {astro-ph.IM},
       adsurl = {https://ui.adsabs.harvard.edu/abs/2009PASP..121.1395L}
}

@article{10.1093/mnras/stt1306,
    author = {Brink, Henrik and Richards, Joseph W. and Poznanski, Dovi and Bloom, Joshua S. and Rice, John and Negahban, Sahand and Wainwright, Martin},
    title = "{Using machine learning for discovery in synoptic survey imaging data}",
    journal = {Monthly Notices of the Royal Astronomical Society},
    volume = {435},
    number = {2},
    pages = {1047-1060},
    year = {2013},
    month = {08},
    issn = {0035-8711},
    doi = {10.1093/mnras/stt1306},
    url = {https://doi.org/10.1093/mnras/stt1306},
    eprint = {https://academic.oup.com/mnras/article-pdf/435/2/1047/3460716/stt1306.pdf},
}

@ARTICLE{2017PASA...34...37A,
       author = {{Andreoni}, I. and {Jacobs}, C. and {Hegarty}, S. and {Pritchard}, T. and {Cooke}, J. and {Ryder}, S.},
        title = "{Mary, a Pipeline to Aid Discovery of Optical Transients}",
      journal = {\pasa},
         year = 2017,
        month = sep,
       volume = {34},
          eid = {e037},
        pages = {e037},
          doi = {10.1017/pasa.2017.33},
archivePrefix = {arXiv},
       eprint = {1708.04629},
 primaryClass = {astro-ph.IM},
       adsurl = {https://ui.adsabs.harvard.edu/abs/2017PASA...34...37A}
}

@article{JMLR:v15:srivastava14a,
  author  = {Nitish Srivastava and Geoffrey Hinton and Alex Krizhevsky and Ilya Sutskever and Ruslan Salakhutdinov},
  title   = {Dropout: A Simple Way to Prevent Neural Networks from Overfitting},
  journal = {Journal of Machine Learning Research},
  year    = {2014},
  volume  = {15},
  number  = {56},
  pages   = {1929--1958},
  url     = {http://jmlr.org/papers/v15/srivastava14a.html}
}

@inproceedings{inproceedings_nair,
author = {Nair, Vinod and Hinton, Geoffrey},
year = {2010},
month = {06},
pages = {807-814},
title = {Rectified Linear Units Improve Restricted Boltzmann Machines Vinod Nair},
volume = {27},
journal = {Proceedings of ICML}
}

@article{adam,
author = {Kingma, Diederik and Ba, Jimmy},
year = {2014},
month = {12},
pages = {},
title = {Adam: A Method for Stochastic Optimization},
journal = {International Conference on Learning Representations}
}

@article{S-PLUS,
    author = {Santos, A and Kilpatrick, C D and Bom, C R and Darc, P and Herpich, F R and Lacerda, E A D and Sartori, M J and Alvarez-Candal, A and Mendes de Oliveira, C and Kanaan, A and Ribeiro, T and Schoenell, W},
    title = "{The S-PLUS Transient Extension Program: imaging pipeline, transient identification, and survey optimization for multimessenger astronomy}",
    journal = {Monthly Notices of the Royal Astronomical Society},
    volume = {529},
    number = {1},
    pages = {59-73},
    year = {2024},
    month = {02},
    issn = {0035-8711},
    doi = {10.1093/mnras/stae466},
    url = {https://doi.org/10.1093/mnras/stae466},
    eprint = {https://academic.oup.com/mnras/article-pdf/529/1/59/56792801/stae466.pdf},
}

@ARTICLE{2019PASP..131c8002M,
       author = {{Mahabal}, Ashish and {Rebbapragada}, Umaa and {Walters}, Richard and {Masci}, Frank J. and {Blagorodnova}, Nadejda and {van Roestel}, Jan and {Ye}, Quan-Zhi and {Biswas}, Rahul and {Burdge}, Kevin and {Chang}, Chan-Kao and {Duev}, Dmitry A. and {Golkhou}, V. Zach and {Miller}, Adam A. and {Nordin}, Jakob and {Ward}, Charlotte and {Adams}, Scott and {Bellm}, Eric C. and {Branton}, Doug and {Bue}, Brian and {Cannella}, Chris and {Connolly}, Andrew and {Dekany}, Richard and {Feindt}, Ulrich and {Hung}, Tiara and {Fortson}, Lucy and {Frederick}, Sara and {Fremling}, C. and {Gezari}, Suvi and {Graham}, Matthew and {Groom}, Steven and {Kasliwal}, Mansi M. and {Kulkarni}, Shrinivas and {Kupfer}, Thomas and {Lin}, Hsing Wen and {Lintott}, Chris and {Lunnan}, Ragnhild and {Parejko}, John and {Prince}, Thomas A. and {Riddle}, Reed and {Rusholme}, Ben and {Saunders}, Nicholas and {Sedaghat}, Nima and {Shupe}, David L. and {Singer}, Leo P. and {Soumagnac}, Maayane T. and {Szkody}, Paula and {Tachibana}, Yutaro and {Tirumala}, Kushal and {van Velzen}, Sjoert and {Wright}, Darryl},
        title = "{Machine Learning for the Zwicky Transient Facility}",
      journal = {\pasp},
         year = 2019,
        month = mar,
       volume = {131},
       number = {997},
        pages = {038002},
          doi = {10.1088/1538-3873/aaf3fa},
archivePrefix = {arXiv},
       eprint = {1902.01936},
 primaryClass = {astro-ph.IM},
       adsurl = {https://ui.adsabs.harvard.edu/abs/2019PASP..131c8002M}
}

@ARTICLE{2016arXiv161205560C,
       author = {{Chambers}, K.~C. and {Magnier}, E.~A. and {Metcalfe}, N. and {Flewelling}, H.~A. and {Huber}, M.~E. and {Waters}, C.~Z. and {Denneau}, L. and {Draper}, P.~W. and {Farrow}, D. and {Finkbeiner}, D.~P. and {Holmberg}, C. and {Koppenhoefer}, J. and {Price}, P.~A. and {Rest}, A. and {Saglia}, R.~P. and {Schlafly}, E.~F. and {Smartt}, S.~J. and {Sweeney}, W. and {Wainscoat}, R.~J. and {Burgett}, W.~S. and {Chastel}, S. and {Grav}, T. and {Heasley}, J.~N. and {Hodapp}, K.~W. and {Jedicke}, R. and {Kaiser}, N. and {Kudritzki}, R. -P. and {Luppino}, G.~A. and {Lupton}, R.~H. and {Monet}, D.~G. and {Morgan}, J.~S. and {Onaka}, P.~M. and {Shiao}, B. and {Stubbs}, C.~W. and {Tonry}, J.~L. and {White}, R. and {Ba{\~n}ados}, E. and {Bell}, E.~F. and {Bender}, R. and {Bernard}, E.~J. and {Boegner}, M. and {Boffi}, F. and {Botticella}, M.~T. and {Calamida}, A. and {Casertano}, S. and {Chen}, W. -P. and {Chen}, X. and {Cole}, S. and {Deacon}, N. and {Frenk}, C. and {Fitzsimmons}, A. and {Gezari}, S. and {Gibbs}, V. and {Goessl}, C. and {Goggia}, T. and {Gourgue}, R. and {Goldman}, B. and {Grant}, P. and {Grebel}, E.~K. and {Hambly}, N.~C. and {Hasinger}, G. and {Heavens}, A.~F. and {Heckman}, T.~M. and {Henderson}, R. and {Henning}, T. and {Holman}, M. and {Hopp}, U. and {Ip}, W. -H. and {Isani}, S. and {Jackson}, M. and {Keyes}, C.~D. and {Koekemoer}, A.~M. and {Kotak}, R. and {Le}, D. and {Liska}, D. and {Long}, K.~S. and {Lucey}, J.~R. and {Liu}, M. and {Martin}, N.~F. and {Masci}, G. and {McLean}, B. and {Mindel}, E. and {Misra}, P. and {Morganson}, E. and {Murphy}, D.~N.~A. and {Obaika}, A. and {Narayan}, G. and {Nieto-Santisteban}, M.~A. and {Norberg}, P. and {Peacock}, J.~A. and {Pier}, E.~A. and {Postman}, M. and {Primak}, N. and {Rae}, C. and {Rai}, A. and {Riess}, A. and {Riffeser}, A. and {Rix}, H.~W. and {R{\"o}ser}, S. and {Russel}, R. and {Rutz}, L. and {Schilbach}, E. and {Schultz}, A.~S.~B. and {Scolnic}, D. and {Strolger}, L. and {Szalay}, A. and {Seitz}, S. and {Small}, E. and {Smith}, K.~W. and {Soderblom}, D.~R. and {Taylor}, P. and {Thomson}, R. and {Taylor}, A.~N. and {Thakar}, A.~R. and {Thiel}, J. and {Thilker}, D. and {Unger}, D. and {Urata}, Y. and {Valenti}, J. and {Wagner}, J. and {Walder}, T. and {Walter}, F. and {Watters}, S.~P. and {Werner}, S. and {Wood-Vasey}, W.~M. and {Wyse}, R.},
        title = "{The Pan-STARRS1 Surveys}",
      journal = {arXiv e-prints},
         year = 2016,
        month = dec,
          eid = {arXiv:1612.05560},
        pages = {arXiv:1612.05560},
          doi = {10.48550/arXiv.1612.05560},
archivePrefix = {arXiv},
       eprint = {1612.05560},
 primaryClass = {astro-ph.IM},
       adsurl = {https://ui.adsabs.harvard.edu/abs/2016arXiv161205560C}
}

@ARTICLE{2019ApJ...873..111I,
       author = {{Ivezi{\'c}}, {\v{Z}}eljko and {Kahn}, Steven M. and {Tyson}, J. Anthony and {Abel}, Bob and {Acosta}, Emily and {Allsman}, Robyn and {Alonso}, David and {AlSayyad}, Yusra and {Anderson}, Scott F. and {Andrew}, John and {Angel}, James Roger P. and {Angeli}, George Z. and {Ansari}, Reza and {Antilogus}, Pierre and {Araujo}, Constanza and {Armstrong}, Robert and {Arndt}, Kirk T. and {Astier}, Pierre and {Aubourg}, {\'E}ric and {Auza}, Nicole and {Axelrod}, Tim S. and {Bard}, Deborah J. and {Barr}, Jeff D. and {Barrau}, Aurelian and {Bartlett}, James G. and {Bauer}, Amanda E. and {Bauman}, Brian J. and {Baumont}, Sylvain and {Bechtol}, Ellen and {Bechtol}, Keith and {Becker}, Andrew C. and {Becla}, Jacek and {Beldica}, Cristina and {Bellavia}, Steve and {Bianco}, Federica B. and {Biswas}, Rahul and {Blanc}, Guillaume and {Blazek}, Jonathan and {Blandford}, Roger D. and {Bloom}, Josh S. and {Bogart}, Joanne and {Bond}, Tim W. and {Booth}, Michael T. and {Borgland}, Anders W. and {Borne}, Kirk and {Bosch}, James F. and {Boutigny}, Dominique and {Brackett}, Craig A. and {Bradshaw}, Andrew and {Brandt}, William Nielsen and {Brown}, Michael E. and {Bullock}, James S. and {Burchat}, Patricia and {Burke}, David L. and {Cagnoli}, Gianpietro and {Calabrese}, Daniel and {Callahan}, Shawn and {Callen}, Alice L. and {Carlin}, Jeffrey L. and {Carlson}, Erin L. and {Chandrasekharan}, Srinivasan and {Charles-Emerson}, Glenaver and {Chesley}, Steve and {Cheu}, Elliott C. and {Chiang}, Hsin-Fang and {Chiang}, James and {Chirino}, Carol and {Chow}, Derek and {Ciardi}, David R. and {Claver}, Charles F. and {Cohen-Tanugi}, Johann and {Cockrum}, Joseph J. and {Coles}, Rebecca and {Connolly}, Andrew J. and {Cook}, Kem H. and {Cooray}, Asantha and {Covey}, Kevin R. and {Cribbs}, Chris and {Cui}, Wei and {Cutri}, Roc and {Daly}, Philip N. and {Daniel}, Scott F. and {Daruich}, Felipe and {Daubard}, Guillaume and {Daues}, Greg and {Dawson}, William and {Delgado}, Francisco and {Dellapenna}, Alfred and {de Peyster}, Robert and {de Val-Borro}, Miguel and {Digel}, Seth W. and {Doherty}, Peter and {Dubois}, Richard and {Dubois-Felsmann}, Gregory P. and {Durech}, Josef and {Economou}, Frossie and {Eifler}, Tim and {Eracleous}, Michael and {Emmons}, Benjamin L. and {Fausti Neto}, Angelo and {Ferguson}, Henry and {Figueroa}, Enrique and {Fisher-Levine}, Merlin and {Focke}, Warren and {Foss}, Michael D. and {Frank}, James and {Freemon}, Michael D. and {Gangler}, Emmanuel and {Gawiser}, Eric and {Geary}, John C. and {Gee}, Perry and {Geha}, Marla and {Gessner}, Charles J.~B. and {Gibson}, Robert R. and {Gilmore}, D. Kirk and {Glanzman}, Thomas and {Glick}, William and {Goldina}, Tatiana and {Goldstein}, Daniel A. and {Goodenow}, Iain and {Graham}, Melissa L. and {Gressler}, William J. and {Gris}, Philippe and {Guy}, Leanne P. and {Guyonnet}, Augustin and {Haller}, Gunther and {Harris}, Ron and {Hascall}, Patrick A. and {Haupt}, Justine and {Hernandez}, Fabio and {Herrmann}, Sven and {Hileman}, Edward and {Hoblitt}, Joshua and {Hodgson}, John A. and {Hogan}, Craig and {Howard}, James D. and {Huang}, Dajun and {Huffer}, Michael E. and {Ingraham}, Patrick and {Innes}, Walter R. and {Jacoby}, Suzanne H. and {Jain}, Bhuvnesh and {Jammes}, Fabrice and {Jee}, M. James and {Jenness}, Tim and {Jernigan}, Garrett and {Jevremovi{\'c}}, Darko and {Johns}, Kenneth and {Johnson}, Anthony S. and {Johnson}, Margaret W.~G. and {Jones}, R. Lynne and {Juramy-Gilles}, Claire and {Juri{\'c}}, Mario and {Kalirai}, Jason S. and {Kallivayalil}, Nitya J. and {Kalmbach}, Bryce and {Kantor}, Jeffrey P. and {Karst}, Pierre and {Kasliwal}, Mansi M. and {Kelly}, Heather and {Kessler}, Richard and {Kinnison}, Veronica and {Kirkby}, David and {Knox}, Lloyd and {Kotov}, Ivan V. and {Krabbendam}, Victor L. and {Krughoff}, K. Simon and {Kub{\'a}nek}, Petr and {Kuczewski}, John and {Kulkarni}, Shri and {Ku}, John and {Kurita}, Nadine R. and {Lage}, Craig S. and {Lambert}, Ron and {Lange}, Travis and {Langton}, J. Brian and {Le Guillou}, Laurent and {Levine}, Deborah and {Liang}, Ming and {Lim}, Kian-Tat and {Lintott}, Chris J. and {Long}, Kevin E. and {Lopez}, Margaux and {Lotz}, Paul J. and {Lupton}, Robert H. and {Lust}, Nate B. and {MacArthur}, Lauren A. and {Mahabal}, Ashish and {Mandelbaum}, Rachel and {Markiewicz}, Thomas W. and {Marsh}, Darren S. and {Marshall}, Philip J. and {Marshall}, Stuart and {May}, Morgan and {McKercher}, Robert and {McQueen}, Michelle and {Meyers}, Joshua and {Migliore}, Myriam and {Miller}, Michelle and {Mills}, David J. and {Miraval}, Connor and {Moeyens}, Joachim and {Moolekamp}, Fred E. and {Monet}, David G. and {Moniez}, Marc and {Monkewitz}, Serge and {Montgomery}, Christopher and {Morrison}, Christopher B. and {Mueller}, Fritz and {Muller}, Gary P. and {Mu{\~n}oz Arancibia}, Freddy and {Neill}, Douglas R. and {Newbry}, Scott P. and {Nief}, Jean-Yves and {Nomerotski}, Andrei and {Nordby}, Martin and {O'Connor}, Paul and {Oliver}, John and {Olivier}, Scot S. and {Olsen}, Knut and {O'Mullane}, William and {Ortiz}, Sandra and {Osier}, Shawn and {Owen}, Russell E. and {Pain}, Reynald and {Palecek}, Paul E. and {Parejko}, John K. and {Parsons}, James B. and {Pease}, Nathan M. and {Peterson}, J. Matt and {Peterson}, John R. and {Petravick}, Donald L. and {Libby Petrick}, M.~E. and {Petry}, Cathy E. and {Pierfederici}, Francesco and {Pietrowicz}, Stephen and {Pike}, Rob and {Pinto}, Philip A. and {Plante}, Raymond and {Plate}, Stephen and {Plutchak}, Joel P. and {Price}, Paul A. and {Prouza}, Michael and {Radeka}, Veljko and {Rajagopal}, Jayadev and {Rasmussen}, Andrew P. and {Regnault}, Nicolas and {Reil}, Kevin A. and {Reiss}, David J. and {Reuter}, Michael A. and {Ridgway}, Stephen T. and {Riot}, Vincent J. and {Ritz}, Steve and {Robinson}, Sean and {Roby}, William and {Roodman}, Aaron and {Rosing}, Wayne and {Roucelle}, Cecille and {Rumore}, Matthew R. and {Russo}, Stefano and {Saha}, Abhijit and {Sassolas}, Benoit and {Schalk}, Terry L. and {Schellart}, Pim and {Schindler}, Rafe H. and {Schmidt}, Samuel and {Schneider}, Donald P. and {Schneider}, Michael D. and {Schoening}, William and {Schumacher}, German and {Schwamb}, Megan E. and {Sebag}, Jacques and {Selvy}, Brian and {Sembroski}, Glenn H. and {Seppala}, Lynn G. and {Serio}, Andrew and {Serrano}, Eduardo and {Shaw}, Richard A. and {Shipsey}, Ian and {Sick}, Jonathan and {Silvestri}, Nicole and {Slater}, Colin T. and {Smith}, J. Allyn and {Smith}, R. Chris and {Sobhani}, Shahram and {Soldahl}, Christine and {Storrie-Lombardi}, Lisa and {Stover}, Edward and {Strauss}, Michael A. and {Street}, Rachel A. and {Stubbs}, Christopher W. and {Sullivan}, Ian S. and {Sweeney}, Donald and {Swinbank}, John D. and {Szalay}, Alexander and {Takacs}, Peter and {Tether}, Stephen A. and {Thaler}, Jon J. and {Thayer}, John Gregg and {Thomas}, Sandrine and {Thornton}, Adam J. and {Thukral}, Vaikunth and {Tice}, Jeffrey and {Trilling}, David E. and {Turri}, Max and {Van Berg}, Richard and {Vanden Berk}, Daniel and {Vetter}, Kurt and {Virieux}, Francoise and {Vucina}, Tomislav and {Wahl}, William and {Walkowicz}, Lucianne and {Walsh}, Brian and {Walter}, Christopher W. and {Wang}, Daniel L. and {Wang}, Shin-Yawn and {Warner}, Michael and {Wiecha}, Oliver and {Willman}, Beth and {Winters}, Scott E. and {Wittman}, David and {Wolff}, Sidney C. and {Wood-Vasey}, W. Michael and {Wu}, Xiuqin and {Xin}, Bo and {Yoachim}, Peter and {Zhan}, Hu},
        title = "{LSST: From Science Drivers to Reference Design and Anticipated Data Products}",
      journal = {\apj},
         year = 2019,
        month = mar,
       volume = {873},
       number = {2},
          eid = {111},
        pages = {111},
          doi = {10.3847/1538-4357/ab042c},
archivePrefix = {arXiv},
       eprint = {0805.2366},
 primaryClass = {astro-ph},
       adsurl = {https://ui.adsabs.harvard.edu/abs/2019ApJ...873..111I}
}

@ARTICLE{5288526,
  author={Pan, Sinno Jialin and Yang, Qiang},
  journal={IEEE Transactions on Knowledge and Data Engineering}, 
  title={A Survey on Transfer Learning}, 
  year={2010},
  volume={22},
  number={10},
  pages={1345-1359},
  doi={10.1109/TKDE.2009.191}}

@ARTICLE{2018PASP..130f4505T,
       author = {{Tonry}, J.~L. and {Denneau}, L. and {Heinze}, A.~N. and {Stalder}, B. and {Smith}, K.~W. and {Smartt}, S.~J. and {Stubbs}, C.~W. and {Weiland}, H.~J. and {Rest}, A.},
        title = "{ATLAS: A High-cadence All-sky Survey System}",
      journal = {\pasp},
         year = 2018,
        month = jun,
       volume = {130},
       number = {988},
        pages = {064505},
          doi = {10.1088/1538-3873/aabadf},
archivePrefix = {arXiv},
       eprint = {1802.00879},
 primaryClass = {astro-ph.IM},
       adsurl = {https://ui.adsabs.harvard.edu/abs/2018PASP..130f4505T}
}

@article{10.1093/mnras/stac013,
    author = {Steeghs, D and Galloway, D K and Ackley, K and Dyer, M J and Lyman, J and Ulaczyk, K and Cutter, R and Mong, Y-L and Dhillon, V and O’Brien, P and Ramsay, G and Poshyachinda, S and Kotak, R and Nuttall, L K and Pallé, E and Breton, R P and Pollacco, D and Thrane, E and Aukkaravittayapun, S and Awiphan, S and Burhanudin, U and Chote, P and Chrimes, A and Daw, E and Duffy, C and Eyles-Ferris, R and Gompertz, B and Heikkilä, T and Irawati, P and Kennedy, M R and Killestein, T and Kuncarayakti, H and Levan, A J and Littlefair, S and Makrygianni, L and Marsh, T and Mata-Sanchez, D and Mattila, S and Maund, J and McCormac, J and Mkrtichian, D and Mullaney, J and Noysena, K and Patel, M and Rol, E and Sawangwit, U and Stanway, E R and Starling, R and Strøm, P and Tooke, S and West, R and White, D J and Wiersema, K},
    title = "{The Gravitational-wave Optical Transient Observer (GOTO): prototype performance and prospects for transient science}",
    journal = {Monthly Notices of the Royal Astronomical Society},
    volume = {511},
    number = {2},
    pages = {2405-2422},
    year = {2022},
    month = {01},
    issn = {0035-8711},
    doi = {10.1093/mnras/stac013},
    url = {https://doi.org/10.1093/mnras/stac013},
    eprint = {https://academic.oup.com/mnras/article-pdf/511/2/2405/48413143/stac013.pdf},
}

@article{Richards_2011,
doi = {10.1088/0004-637X/733/1/10},
url = {https://dx.doi.org/10.1088/0004-637X/733/1/10},
year = {2011},
month = {apr},
publisher = {The American Astronomical Society},
volume = {733},
number = {1},
pages = {10},
author = {Joseph W. Richards and Dan L. Starr and Nathaniel R. Butler and Joshua S. Bloom and John M. Brewer and Arien Crellin-Quick and Justin Higgins and Rachel Kennedy and Maxime Rischard},
title = {ON MACHINE-LEARNED CLASSIFICATION OF VARIABLE STARS WITH SPARSE AND NOISY TIME-SERIES DATA},
journal = {The Astrophysical Journal}
}

@article{Martínez-Palomera_2018,
doi = {10.3847/1538-3881/aadfd8},
url = {https://dx.doi.org/10.3847/1538-3881/aadfd8},
year = {2018},
month = {oct},
publisher = {The American Astronomical Society},
volume = {156},
number = {5},
pages = {186},
author = {Jorge Martínez-Palomera and Francisco Förster and Pavlos Protopapas and Juan Carlos Maureira and Paulina Lira and Guillermo Cabrera-Vives and Pablo Huijse and Lluis Galbany and Thomas de Jaeger and Santiago González-Gaitán and Gustavo Medina and Giuliano Pignata and Jaime San Martín and Mario Hamuy and Ricardo R. Muñoz},
title = {The High Cadence Transit Survey (HiTS): Compilation and Characterization of Light-curve Catalogs},
journal = {The Astronomical Journal}
}

@article{Boone_2019,
doi = {10.3847/1538-3881/ab5182},
url = {https://dx.doi.org/10.3847/1538-3881/ab5182},
year = {2019},
month = {dec},
publisher = {The American Astronomical Society},
volume = {158},
number = {6},
pages = {257},
author = {Kyle Boone},
title = {Avocado: Photometric Classification of Astronomical Transients with Gaussian Process Augmentation},
journal = {The Astronomical Journal}
}

@article{Pichara_2016,
doi = {10.3847/0004-637X/819/1/18},
url = {https://dx.doi.org/10.3847/0004-637X/819/1/18},
year = {2016},
month = {feb},
publisher = {The American Astronomical Society},
volume = {819},
number = {1},
pages = {18},
author = {Karim Pichara and Pavlos Protopapas and Daniel León},
title = {META-CLASSIFICATION FOR VARIABLE STARS},
journal = {The Astrophysical Journal}
}

@INPROCEEDINGS{8280984,
  author={Mahabal, A and Sheth, K and Gieseke, F and Pai, A and Djorgovski, S G and Drake, A J and Graham, M J},
  booktitle={2017 IEEE Symposium Series on Computational Intelligence (SSCI)}, 
  title={Deep-learnt classification of light curves}, 
  year={2017},
  volume={},
  number={},
  pages={1-8},
  doi={10.1109/SSCI.2017.8280984}}

@ARTICLE{2018NatAs...2..151N,
       author = {{Naul}, Brett and {Bloom}, Joshua S. and {P{\'e}rez}, Fernando and {van der Walt}, St{\'e}fan},
        title = "{A recurrent neural network for classification of unevenly sampled variable stars}",
      journal = {Nature Astronomy},
         year = 2018,
        month = nov,
       volume = {2},
        pages = {151-155},
          doi = {10.1038/s41550-017-0321-z},
archivePrefix = {arXiv},
       eprint = {1711.10609},
 primaryClass = {astro-ph.IM},
       adsurl = {https://ui.adsabs.harvard.edu/abs/2018NatAs...2..151N}
}

@ARTICLE{2019PASP..131k8002M,
       author = {{Muthukrishna}, Daniel and {Narayan}, Gautham and {Mandel}, Kaisey S. and {Biswas}, Rahul and {Hlo{\v{z}}ek}, Ren{\'e}e},
        title = "{RAPID: Early Classification of Explosive Transients Using Deep Learning}",
      journal = {\pasp},
         year = 2019,
        month = nov,
       volume = {131},
       number = {1005},
        pages = {118002},
          doi = {10.1088/1538-3873/ab1609},
archivePrefix = {arXiv},
       eprint = {1904.00014},
 primaryClass = {astro-ph.IM},
       adsurl = {https://ui.adsabs.harvard.edu/abs/2019PASP..131k8002M}
}

@article{
doi:10.1073/pnas.1611835114,
author = {James Kirkpatrick  and Razvan Pascanu  and Neil Rabinowitz  and Joel Veness  and Guillaume Desjardins  and Andrei A. Rusu  and Kieran Milan  and John Quan  and Tiago Ramalho  and Agnieszka Grabska-Barwinska  and Demis Hassabis  and Claudia Clopath  and Dharshan Kumaran  and Raia Hadsell },
title = {Overcoming catastrophic forgetting in neural networks},
journal = {Proceedings of the National Academy of Sciences},
volume = {114},
number = {13},
pages = {3521-3526},
year = {2017},
doi = {10.1073/pnas.1611835114},
URL = {https://www.pnas.org/doi/abs/10.1073/pnas.1611835114},
eprint = {https://www.pnas.org/doi/pdf/10.1073/pnas.1611835114}}

@inproceedings{Ribani_TL_CNN,
author = {Ribani, Ricardo and Marengoni, Maurício},
year = {2019},
month = {10},
pages = {47-57},
title = {A Survey of Transfer Learning for Convolutional Neural Networks},
doi = {10.1109/SIBGRAPI-T.2019.00010}
}

@article{Soares_Santos_2017,
   title={The Electromagnetic Counterpart of the Binary Neutron Star Merger LIGO/Virgo GW170817. I. Discovery of the Optical Counterpart Using the Dark Energy Camera},
   volume={848},
   ISSN={2041-8213},
   url={http://dx.doi.org/10.3847/2041-8213/aa9059},
   DOI={10.3847/2041-8213/aa9059},
   number={2},
   journal={The Astrophysical Journal Letters},
   publisher={American Astronomical Society},
   author={Soares-Santos, M. and Holz, D. E. and Annis, J. and Chornock, R. and Herner, K. and Berger, E. and Brout, D. and Chen, H.-Y. and Kessler, R. and Sako, M. and Allam, S. and Tucker, D. L. and Butler, R. E. and Palmese, A. and Doctor, Z. and Diehl, H. T. and Frieman, J. and Yanny, B. and Lin, H. and Scolnic, D. and Cowperthwaite, P. and Neilsen, E. and Marriner, J. and Kuropatkin, N. and Hartley, W. G. and Paz-Chinchón, F. and Alexander, K. D. and Balbinot, E. and Blanchard, P. and Brown, D. A. and Carlin, J. L. and Conselice, C. and Cook, E. R. and Drlica-Wagner, A. and Drout, M. R. and Durret, F. and Eftekhari, T. and Farr, B. and Finley, D. A. and Foley, R. J. and Fong, W. and Fryer, C. L. and García-Bellido, J. and Gill, M. S . S. and Gruendl, R. A. and Hanna, C. and Kasen, D. and Li, T. S. and Lopes, P. A. A. and Lourenço, A. C. C. and Margutti, R. and Marshall, J. L. and Matheson, T. and Medina, G. E. and Metzger, B. D. and Muñoz, R. R. and Muir, J. and Nicholl, M. and Quataert, E. and Rest, A. and Sauseda, M. and Schlegel, D. J. and Secco, L. F. and Sobreira, F. and Stebbins, A. and Villar, V. A. and Vivas, K. and Walker, A. R. and Wester, W. and Williams, P. K. G. and Zenteno, A. and Zhang, Y. and Abbott, T. M. C. and Abdalla, F. B. and Banerji, M. and Bechtol, K. and Benoit-Lévy, A. and Bertin, E. and Brooks, D. and Buckley-Geer, E. and Burke, D. L. and Rosell, A. Carnero and Kind, M. Carrasco and Carretero, J. and Castander, F. J. and Crocce, M. and Cunha, C. E. and D’Andrea, C. B. and Costa, L. N. da and Davis, C. and Desai, S. and Dietrich, J. P. and Doel, P. and Eifler, T. F. and Fernandez, E. and Flaugher, B. and Fosalba, P. and Gaztanaga, E. and Gerdes, D. W. and Giannantonio, T. and Goldstein, D. A. and Gruen, D. and Gschwend, J. and Gutierrez, G. and Honscheid, K. and Jain, B. and James, D. J. and Jeltema, T. and Johnson, M. W. G. and Johnson, M. D. and Kent, S. and Krause, E. and Kron, R. and Kuehn, K. and Kuhlmann, S. and Lahav, O. and Lima, M. and Maia, M. A. G. and March, M. and McMahon, R. G. and Menanteau, F. and Miquel, R. and Mohr, J. J. and Nichol, R. C. and Nord, B. and Ogando, R. L. C. and Petravick, D. and Plazas, A. A. and Romer, A. K. and Roodman, A. and Rykoff, E. S. and Sanchez, E. and Scarpine, V. and Schubnell, M. and Sevilla-Noarbe, I. and Smith, M. and Smith, R. C. and Suchyta, E. and Swanson, M. E. C. and Tarle, G. and Thomas, D. and Thomas, R. C. and Troxel, M. A. and Vikram, V. and Wechsler, R. H. and Weller, J.},
   year={2017},
   month=oct, pages={L16} }

@ARTICLE{2025MNRAS.538..133P,
       author = {{Pranshu}, Kumar and {Misra}, Kuntal and {Ailawadhi}, Bhavya and {Dubey}, Monalisa and {Dukiya}, Naveen and {Filali}, Sara and {Hickson}, Paul and {Kumar}, Brajesh and {Negi}, Vibhore and {Surdej}, Jean},
        title = "{PYLMT : a transient detection pipeline for the 4-m International Liquid Mirror Telescope}",
      journal = {\mnras},
         year = 2025,
        month = mar,
       volume = {538},
       number = {1},
        pages = {133-152},
          doi = {10.1093/mnras/staf206},
archivePrefix = {arXiv},
       eprint = {2502.00556},
 primaryClass = {astro-ph.IM},
       adsurl = {https://ui.adsabs.harvard.edu/abs/2025MNRAS.538..133P}
}

@article{Cabrera_Vives_2017,
   title={Deep-HiTS: Rotation Invariant Convolutional Neural Network for Transient Detection∗},
   volume={836},
   ISSN={1538-4357},
   url={http://dx.doi.org/10.3847/1538-4357/836/1/97},
   DOI={10.3847/1538-4357/836/1/97},
   number={1},
   journal={The Astrophysical Journal},
   publisher={American Astronomical Society},
   author={Cabrera-Vives, Guillermo and Reyes, Ignacio and Förster, Francisco and Estévez, Pablo A. and Maureira, Juan-Carlos},
   year={2017},
   month=feb, pages={97} }

@article{10.1093/mnras/stx2161,
    author = {Gieseke, Fabian and Bloemen, Steven and van den Bogaard, Cas and Heskes, Tom and Kindler, Jonas and Scalzo, Richard A. and Ribeiro, Valério A. R. M. and van Roestel, Jan and Groot, Paul J. and Yuan, Fang and Möller, Anais and Tucker, Brad E.},
    title = {Convolutional neural networks for transient candidate vetting in large-scale surveys},
    journal = {Monthly Notices of the Royal Astronomical Society},
    volume = {472},
    number = {3},
    pages = {3101-3114},
    year = {2017},
    month = {08},
    issn = {0035-8711},
    doi = {10.1093/mnras/stx2161},
    url = {https://doi.org/10.1093/mnras/stx2161},
    eprint = {https://academic.oup.com/mnras/article-pdf/472/3/3101/20131479/stx2161.pdf},
}

@ARTICLE{2020MNRAS.497.2641T,
       author = {{Turpin}, Damien and {Ganet}, M. and {Antier}, S. and {Bertin}, E. and {Xin}, L.~P. and {Leroy}, N. and {Wu}, C. and {Xu}, Y. and {Han}, X.~H. and {Cai}, H.~B. and {Li}, H.~L. and {Lu}, X.~M. and {Feng}, Q.~C. and {Wei}, J.~Y.},
        title = "{Vetting the optical transient candidates detected by the GWAC network using convolutional neural networks}",
      journal = {\mnras},
         year = 2020,
        month = sep,
       volume = {497},
       number = {3},
        pages = {2641-2650},
          doi = {10.1093/mnras/staa2046},
archivePrefix = {arXiv},
       eprint = {2001.03424},
 primaryClass = {astro-ph.IM},
       adsurl = {https://ui.adsabs.harvard.edu/abs/2020MNRAS.497.2641T}
}

@article{refId0,
	author = {{Makhlouf, K.} and {Turpin, D.} and {Corre, D.} and {Karpov, S.} and {Kann, D. A.} and {Klotz, A.}},
	title = {O’TRAIN: A robust and flexible ‘real or bogus’ classifier for the study of the optical transient sky★},
	DOI= "10.1051/0004-6361/202142952",
	url= "https://doi.org/10.1051/0004-6361/202142952",
	journal = {A&A},
	year = 2022,
	volume = 664,
	pages = "A81",
}

@ARTICLE{2023AJ....166..115A,
       author = {{Acero-Cuellar}, Tatiana and {Bianco}, Federica and {Dobler}, Gregory and {Sako}, Masao and {Qu}, Helen and {LSST Dark Energy Science Collaboration}},
        title = "{What's the Difference? The Potential for Convolutional Neural Networks for Transient Detection without Template Subtraction}",
      journal = {\aj},
         year = 2023,
        month = sep,
       volume = {166},
       number = {3},
          eid = {115},
        pages = {115},
          doi = {10.3847/1538-3881/ace9d8},
archivePrefix = {arXiv},
       eprint = {2203.07390},
 primaryClass = {cs.CV},
       adsurl = {https://ui.adsabs.harvard.edu/abs/2023AJ....166..115A}
}

@article{Geman_1992,
author = {Geman, Stuart and Bienenstock, Elie and Doursat, René},
year = {1992},
month = {01},
pages = {1-58},
title = {Neural Networks and the Bias/Variance Dilemma},
volume = {4},
journal = {Neural Computation},
doi = {10.1162/neco.1992.4.1.1}
}

@article{PARISI201954,
title = {Continual lifelong learning with neural networks: A review},
journal = {Neural Networks},
volume = {113},
pages = {54-71},
year = {2019},
issn = {0893-6080},
doi = {https://doi.org/10.1016/j.neunet.2019.01.012},
url = {https://www.sciencedirect.com/science/article/pii/S0893608019300231},
author = {German I. Parisi and Ronald Kemker and Jose L. Part and Christopher Kanan and Stefan Wermter}
}

@article{10.1093/mnras/sty3497,
    author = {Domínguez Sánchez, H and Huertas-Company, M and Bernardi, M and Kaviraj, S and Fischer, J L and Abbott, T M C and Abdalla, F B and Annis, J and Avila, S and Brooks, D and Buckley-Geer, E and Carnero Rosell, A and Carrasco Kind, M and Carretero, J and Cunha, C E and D’Andrea, C B and da Costa, L N and Davis, C and De Vicente, J and Doel, P and Evrard, A E and Fosalba, P and Frieman, J and García-Bellido, J and Gaztanaga, E and Gerdes, D W and Gruen, D and Gruendl, R A and Gschwend, J and Gutierrez, G and Hartley, W G and Hollowood, D L and Honscheid, K and Hoyle, B and James, D J and Kuehn, K and Kuropatkin, N and Lahav, O and Maia, M A G and March, M and Melchior, P and Menanteau, F and Miquel, R and Nord, B and Plazas, A A and Sanchez, E and Scarpine, V and Schindler, R and Schubnell, M and Smith, M and Smith, R C and Soares-Santos, M and Sobreira, F and Suchyta, E and Swanson, M E C and Tarle, G and Thomas, D and Walker, A R and Zuntz, J},
    title = {Transfer learning for galaxy morphology from one survey to another},
    journal = {Monthly Notices of the Royal Astronomical Society},
    volume = {484},
    number = {1},
    pages = {93-100},
    year = {2018},
    month = {12},
    issn = {0035-8711},
    doi = {10.1093/mnras/sty3497},
    url = {https://doi.org/10.1093/mnras/sty3497},
    eprint = {https://academic.oup.com/mnras/article-pdf/484/1/93/27507112/sty3497.pdf},
}

@article{Vilalta_2018,
doi = {10.1088/1742-6596/1085/5/052014},
url = {https://dx.doi.org/10.1088/1742-6596/1085/5/052014},
year = {2018},
month = {sep},
publisher = {IOP Publishing},
volume = {1085},
number = {5},
pages = {052014},
author = {Vilalta, Ricardo},
title = {Transfer Learning in Astronomy: A New Machine-Learning Paradigm},
journal = {Journal of Physics: Conference Series}
}

@ARTICLE{2021A&A...653A..22K,
       author = {{Kim}, Dae-Won and {Yeo}, Doyeob and {Bailer-Jones}, Coryn A.~L. and {Lee}, Giyoung},
        title = "{Deep transfer learning for the classification of variable sources}",
      journal = {\aap},
         year = 2021,
        month = sep,
       volume = {653},
          eid = {A22},
        pages = {A22},
          doi = {10.1051/0004-6361/202140369},
archivePrefix = {arXiv},
       eprint = {2106.00187},
 primaryClass = {astro-ph.IM},
       adsurl = {https://ui.adsabs.harvard.edu/abs/2021A&A...653A..22K}
}

@article{10.1093/mnras/stad2238,
    author = {Hannon, Stephen and Whitmore, Bradley C and Lee, Janice C and Thilker, David A and Deger, Sinan and Huerta, E A and Wei, Wei and Mobasher, Bahram and Klessen, Ralf and Boquien, Médéric and Dale, Daniel A and Chevance, Mélanie and Grasha, Kathryn and Sanchez-Blazquez, Patricia and Williams, Thomas and Scheuermann, Fabian and Groves, Brent and Kim, Hwihyun and Kruijssen, J M Diederik and the PHANGS-HST Team},
    title = {Star cluster classification using deep transfer learning with PHANGS-HST},
    journal = {Monthly Notices of the Royal Astronomical Society},
    volume = {526},
    number = {2},
    pages = {2991-3006},
    year = {2023},
    month = {08},
    issn = {0035-8711},
    doi = {10.1093/mnras/stad2238},
    url = {https://doi.org/10.1093/mnras/stad2238},
    eprint = {https://academic.oup.com/mnras/article-pdf/526/2/2991/51991998/stad2238.pdf},
}

@ARTICLE{2015arXiv150203167I,
       author = {{Ioffe}, Sergey and {Szegedy}, Christian},
        title = "{Batch Normalization: Accelerating Deep Network Training by Reducing Internal Covariate Shift}",
      journal = {arXiv e-prints},
         year = 2015,
        month = feb,
          eid = {arXiv:1502.03167},
        pages = {arXiv:1502.03167},
          doi = {10.48550/arXiv.1502.03167},
archivePrefix = {arXiv},
       eprint = {1502.03167},
 primaryClass = {cs.LG},
       adsurl = {https://ui.adsabs.harvard.edu/abs/2015arXiv150203167I}
}

@ARTICLE{2018ApJ...865L...3P,
       author = {{Prentice}, S.~J. and {Maguire}, K. and {Smartt}, S.~J. and {Magee}, M.~R. and {Schady}, P. and {Sim}, S. and {Chen}, T. -W. and {Clark}, P. and {Colin}, C. and {Fulton}, M. and {McBrien}, O. and {O'Neill}, D. and {Smith}, K.~W. and {Ashall}, C. and {Chambers}, K.~C. and {Denneau}, L. and {Flewelling}, H.~A. and {Heinze}, A. and {Holoien}, T.~W. -S. and {Huber}, M.~E. and {Kochanek}, C.~S. and {Mazzali}, P.~A. and {Prieto}, J.~L. and {Rest}, A. and {Shappee}, B.~J. and {Stalder}, B. and {Stanek}, K.~Z. and {Stritzinger}, M.~D. and {Thompson}, T.~A. and {Tonry}, J.~L.},
        title = "{The Cow: Discovery of a Luminous, Hot, and Rapidly Evolving Transient}",
      journal = {\apjl},
         year = 2018,
        month = sep,
       volume = {865},
       number = {1},
          eid = {L3},
        pages = {L3},
          doi = {10.3847/2041-8213/aadd90},
archivePrefix = {arXiv},
       eprint = {1807.05965},
 primaryClass = {astro-ph.HE},
       adsurl = {https://ui.adsabs.harvard.edu/abs/2018ApJ...865L...3P}
}

@ARTICLE{2014ApJ...794...23D,
       author = {{Drout}, M.~R. and {Chornock}, R. and {Soderberg}, A.~M. and {Sanders}, N.~E. and {McKinnon}, R. and {Rest}, A. and {Foley}, R.~J. and {Milisavljevic}, D. and {Margutti}, R. and {Berger}, E. and {Calkins}, M. and {Fong}, W. and {Gezari}, S. and {Huber}, M.~E. and {Kankare}, E. and {Kirshner}, R.~P. and {Leibler}, C. and {Lunnan}, R. and {Mattila}, S. and {Marion}, G.~H. and {Narayan}, G. and {Riess}, A.~G. and {Roth}, K.~C. and {Scolnic}, D. and {Smartt}, S.~J. and {Tonry}, J.~L. and {Burgett}, W.~S. and {Chambers}, K.~C. and {Hodapp}, K.~W. and {Jedicke}, R. and {Kaiser}, N. and {Magnier}, E.~A. and {Metcalfe}, N. and {Morgan}, J.~S. and {Price}, P.~A. and {Waters}, C.},
        title = "{Rapidly Evolving and Luminous Transients from Pan-STARRS1}",
      journal = {\apj},
         year = 2014,
        month = oct,
       volume = {794},
       number = {1},
          eid = {23},
        pages = {23},
          doi = {10.1088/0004-637X/794/1/23},
archivePrefix = {arXiv},
       eprint = {1405.3668},
 primaryClass = {astro-ph.HE},
       adsurl = {https://ui.adsabs.harvard.edu/abs/2014ApJ...794...23D}
}

@ARTICLE{2019MNRAS.484.1031P,
       author = {{Perley}, Daniel A. and {Mazzali}, Paolo A. and {Yan}, Lin and {Cenko}, S. Bradley and {Gezari}, Suvi and {Taggart}, Kirsty and {Blagorodnova}, Nadia and {Fremling}, Christoffer and {Mockler}, Brenna and {Singh}, Avinash and {Tominaga}, Nozomu and {Tanaka}, Masaomi and {Watson}, Alan M. and {Ahumada}, Tom{\'a}s and {Anupama}, G.~C. and {Ashall}, Chris and {Becerra}, Rosa L. and {Bersier}, David and {Bhalerao}, Varun and {Bloom}, Joshua S. and {Butler}, Nathaniel R. and {Copperwheat}, Chris and {Coughlin}, Michael W. and {De}, Kishalay and {Drake}, Andrew J. and {Duev}, Dmitry A. and {Frederick}, Sara and {Gonz{\'a}lez}, J. Jes{\'u}s and {Goobar}, Ariel and {Heida}, Marianne and {Ho}, Anna Y.~Q. and {Horst}, John and {Hung}, Tiara and {Itoh}, Ryosuke and {Jencson}, Jacob E. and {Kasliwal}, Mansi M. and {Kawai}, Nobuyuki and {Khanam}, Tanazza and {Kulkarni}, Shrinivas R. and {Kumar}, Brajesh and {Kumar}, Harsh and {Kutyrev}, Alexander S. and {Lee}, William H. and {Maeda}, Keiichi and {Mahabal}, Ashish and {Murata}, Katsuhiro L. and {Neill}, James D. and {Ngeow}, Chow-Choong and {Penprase}, Bryan and {Pian}, Elena and {Quimby}, Robert and {Ramirez-Ruiz}, Enrico and {Richer}, Michael G. and {Rom{\'a}n-Z{\'u}{\~n}iga}, Carlos G. and {Sahu}, D.~K. and {Srivastav}, Shubham and {Socia}, Quentin and {Sollerman}, Jesper and {Tachibana}, Yutaro and {Taddia}, Francesco and {Tinyanont}, Samaporn and {Troja}, Eleonora and {Ward}, Charlotte and {Wee}, Jerrick and {Yu}, Po-Chieh},
        title = "{The fast, luminous ultraviolet transient AT2018cow: extreme supernova, or disruption of a star by an intermediate-mass black hole?}",
      journal = {\mnras},
         year = 2019,
        month = mar,
       volume = {484},
       number = {1},
        pages = {1031-1049},
          doi = {10.1093/mnras/sty3420},
archivePrefix = {arXiv},
       eprint = {1808.00969},
 primaryClass = {astro-ph.HE},
       adsurl = {https://ui.adsabs.harvard.edu/abs/2019MNRAS.484.1031P}
}

@ARTICLE{2007Sci...318..777L,
       author = {{Lorimer}, D.~R. and {Bailes}, M. and {McLaughlin}, M.~A. and {Narkevic}, D.~J. and {Crawford}, F.},
        title = "{A Bright Millisecond Radio Burst of Extragalactic Origin}",
      journal = {Science},
         year = 2007,
        month = nov,
       volume = {318},
       number = {5851},
        pages = {777},
          doi = {10.1126/science.1147532},
archivePrefix = {arXiv},
       eprint = {0709.4301},
 primaryClass = {astro-ph},
       adsurl = {https://ui.adsabs.harvard.edu/abs/2007Sci...318..777L}
}

@ARTICLE{2013Sci...341...53T,
       author = {{Thornton}, D. and {Stappers}, B. and {Bailes}, M. and {Barsdell}, B. and {Bates}, S. and {Bhat}, N.~D.~R. and {Burgay}, M. and {Burke-Spolaor}, S. and {Champion}, D.~J. and {Coster}, P. and {D'Amico}, N. and {Jameson}, A. and {Johnston}, S. and {Keith}, M. and {Kramer}, M. and {Levin}, L. and {Milia}, S. and {Ng}, C. and {Possenti}, A. and {van Straten}, W.},
        title = "{A Population of Fast Radio Bursts at Cosmological Distances}",
      journal = {Science},
         year = 2013,
        month = jul,
       volume = {341},
       number = {6141},
        pages = {53-56},
          doi = {10.1126/science.1236789},
archivePrefix = {arXiv},
       eprint = {1307.1628},
 primaryClass = {astro-ph.HE},
       adsurl = {https://ui.adsabs.harvard.edu/abs/2013Sci...341...53T}
}

@ARTICLE{1997Natur.387..783C,
       author = {{Costa}, E. and {Frontera}, F. and {Heise}, J. and {Feroci}, M. and {in't Zand}, J. and {Fiore}, F. and {Cinti}, M.~N. and {Dal Fiume}, D. and {Nicastro}, L. and {Orlandini}, M. and {Palazzi}, E. and {Rapisarda\#}, M. and {Zavattini}, G. and {Jager}, R. and {Parmar}, A. and {Owens}, A. and {Molendi}, S. and {Cusumano}, G. and {Maccarone}, M.~C. and {Giarrusso}, S. and {Coletta}, A. and {Antonelli}, L.~A. and {Giommi}, P. and {Muller}, J.~M. and {Piro}, L. and {Butler}, R.~C.},
        title = "{Discovery of an X-ray afterglow associated with the {\ensuremath{\gamma}}-ray burst of 28 February 1997}",
      journal = {\nat},
         year = 1997,
        month = jun,
       volume = {387},
       number = {6635},
        pages = {783-785},
          doi = {10.1038/42885},
archivePrefix = {arXiv},
       eprint = {astro-ph/9706065},
 primaryClass = {astro-ph},
       adsurl = {https://ui.adsabs.harvard.edu/abs/1997Natur.387..783C}
}

@ARTICLE{2025A&A...694A..80S,
       author = {{Surdej}, J. and {Hickson}, P. and {Misra}, K. and {Banerjee}, D. and {Ailawadhi}, B. and {Akhunov}, T. and {Borra}, E. and {Dubey}, M. and {Dukiya}, N. and {Filali}, S. and {Hellemeier}, J. and {Kharayat}, M. and {Kumar}, B. and {Kumar}, H. and {Kumar}, M. and {Kumar}, T.~S. and {Kumari}, P. and {Negi}, V. and {Pospieszalska-Surdej}, A. and {Prabhavu}, S. and {Pradhan}, B. and {Pranshu}, K. and {Rawat}, H. and {Reddy}, B.~K. and {Sasidharan Pillai}, A. and {Singh}, K. and {Tremblay}, S. and {Turakhia}, S. and {Vijay}, S.},
        title = "{The 4 m International Liquid Mirror Telescope: Construction, operation, and science}",
      journal = {\aap},
         year = 2025,
        month = feb,
       volume = {694},
          eid = {A80},
        pages = {A80},
          doi = {10.1051/0004-6361/202452667},
archivePrefix = {arXiv},
       eprint = {2502.00564},
 primaryClass = {astro-ph.IM},
       adsurl = {https://ui.adsabs.harvard.edu/abs/2025A&A...694A..80S}
}

@ARTICLE{2025MNRAS.542L.132G,
       author = {{Gupta}, Rithwik and {Muthukrishna}, Daniel and {Rehemtulla}, Nabeel and {Shah}, Ved},
        title = "{Transfer learning for transient classification: from simulations to real data and ZTF to LSST}",
      journal = {\mnras},
         year = 2025,
        month = sep,
       volume = {542},
       number = {1},
        pages = {L132-L138},
          doi = {10.1093/mnrasl/slaf074},
archivePrefix = {arXiv},
       eprint = {2502.18558},
 primaryClass = {astro-ph.IM},
       adsurl = {https://ui.adsabs.harvard.edu/abs/2025MNRAS.542L.132G}
}

@ARTICLE{1973ApJ...182L..85K,
       author = {{Klebesadel}, Ray W. and {Strong}, Ian B. and {Olson}, Roy A.},
        title = "{Observations of Gamma-Ray Bursts of Cosmic Origin}",
      journal = {\apjl},
         year = 1973,
        month = jun,
       volume = {182},
        pages = {L85},
          doi = {10.1086/181225},
       adsurl = {https://ui.adsabs.harvard.edu/abs/1973ApJ...182L..85K}
}

@ARTICLE{1992Natur.355..143M,
       author = {{Meegan}, C.~A. and {Fishman}, G.~J. and {Wilson}, R.~B. and {Paciesas}, W.~S. and {Pendleton}, G.~N. and {Horack}, J.~M. and {Brock}, M.~N. and {Kouveliotou}, C.},
        title = "{Spatial distribution of {\ensuremath{\gamma}}-ray bursts observed by BATSE}",
      journal = {\nat},
         year = 1992,
        month = jan,
       volume = {355},
       number = {6356},
        pages = {143-145},
          doi = {10.1038/355143a0},
       adsurl = {https://ui.adsabs.harvard.edu/abs/1992Natur.355..143M}
}

@ARTICLE{1997Natur.386..686V,
       author = {{van Paradijs}, J. and {Groot}, P.~J. and {Galama}, T. and {Kouveliotou}, C. and {Strom}, R.~G. and {Telting}, J. and {Rutten}, R.~G.~M. and {Fishman}, G.~J. and {Meegan}, C.~A. and {Pettini}, M. and {Tanvir}, N. and {Bloom}, J. and {Pedersen}, H. and {N{\o}rdgaard-Nielsen}, H.~U. and {Linden-V{\o}rnle}, M. and {Melnick}, J. and {Van der Steene}, G. and {Bremer}, M. and {Naber}, R. and {Heise}, J. and {in't Zand}, J. and {Costa}, E. and {Feroci}, M. and {Piro}, L. and {Frontera}, F. and {Zavattini}, G. and {Nicastro}, L. and {Palazzi}, E. and {Bennett}, K. and {Hanlon}, L. and {Parmar}, A.},
        title = "{Transient optical emission from the error box of the {\ensuremath{\gamma}}-ray burst of 28 February 1997}",
      journal = {\nat},
         year = 1997,
        month = apr,
       volume = {386},
       number = {6626},
        pages = {686-689},
          doi = {10.1038/386686a0},
       adsurl = {https://ui.adsabs.harvard.edu/abs/1997Natur.386..686V}
}

@Article{electronics13173509,
AUTHOR = {Goyal, Mandeep and Mahmoud, Qusay H.},
TITLE = {A Systematic Review of Synthetic Data Generation Techniques Using Generative AI},
JOURNAL = {Electronics},
VOLUME = {13},
YEAR = {2024},
NUMBER = {17},
ARTICLE-NUMBER = {3509},
URL = {https://www.mdpi.com/2079-9292/13/17/3509},
ISSN = {2079-9292},
DOI = {10.3390/electronics13173509}
}




\appendix

\section{Training curves}

\begin{figure*}
   \captionsetup{labelformat=empty}
    \centering
    \begin{subfigure}{0.50\textwidth}
        \centering
        \includegraphics[width=1.0\textwidth]{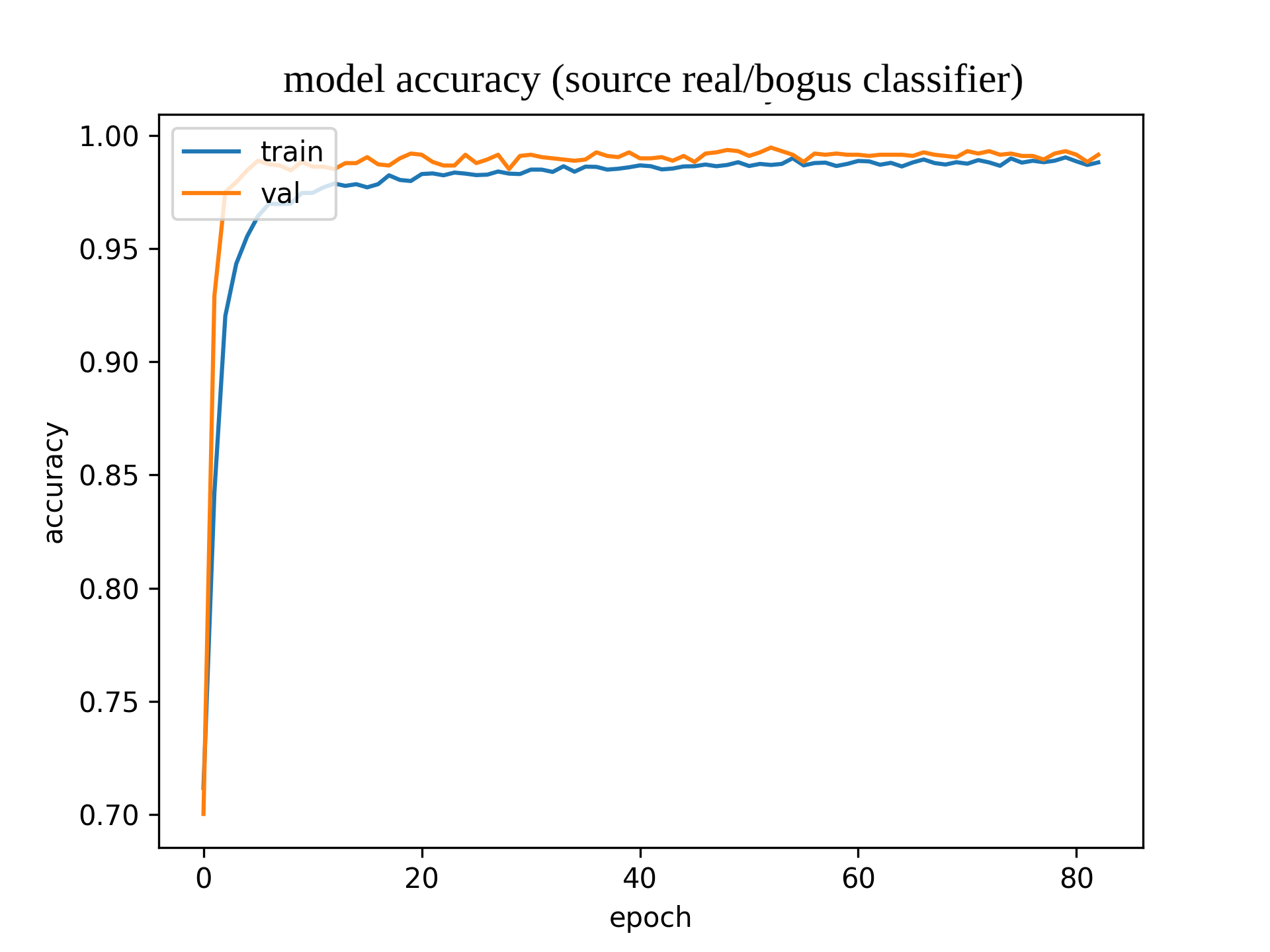} 
        \label{fig:subfig11}
    \end{subfigure}
    \hspace{-0.5cm}
    \begin{subfigure}{0.50\textwidth}
        \centering
        \includegraphics[width=1.0\textwidth]{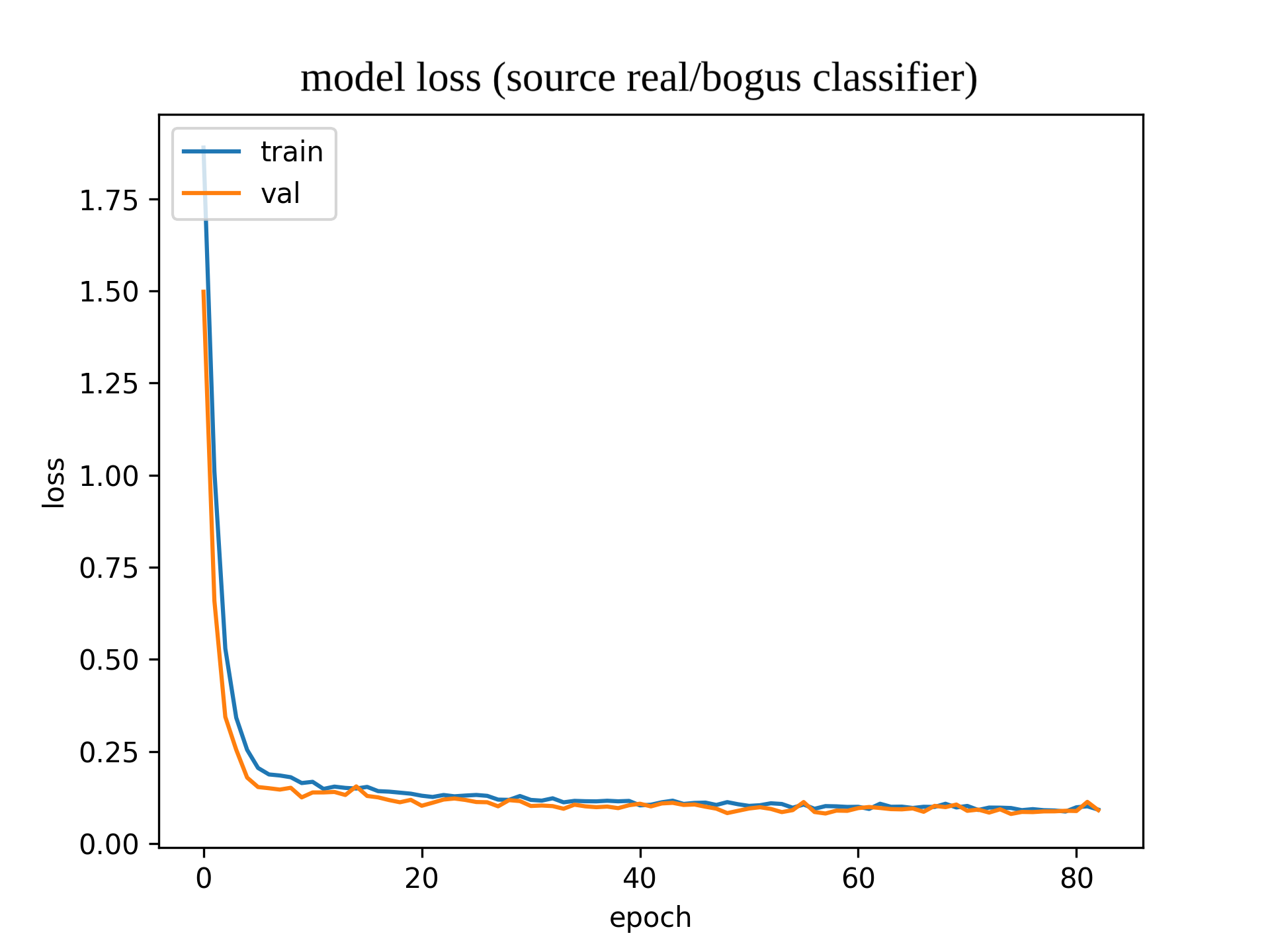} 
        \label{fig:subfig21}
    \end{subfigure}
    \begin{subfigure}{0.50\textwidth}
        \centering
        \includegraphics[width=1.0\textwidth]{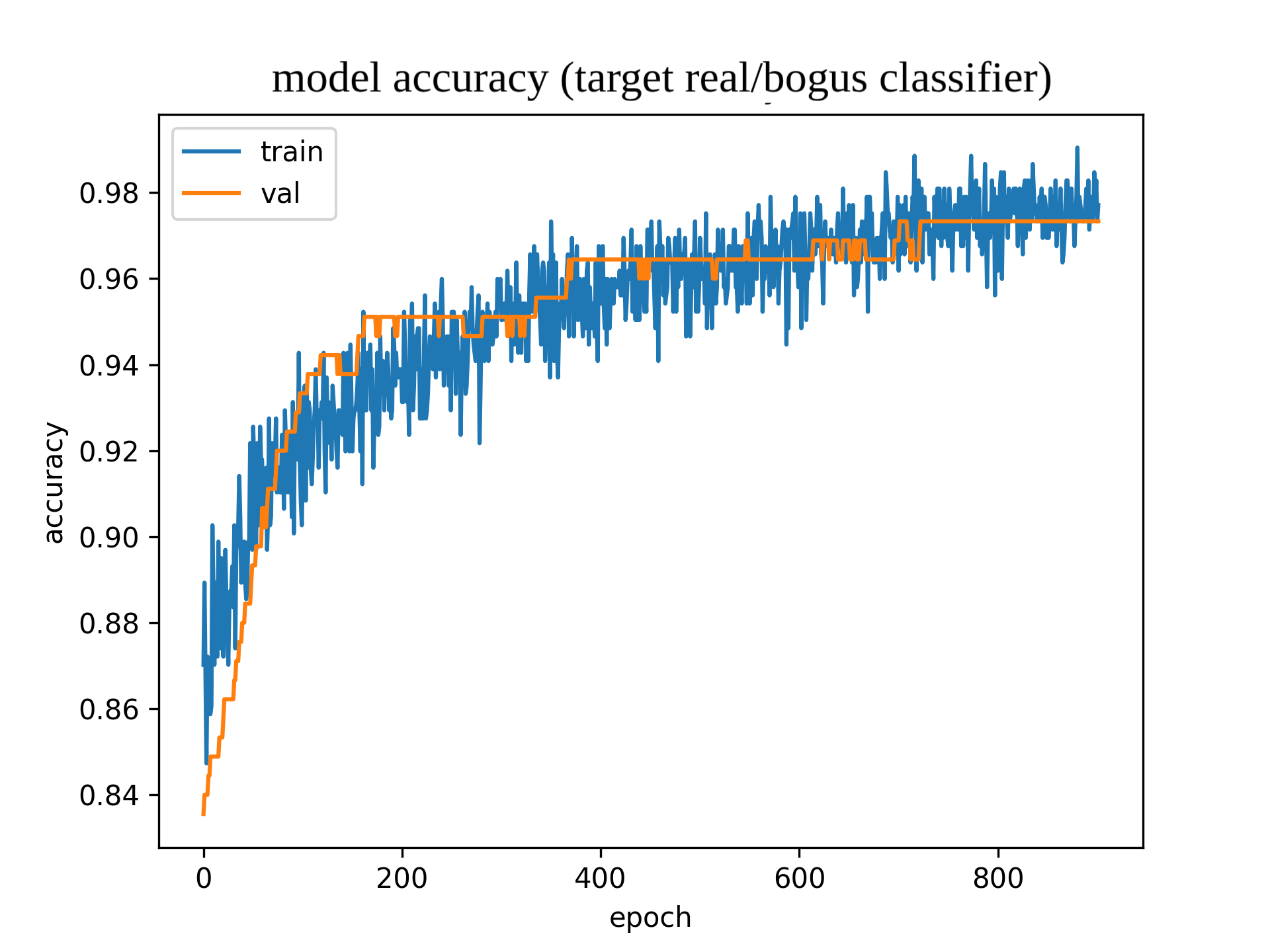} 
        \label{fig:subfig31}
    \end{subfigure}
    \hspace{-0.5cm}
    \begin{subfigure}{0.50\textwidth}
        \centering
        \includegraphics[width=1.0\textwidth]{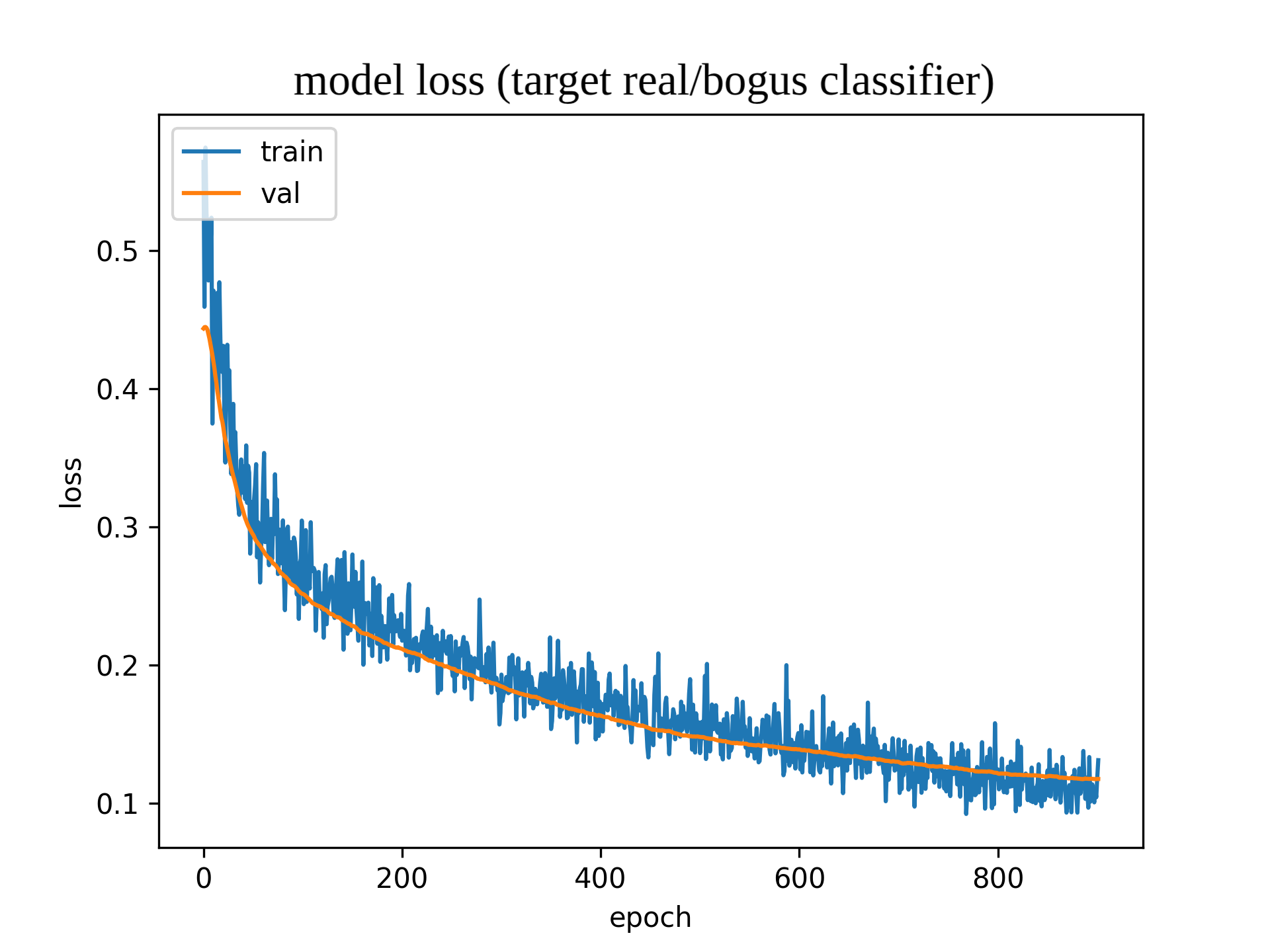} 
        \label{fig:subfig41}
    \end{subfigure}
    \centering
    \begin{subfigure}{0.50\textwidth}
        \centering
        \includegraphics[width=1.0\textwidth]{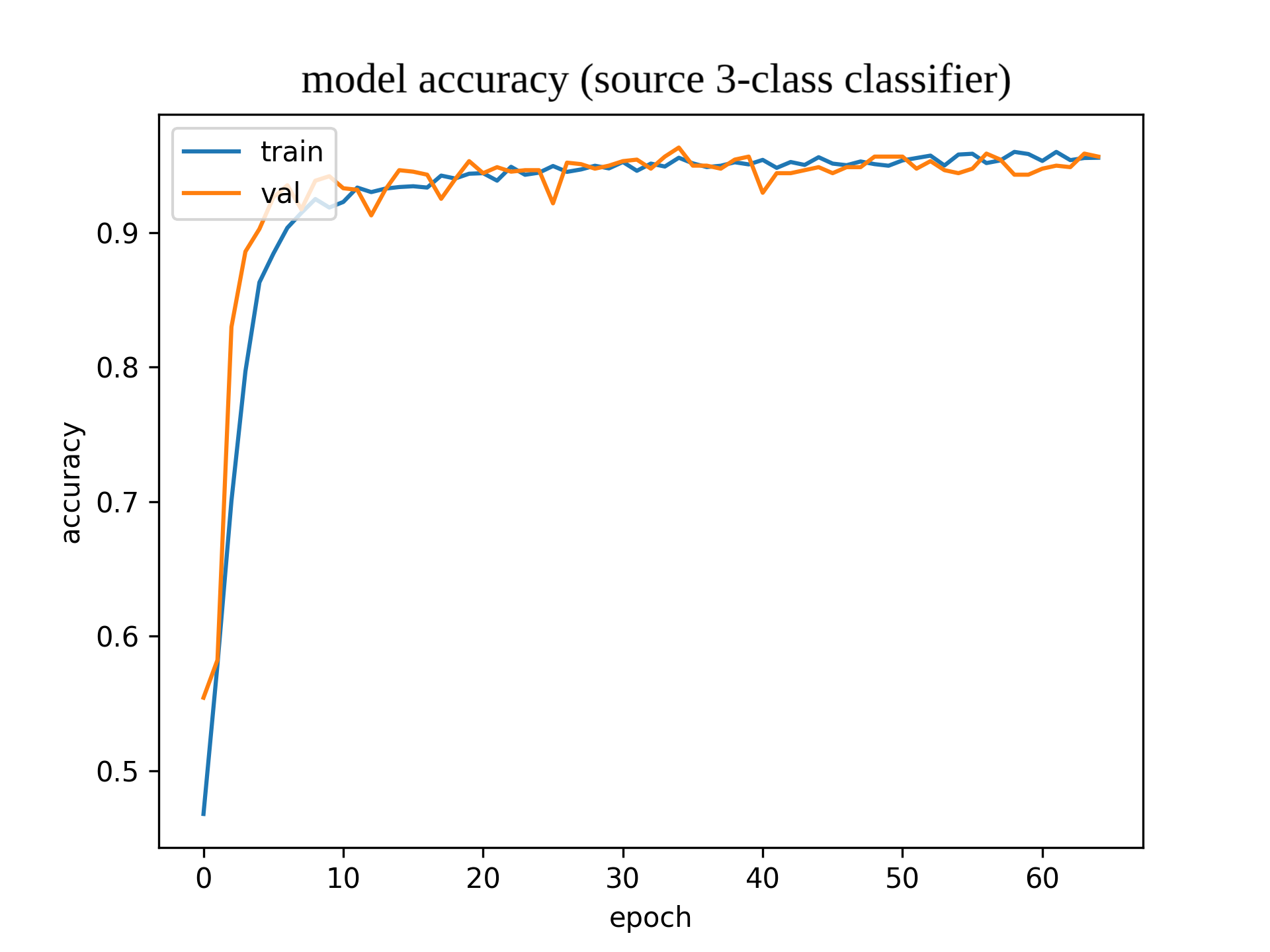} 
        \label{fig:subfig12}
    \end{subfigure}
    \hspace{-0.5cm}
    \begin{subfigure}{0.50\textwidth}
        \centering
        \includegraphics[width=1.0\textwidth]{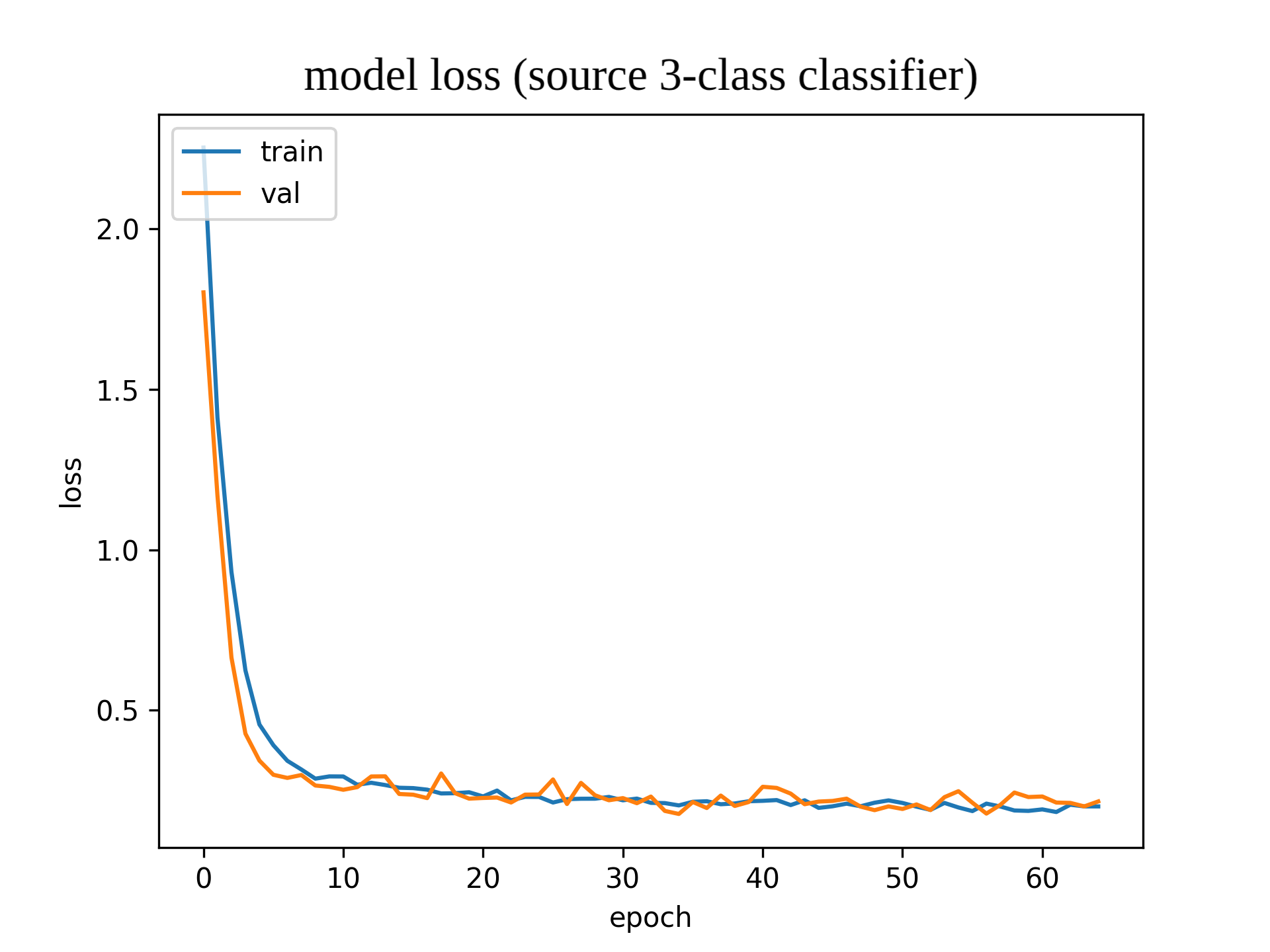} 
        \label{fig:subfig22}
    \end{subfigure}
    \caption*{} 
\end{figure*}

\setcounter{figure}{0}

\begin{figure*}
    \centering
    \begin{subfigure}{0.50\textwidth}
        \centering
        \includegraphics[width=1.0\textwidth]{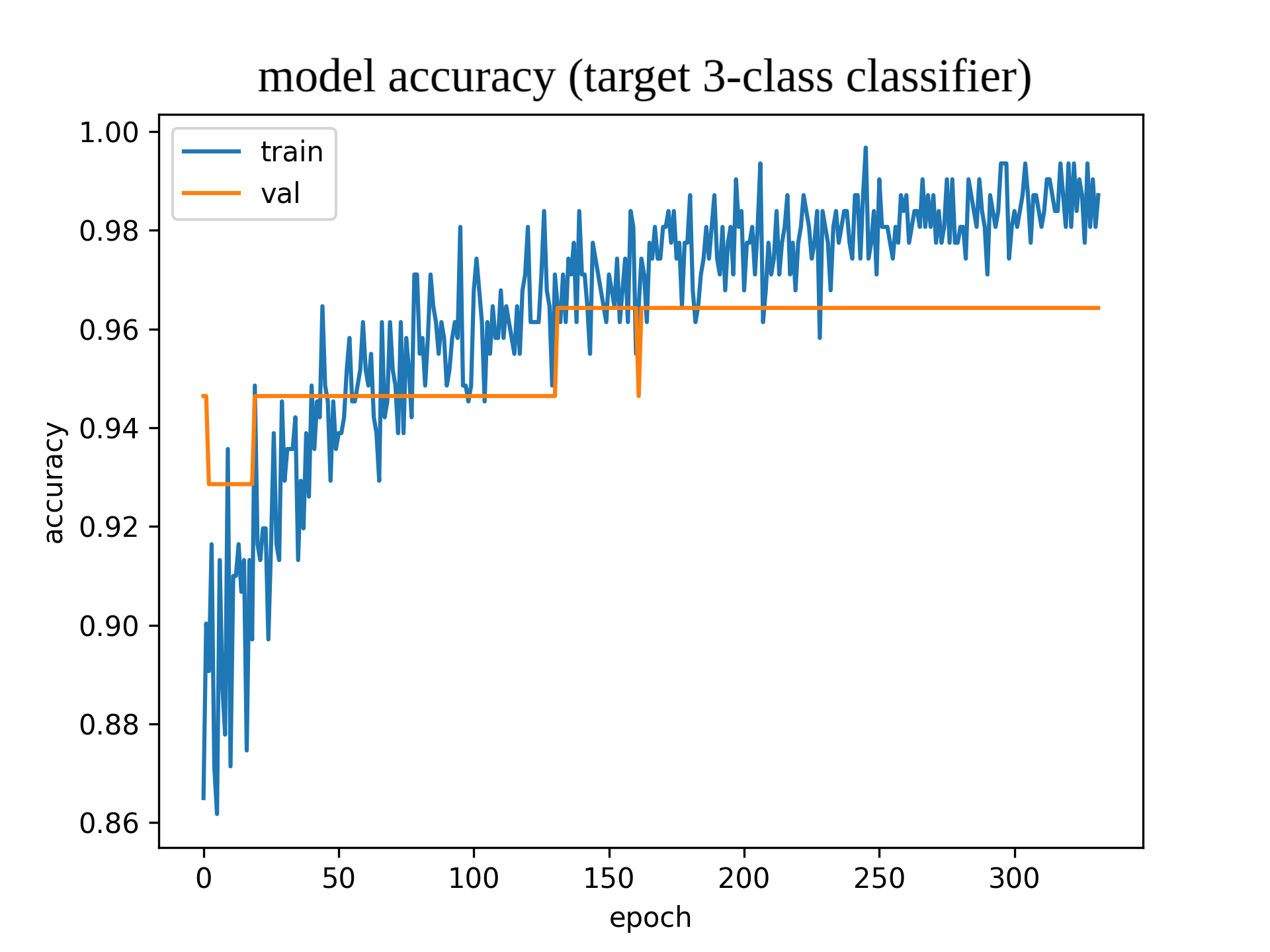} 
        \label{fig:subfig32}
    \end{subfigure}
    \hspace{-0.5cm}
    \begin{subfigure}{0.50\textwidth}
        \centering
        \includegraphics[width=1.0\textwidth]{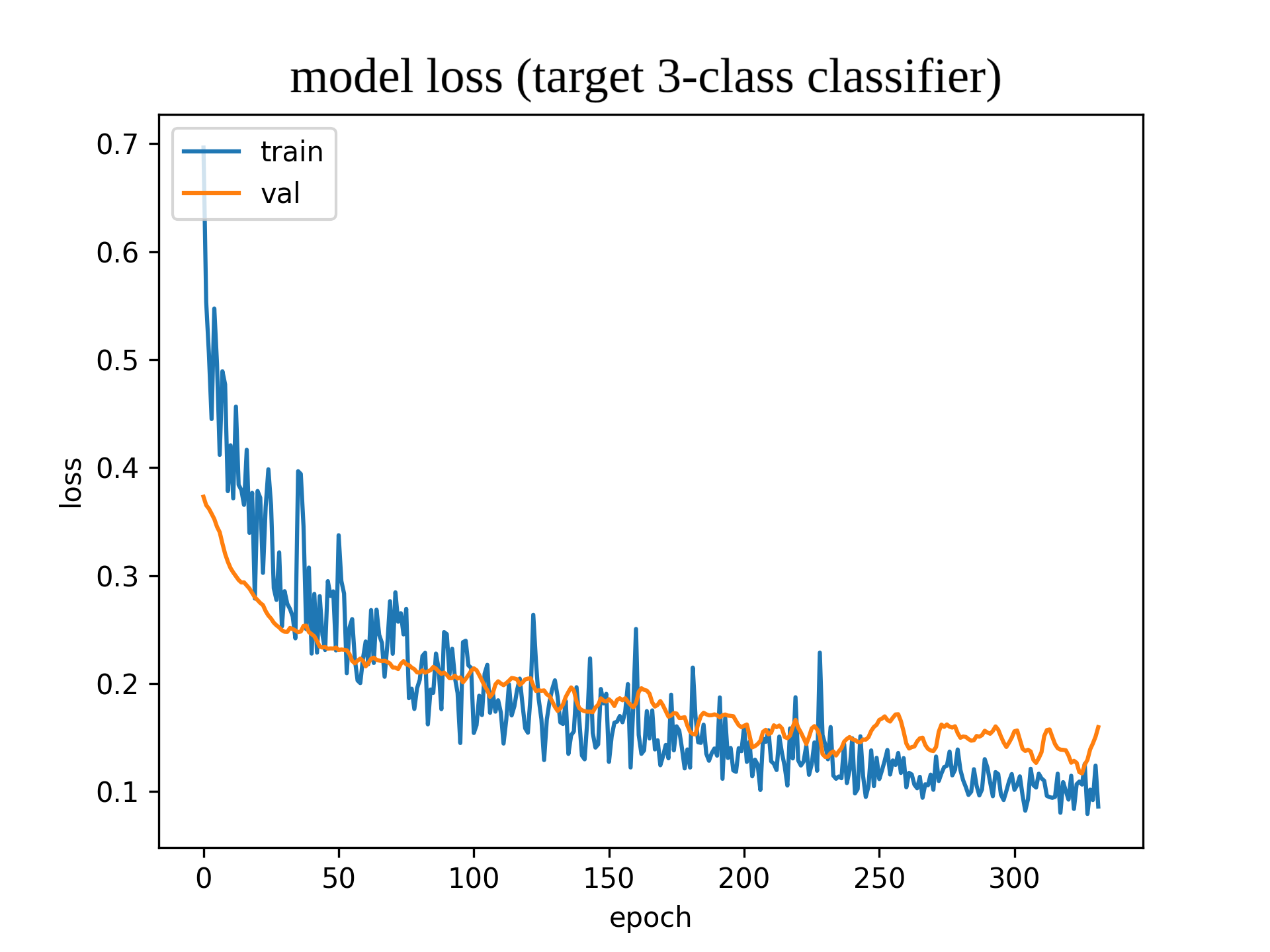} 
        \label{fig:subfig42}
    \end{subfigure}
    \centering
    \begin{subfigure}{0.50\textwidth}
        \centering
        \includegraphics[width=1.0\textwidth]{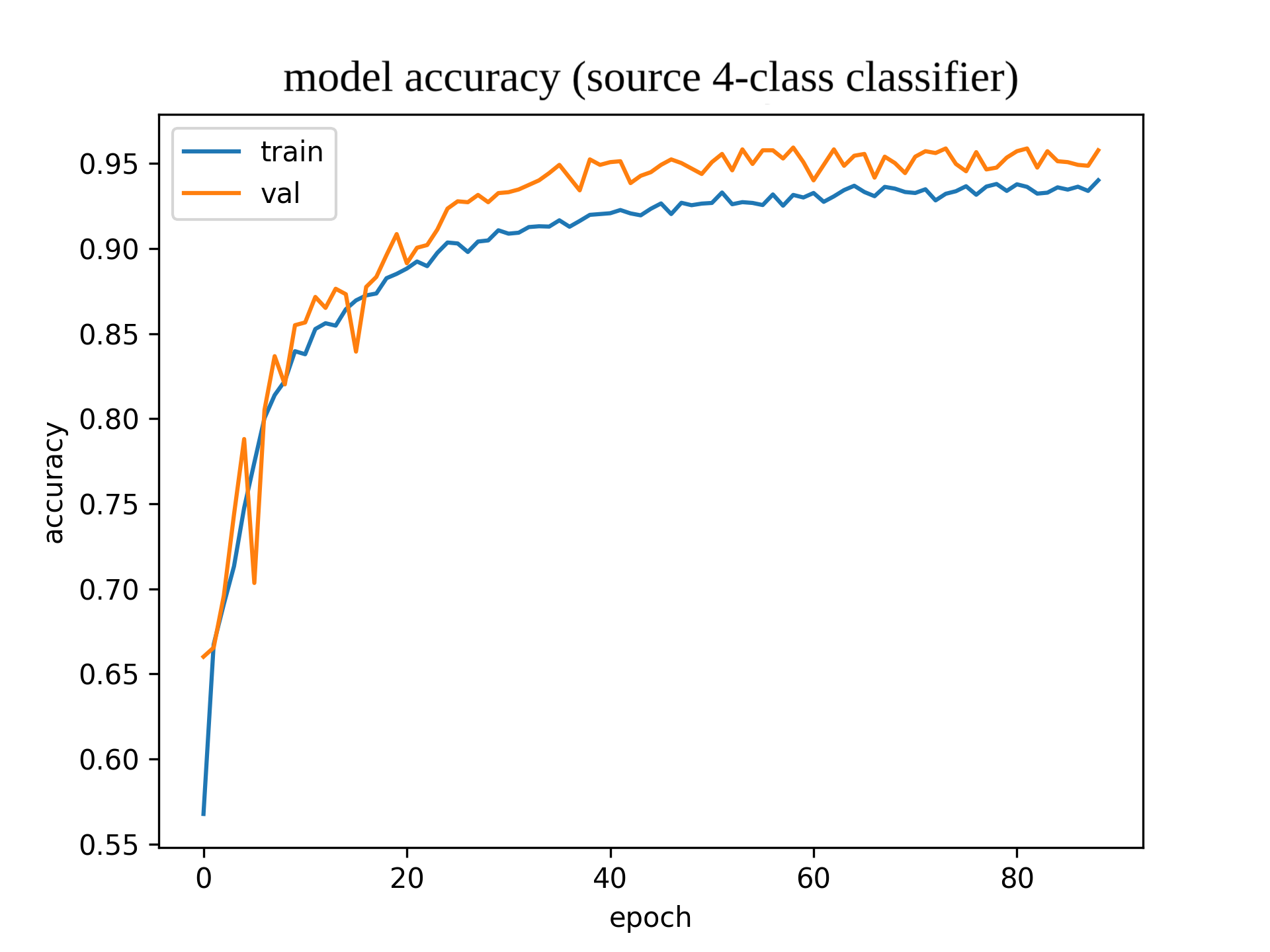} 
        \label{fig:subfig13}
    \end{subfigure}
    \hspace{-0.5cm}
    \begin{subfigure}{0.50\textwidth}
        \centering
        \includegraphics[width=1.0\textwidth]{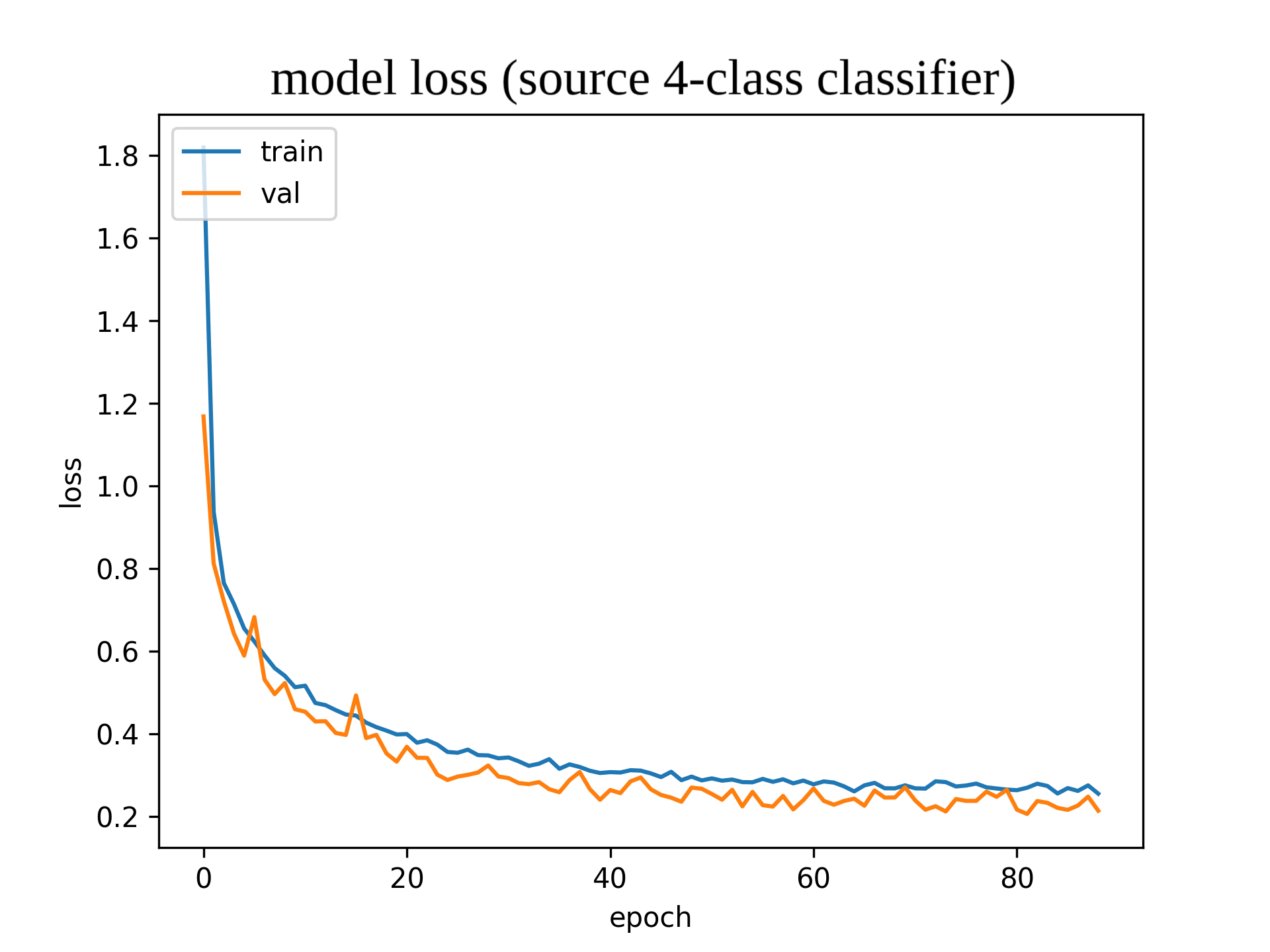} 
        \label{fig:subfig23}
    \end{subfigure}
    \centering
    \begin{subfigure}{0.50\textwidth}
        \centering
        \includegraphics[width=1.0\textwidth]{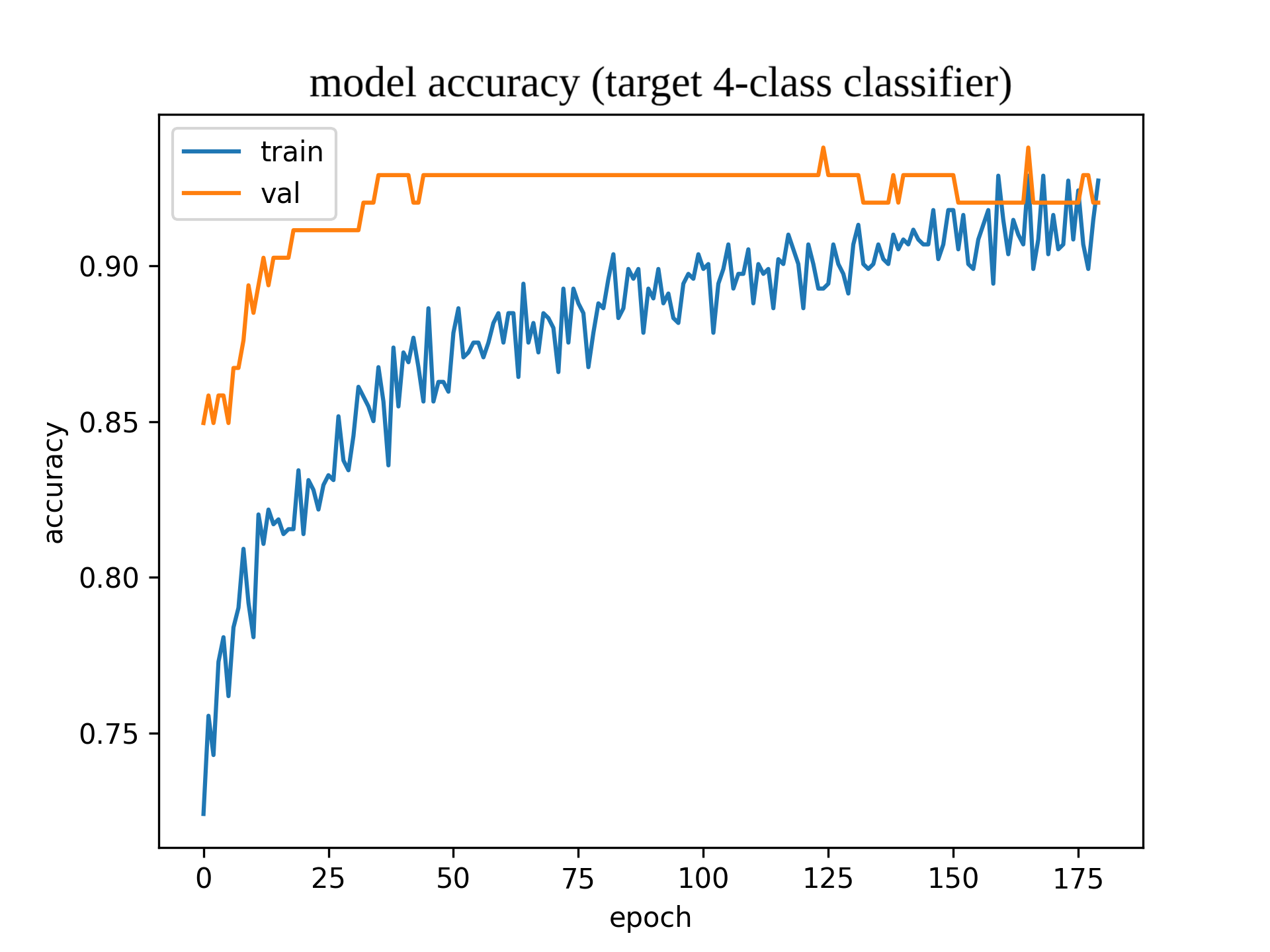} 
        \label{fig:subfig33}
    \end{subfigure}
    \hspace{-0.5cm}
    \begin{subfigure}{0.50\textwidth}
        \centering
        \includegraphics[width=1.0\textwidth]{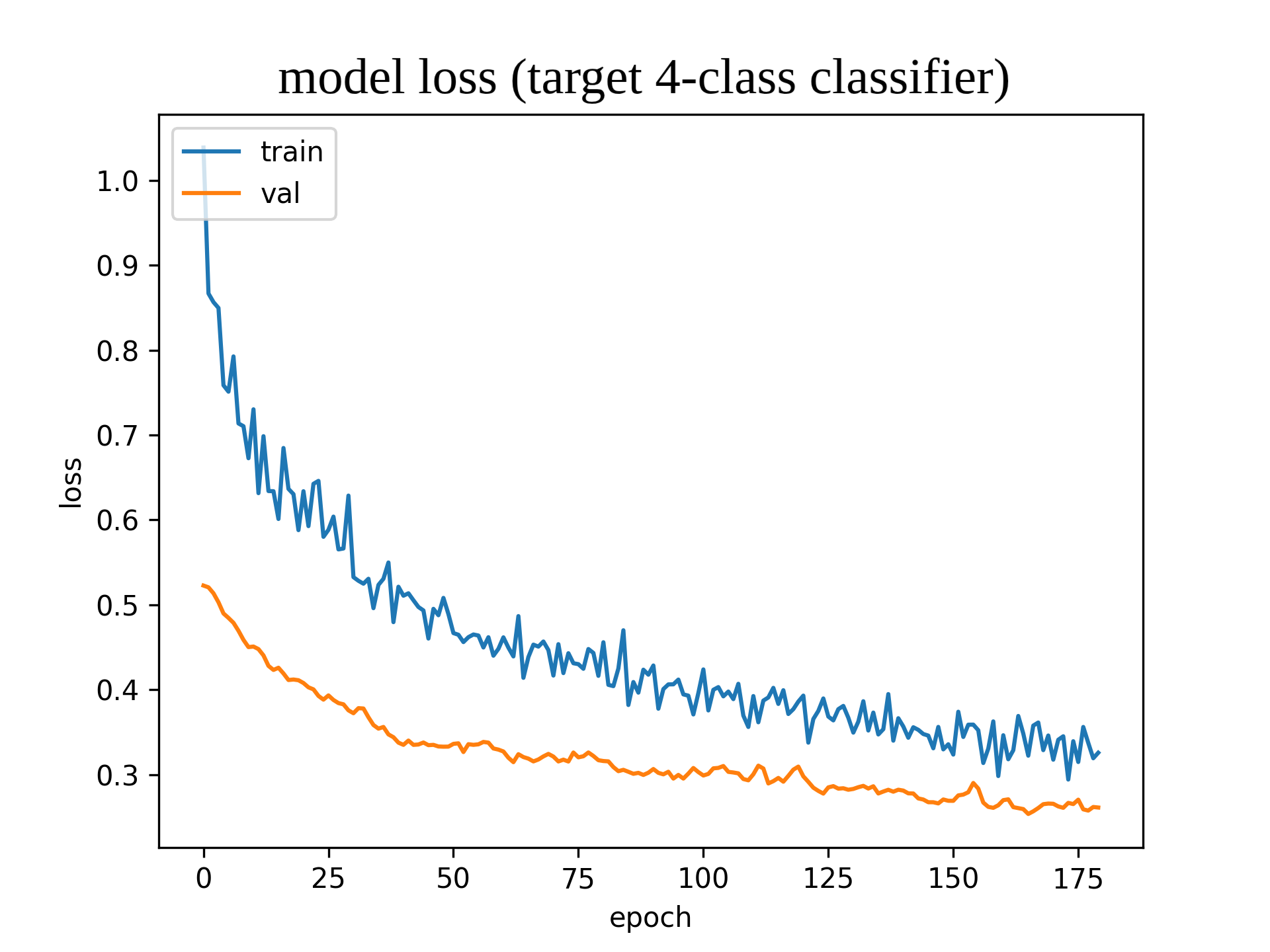} 
        \label{fig:subfig43}
    \end{subfigure}
    \caption{From top to bottom: accuracy (left) and loss (right) curves of source real/bogus classifier, target real/bogus classifier, source 3-class classifier, target 3-class classifier, source 4-class classifier, and target 4-class classifier.}
    \label{fig:CLF4_ACC_LOSS}
\end{figure*}




\bsp	
\label{lastpage}
\end{document}